\documentclass[11pt]{article}
\usepackage{epubtk}
\usepackage{latexsym}
\usepackage{graphicx}
\usepackage{amssymb,amsmath,amsthm,longtable}
\usepackage{epsfig}
\usepackage{epsf}
\usepackage{caption}
\usepackage{cancel}

\usepackage{aas_macros}
\usepackage{hyperref}
\usepackage{subfigure}

\usepackage[top=1in, bottom=1in, left=1in, right=1in]{geometry}

\numberwithin{equation}{section}

\def\be{\begin{equation}}
\def\ee{\end{equation}}
\def\bi{\begin{itemize}}
\def\ei{\end{itemize}}
\def\ben{\begin{enumerate}}
\def\een{\end{enumerate}}
\def\i{\item{}}

\newcommand{\mb}[1]{\mathbf{#1}}

\newcommand{\unit}{\mathsf{1}}

\def\D{{\rm d}}

\begin{document}

\title{Searches for stochastic gravitational-wave backgrounds}
\author{
\epubtkAuthorData{Joseph D.\ Romano}{%
Department of Physics\\
Texas Tech University, Lubbock, TX 79409}{%
joseph.d.romano@ttu.edu}{%
}
}
\date{Les Houches Summer School\\
July 2018}

\maketitle

\begin{abstract}
These lecture notes provide a brief introduction to methods used to
search for a stochastic background of gravitational radiation---a
superposition of gravitational-wave signals that are either too weak
or too numerous to individually detect.  The focus of these notes is
on relevant data analysis techniques, not on the particular
astrophysical or cosmological sources that are responsible for
producing the background.  The lecture notes are divided into two main
parts: (i) an overview, consisting of a description of different types
of gravitational-wave backgrounds and an introduction to the method of
cross-correlating data from multiple detectors, which can be used to
extract the signal from the noise; (ii) details, extending the
previous discussion to non-trivial detector response, non-trivial
overlap functions, and a recently proposed Bayesian method to search
for the gravitational-wave background produced by stellar-mass binary
black hole mergers throughout the universe.  Suggested exercises for
the reader are given throughout the text, and compiled in an appendix.
\end{abstract}

\pagebreak
\tableofcontents
\pagebreak


\part{Overview / Basics}
\label{p:overview}

In the first part of these lecture notes, we describe 
different types of stochastic gravitational-wave 
backgrounds and introduce the
method of cross-correlation for extracting the signal
from the noise.
Interested readers should see e.g., 
\cite{Allen:1997, Allen-Romano:1999, Romano-Cornish:2017}
for more details.

\section{Motivation}
\label{s:motivation}

A stochastic background of gravitational radiation 
is a superposition of gravitational-wave signals that
are either too weak or too numerous to individually detect.
The individual signals making up the background are thus
{\em unresolvable}, unlike the large signal-to-noise ratio
binary black-hole (BBH) and binary neutron-star (BNS)
merger signals that have been recently detected by the 
advanced LIGO and Virgo 
detectors~\cite{TheLIGOScientific:2016-GW150914,
TheLIGOScientific:2017-GW170817,TheLIGOScientific:2018-GWTC-1}.
But despite the fact that one cannot resolve the individual  
signals that comprise the background, the detection of a 
gravitational-wave background (GWB) will provide information 
about the 
{\em statistical} properties (or population properties)
of the source.

\subsection{Gravitational-wave analogue of the cosmic
microwave background}
\label{s:GW_CMB}

\begin{figure}[htbp!]
\begin{center}
\includegraphics[width=0.7\textwidth]{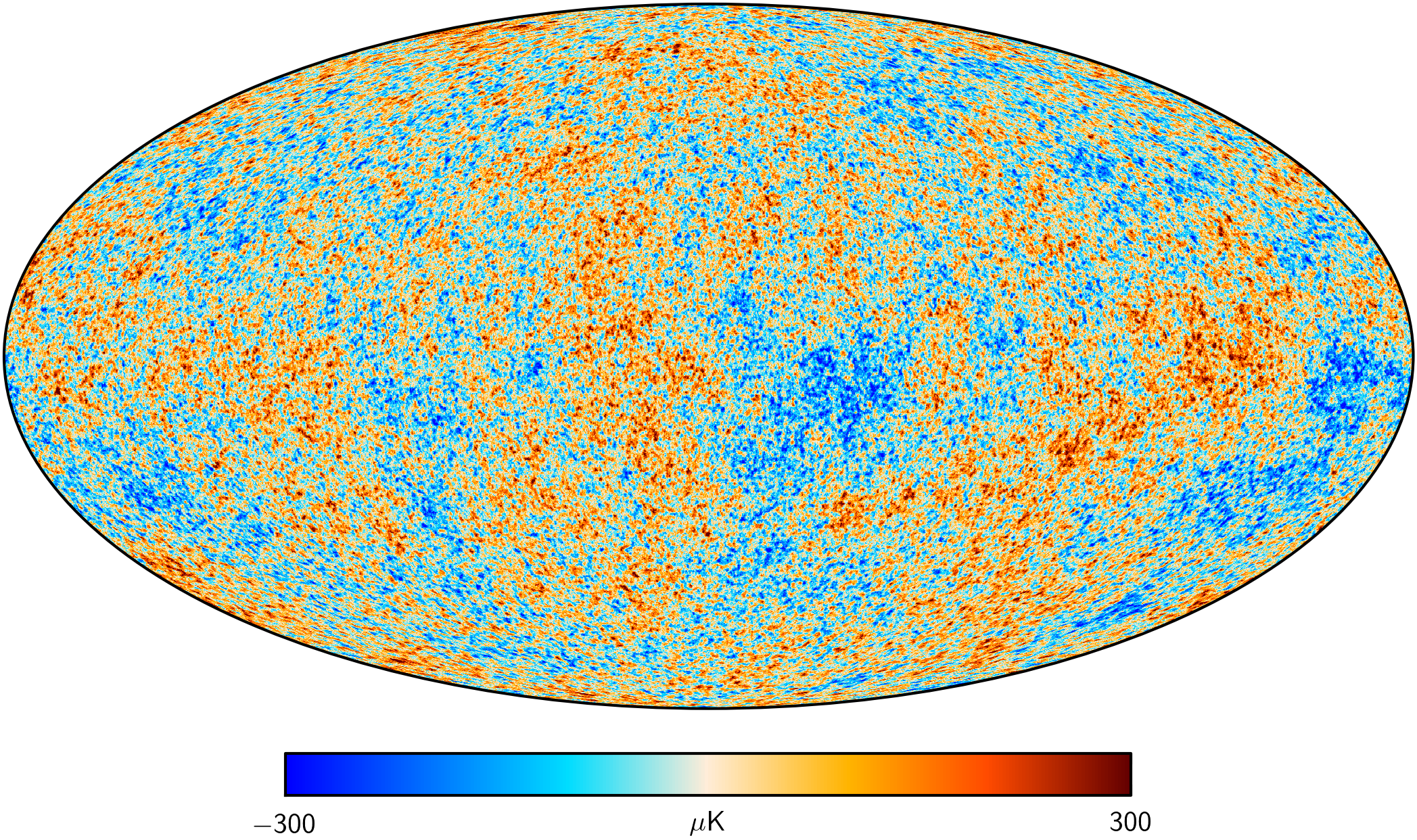}
\caption{Skymap of $\Delta T/T_0$ for the cosmic microwave background
radiation
(\url{https://www.cosmos.esa.int/documents/387566/425793/2015_SMICA_CMB/}).}
\label{f:CMB}
\end{center}
\end{figure}

The ultimate goal of GWB searches is to produce 
the GW analogue of Figure~\ref{f:CMB},
which is a sky map of the temperature fluctuations in 
the cosmic microwave background (CMB) 
blackbody radiation, $\Delta T/T_0$, relative 
to the $T_0 = 2.73~{\rm K}$ isotropic 
component~\cite{Penzias-Wilson:1965, Dicke-et-al:1965}.
(The dipole contribution due to our motion with respect 
to the cosmic rest frame has also been subtracted out.)
The CMB is a background of electromagnetic
radiation, produced roughly 380,000~yr after the Big 
Bang~\cite{Kolb-Turner:1999, Ryden:2003}.
At that time, the universe had a temperature of 
$\sim 3000~{\rm K}$, approximately one thousand times 
larger than the temperature today, but cool enough for 
neutral hydrogen atoms to first form and photons to 
propagate freely.
The temperature fluctuations in the CMB radiation tell
us about the density perturbations at the time of 
last scattering of photons, 
thus giving us a picture of the ``seeds" of 
large-scale structure formation in the early 
universe~\cite{COBE:1992, CMB-anisotropies-Planck-Planck:2019}.
Given the weakness of the gravitational interaction 
compared to the electromagnetic force, the GW analogue 
of Figure~\ref{f:CMB} would provide information about
a much earlier time in the evolution of the universe,
a mere fraction of a second after the Big 
Bang~\cite{Allen:1997, Maggiore:2000} 
(this is explained in a bit more detail in 
Section~\ref{s:different_sources}).  
Detecting the cosmological GWB is thus a ``holy grail" 
for GW astronomy.

For perspective, Figure~\ref{f:CMB} was produced by 
the Planck mission in 2015,
50 years after the CMB radiation was initially
detected by Penzias and Wilson~\cite{Penzias-Wilson:1965}
in 1965.
It took many years and improved experiments
(COBE~\cite{COBE:web}, Boomerang~\cite{boomerang:2002}, 
WMAP~\cite{WMAP:web}, and Planck~\cite{Planck:web} to name a few) 
to get to the high-precision measurements that we have today.
So it is somewhat sobering to realize that now---in the
summer of 2019---we have yet to detect even the isotropic 
component of the GWB.

\subsection{The background of BBH and BNS mergers in the 
LIGO band}
\label{s:BBH-BNS-LIGO}

Even though a detailed map of the primordial GWB is likely
to be out of reach for many years, there are other
sources of GWBs that are much more immediately accessible
to us.
For example, as mentioned above, the advanced LIGO
and Virgo detectors have detected other 
GW signals from several individual BBH 
and BNS mergers~\cite{TheLIGOScientific:2018-GWTC-1}.
These were very strong signals, having 
matched-filter signal-to-noise ratios (SNR) $\gtrsim 10$, 
and false alarm probabilities $<2\times 10^{-7}$,
corresponding to 5-sigma ``gold-plated" events.
Similar large-SNR detections are expected during the 
observing run O3, which started on 1 April 2019.
But we also expect that there are many more signals, 
corresponding to more distant mergers or smaller mass systems, 
which are 
individually undetectable (i.e., {\em subthreshold} events).
This weaker background of gravitational radiation is 
nonetheless detectable as a combined/aggregate signal
via the common influence of the component GWs on 
multiple detectors~\cite{TheLIGOScientific:2016-stochastic,
StochImplications:2018}.

\begin{figure}[htbp!]
\begin{center}
\includegraphics[width=0.65\textwidth]{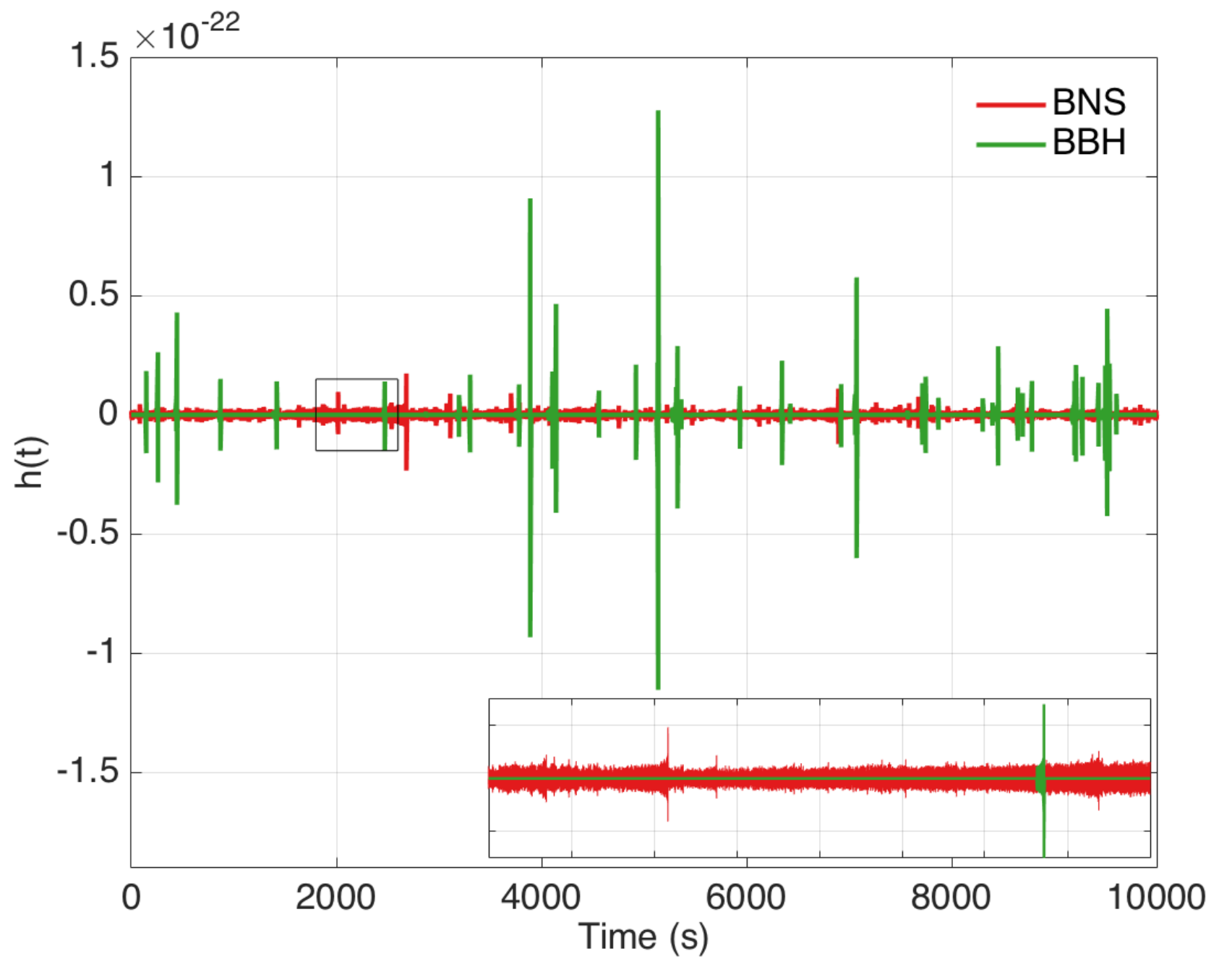}
\caption{Simulated time-domain signal for the predicted BBH and 
BNS backgrounds.  
(Figure taken from \cite{StochImplications:2018}.)}
\label{f:BBH-BNS-timeseries}
\end{center}
\end{figure}

To get an idea of the statistical properties of this
background signal, we can estimate the total rate of 
stellar-mass BBH mergers throughout the universe by
using the local rate estimate from these first detections,
$9$-$240~{\rm Gpc}^{-3}\,{\rm yr}^{-1}$~\cite{TheLIGOScientific:2016-O1-rates}.
This leads to a prediction for the total rate of 
BBH mergers between $\sim\!1$~per minute to a few per hour.
(You are asked in Exercise~\ref{exer:1} to verify
these predictions.)%
\footnote{A more complete description of this and 
all other exercises are given in 
Appendix~\ref{s:exercises}.
The number next to ``Exercise" is a link that brings 
you to the detailed exercise in Appendix~\ref{s:exercises}.}
Since the duration of BBH merger signals in band 
is $\sim\!1~{\rm s}$, which is much smaller than the 
average duration between successive mergers, the 
combined signal will consist of discrete bursts of 
radiation separated by periods of silence 
(i.e., it will be {\em popcorn}-like).
We can perform similar calculations for BNS mergers.
The predicted total rate for such events is roughly 
one event every $15~{\rm s}$, while the duration of 
a BNS signal in band is roughly 100~{\rm s}. 
Thus, the BNS merger signals overlap in time leading to 
a continuous (or {\em confusion-limited}) background. 
Figure~\ref{f:BBH-BNS-timeseries} is a plot of the
expected time-domain signal corresponding to the rate
estimates mentioned above.

\begin{figure}[htbp!]
\begin{center}
\includegraphics[width=0.65\textwidth]{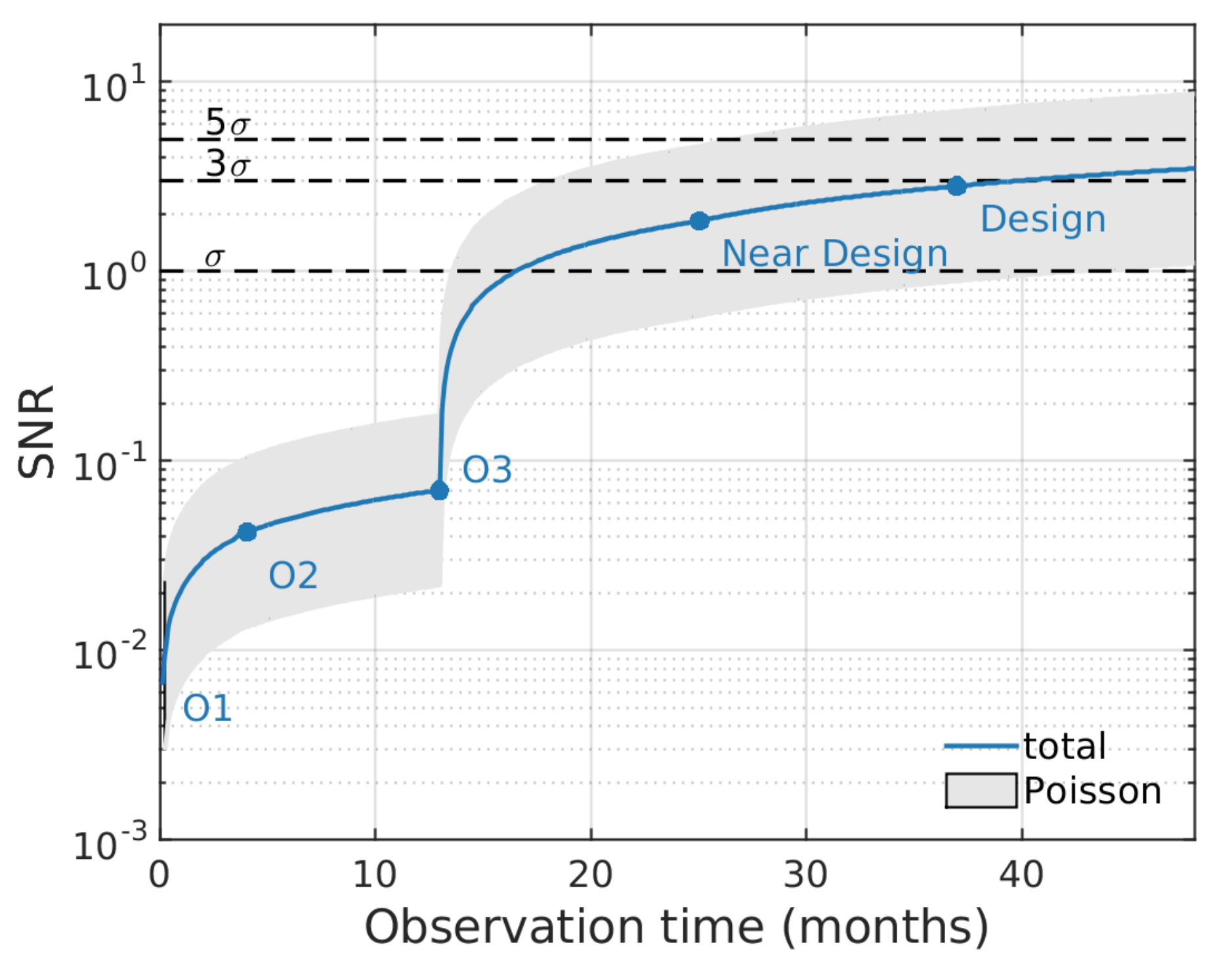}
\caption{Expected signal-to-noise ratio as a function of
observation time for the standard cross-correlation search 
using the advanced LIGO and Virgo detectors.
The points labeled O1, O2, etc., indicate the start of
advanced LIGO's first observation run, second observation
run, etc. 
(Figure taken from \cite{StochImplications:2018}.)}
\label{f:BBH-BNS-SNR}
\end{center}
\end{figure}

The combined signal from BBH and BNS mergers is 
potentially detectable with advanced LIGO and Virgo, 
shortly after reaching design sensitivity~\cite{StochImplications:2018}.
Although the signal-to-noise ratios for the 
individual events are small, the combined 
signal-to-noise ratio of the correlated data 
summed over all events grows like the square-root 
of the observation time, reaching a detectable
level of 3-$\sigma$ after roughly 40~months of
observation (Figure~\ref{f:BBH-BNS-SNR}).
This estimate of time to detection is based on 
the standard cross-correlation search (Section~\ref{s:correlations}), 
which assumes a Gaussian-stationary background.
But there is a new method~\cite{Smith-Thrane:2018}, 
recently proposed by Smith and Thrane,
which can potentially reduce the time to detection by several 
orders of magnitude (factor of $\sim\!1000$), 
meaning that the background would be detectable after only
a few days of operation. 
We will describe this method in more detail in 
Section~\ref{s:nonstationary}.

\section{Different types of stochastic backgrounds}
\label{s:different_types}

\subsection{Different sources}
\label{s:different_sources}

\begin{figure}[htbp!]
\begin{center}
\includegraphics[width=\textwidth]{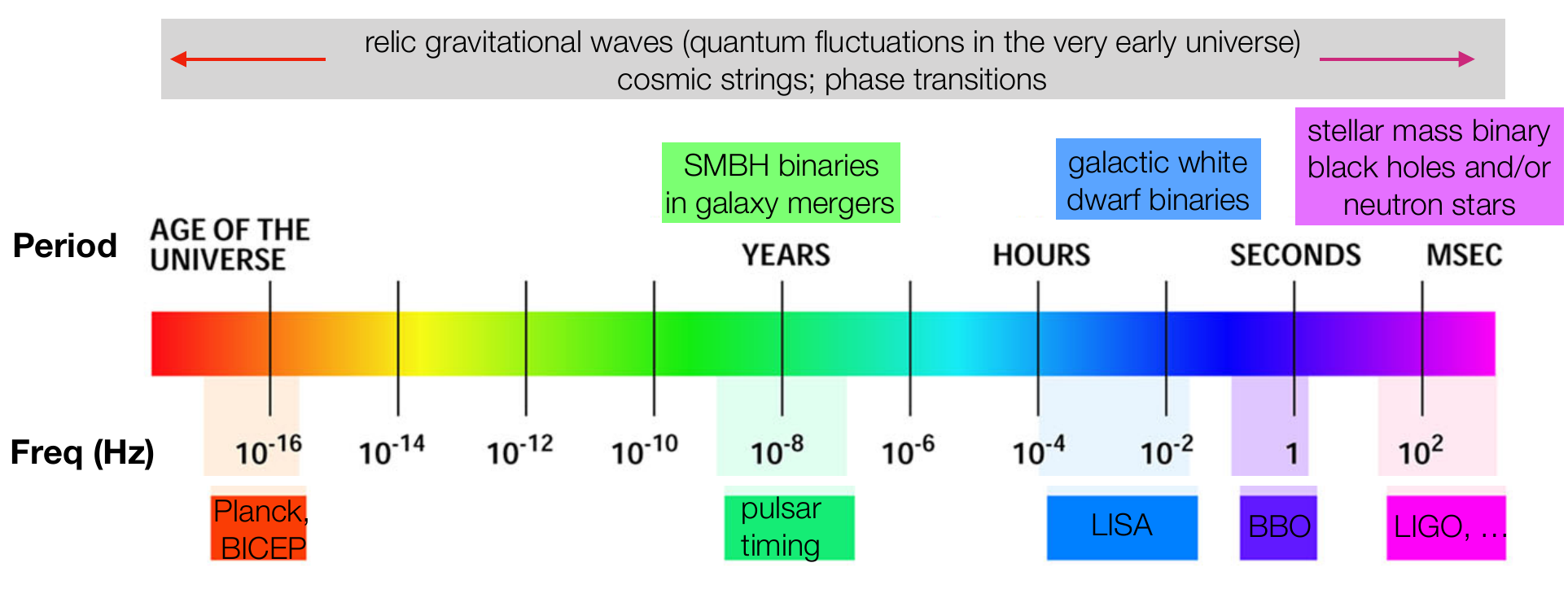}
\caption{Detectors and potential sources of GWBs
across the GW spectrum.
Note that the GWB signal from cosmic strings and phase 
transitions stretch across a broad range of frequencies, 
and peak at basically any frequency depending on 
the parameters that define the string network and the 
energy scale where the phase transition occurs.
Also, the primordial background of relic GWs predicted 
by standard inflation is flat across the whole frequency 
band shown here.
(Figure adapted from \cite{Romano-Cornish:2017}).}
\label{f:GWspectrum}
\end{center}
\end{figure}

The combined signal from stellar-mass BBH and BNS 
mergers throughout the universe is just one way 
of producing a GWB.
Due to the relatively small masses of stellar-mass 
BHs and NSs, the signal is at the high-frequency 
end of the spectrum ($\sim\!10~{\rm Hz}$ to a few kHz), 
which is the sensitive band for the current generation 
of km-scale ground-based laser interferometers like 
LIGO and Virgo.
Heavier-mass systems, which produce lower-frequency 
gravitational waves, are also expected to give rise 
to GWBs that are potentially detectable with other 
existing or proposed detectors.
Figure~\ref{f:GWspectrum} is a plot of the GW 
spectrum, with frequencies ranging
from a few kHz (for ground-based detectors)
to $10^{-17}~{\rm Hz}$ (corresponding to a period
equal to the age of the universe), together with 
potential sources of GWBs and relevant detectors.  

Of particular note is the combined GW 
signal produced by compact white-dwarf binaries in 
the Milky Way, producing a ``confusion-limited" GWB
in the frequency band $\sim 10^{-4}~{\rm Hz}$ to 
$10^{-1}~{\rm Hz}$~\cite{Bender-Hils:1997}.
This is a guaranteed signal for the proposed 
space-based interferometer LISA (expected launch date 2034), 
which consists of three spacecraft in an 
equilaterial-triangle configuration 
in orbit around the Sun~\cite{LISA:web}.
Each spacecraft houses two lasers, two telescopes, and 
two test masses; 
the arms will be several million km long.
The confusion-limited 
white-dwarf binary signal is expected to be 
so strong that it will
dominate the instrumental noise at low frequencies, 
forming a GW ``foreground" that will
have to be contended with when searching for other 
gravitational sources in the LISA band~\cite{Adams-Cornish:2014}.

At lower frequencies between 
$\sim\!10^{-9}~{\rm Hz}$ and $10^{-7}~{\rm Hz}$
(corresponding to periods of order decades to years), 
pulsar timing arrays can be used to 
search for the GWB produced by the inspiral and merger 
of supermassive black-holes (SMBHs) in the centers of
merging galaxies.
A pulsar timing array basically functions as a 
galactic-scale gravitational-wave detector, with the
radio pulses emitted by each pulsar behaving like 
`ticks' of an extremely stable clock.
By carefully monitoring the arrival times of these
radio pulses, one can search for a GWB by looking for 
correlated modulations in the arrival times 
induced by a passing gravitational wave~\cite{Sazhin:1978, 
Detweiler:1979, Hellings-Downs:1983}.

In addition to these {\em astrophysical} GWBs 
associated with stellar-mass or supermassive BHs 
and NSs, one also expects backgrounds of 
{\em cosmological} origin, produced in the 
very early universe~\cite{Grishchuk:1976}, 
much before the formation of stars and galaxies.
Two examples, indicated in Figure~\ref{f:GWspectrum}, 
are cosmic strings (line-like topological defects 
associated with 
phase transistions in the early universe) and
relic gravitational waves (quantum 
fluctuations in the geometry of space-time, 
driven to macroscopic scales by a period of rapid 
expansion---e.g., inflation---a mere 
$\sim\!10^{-32}~{\rm s}$ after the Big Bang); 
see, e.g., \cite{Allen:1997, Maggiore:2000}
for a discussion of these sources.
This relic background is potentially detectable 
by its effect on the polarization of the CMB 
radiation~\cite{Seljak:1996gy}.
This signal has been searched for by CMB experiments
such as Planck and BICEP~\cite{BICEP/Keck:web}, 
and is a core target of many
proposed future experiments, such as PIXIE~\cite{pixie:2011} 
and LiteBIRD~\cite{litebird:web}.

\subsection{Different signal properties}
\label{s:different_signal_properties}

Not surprisingly, different sources of a GWB give
rise, in general, to different properties of the 
observed signal.
These differences are what will allow us to infer 
the source of the background from the measured signal.

(i) Stochastic backgrounds can differ from one another 
in terms of the angular distribution of 
GW power on the sky.
Cosmologically-generated backgrounds, like those from 
cosmic strings or relic GWs,
are expected to be {\em statistically isotropic},
qualitatively similar to the CMB (Figure~\ref{f:CMB}).
The GW power in these backgrounds is {\em anisotropic}, 
following the spatial distribution of the particular
sources that produced it, but has no preferred 
direction when averaged over different realizations 
of the sources.
Different statistically isotropic backgrounds will
be characterized by different angular power spectra,
$C_l$ as a function of multipole moment $l$, where~\cite{Ryden:2003}
\be
C(\theta) = \sum_{l=0}^\infty \frac{2l+1}{4\pi} 
C_l P_l(\cos\theta)\,,
\label{e:legendre_series}
\ee
is the angular correlation between the GW power 
coming from two directions $\hat n$ and $\hat n'$
separated by angle $\theta$.
If all of the $C_l$'s except the monopole, $C_0$, 
are equal to zero, then the GWB is said to be 
``exactly" isotropic.
Exact isotropy is the simplest mathematical model 
for stochastic backgrounds, and will be discussed
further in Section~\ref{s:ensemble_averages}.

\begin{figure}[htbp!]
\begin{center}
\includegraphics[width=\textwidth]{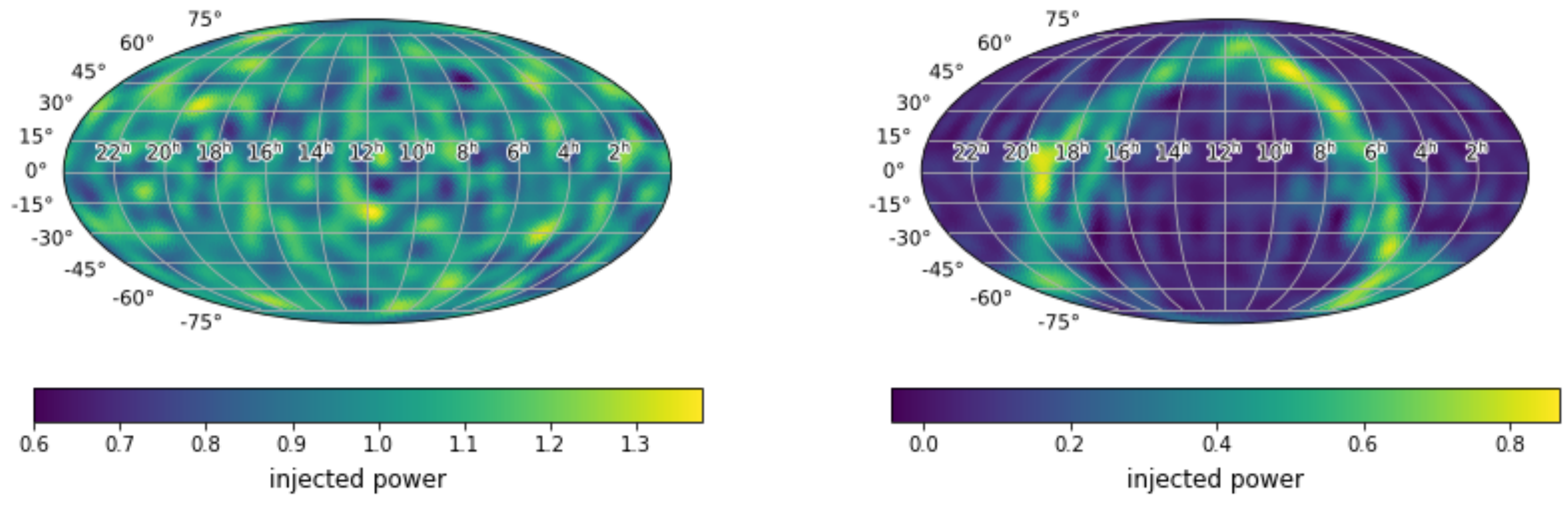}
\caption{Simulated sky maps of GW power
for a statistically isotropic background (left panel) and an 
anisotropic background (right panel).}
\label{f:statiso-vs-aniso}
\end{center}
\end{figure}

Statistically isotropic backgrounds are to be contrasted
with {\em statistically anisotropic} backgrounds, 
whose distribution of power on the sky has preferred 
directions, even when averaged over different 
realizations of the sources that produce it.
For example, the ``confusion-limited" foreground that 
LISA will see from the population of close white-dwarf 
binaries in the Milky Way will have its GW
power concentrated in the direction of the Milky Way.
Figure~\ref{f:statiso-vs-aniso} shows simulated skymaps 
for a statistically isotropic background (left panel) and 
a statistically anisotropic background (right panel). 
The anisotropic background in that figure follows the
galactic plane in equatorial coordinates. 

\begin{figure}[htbp!]
\begin{center}
\includegraphics[width=0.6\textwidth]{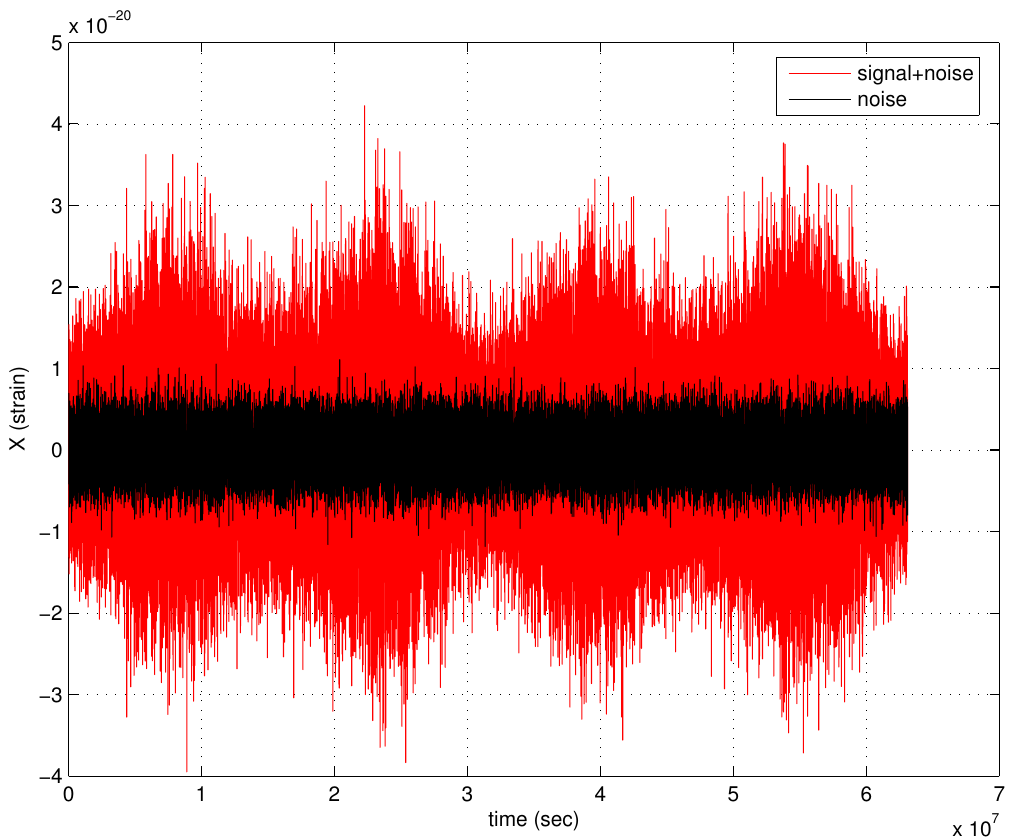}
\caption{Simulated time-domain output of a particular 
combination of the LISA data over a 2-year period.
The modulation of the signal with a 6-month period
is apparent in the data.
(Figure taken from \cite{Romano-Cornish:2017}.)}
\label{f:cyclostationary_data}
\end{center}
\end{figure}

(ii) Stochastic backgrounds can also differ from one 
another in temporal distribution and amplitude.
We have already seen examples of this in 
Figure~\ref{f:BBH-BNS-timeseries}, for the expected
backgrounds from stellar-mass BBH mergers and 
BNS mergers throughout the universe (a LIGO source).
As mentioned earlier, the rate estimates and 
durations of these individual merger signals are such 
that the BBH background is expected to be popcorn-like 
(consisting of non-overlapping mergers), 
while that for the BNS background is expected to be
stationary and confusion-limited 
(consisting of several overlapping BNS mergers at any
instant of time).
Another example of non-trivial temporal dependence
is the confusion-limited signal from close 
white-dwarf binaries in the Milky Way (a LISA source).
This is an amplitude-modulated signal with 
a 6-month period (Figure~\ref{f:cyclostationary_data}), 
due to LISA's ``cartwheeling" orbital motion around 
the Sun.
(The antenna pattern of LISA will point in the direction
of the Galactic center twice every year.) 
From the figure, we also see that the expected 
white-dwarf binary signal will be larger than that 
of the instrumental noise for LISA, thus constituting 
an astrophysical {\em foreground}.
This is atypical, however, as most expected GWBs will sit 
below the instrumental noise (e.g., for advanced LIGO / Virgo,
pulsar timing, CMB polarization experiments), requiring 
observation over long periods of time to confidently detect.

\begin{figure}[htbp!]
\begin{center}
\includegraphics[width=\textwidth]{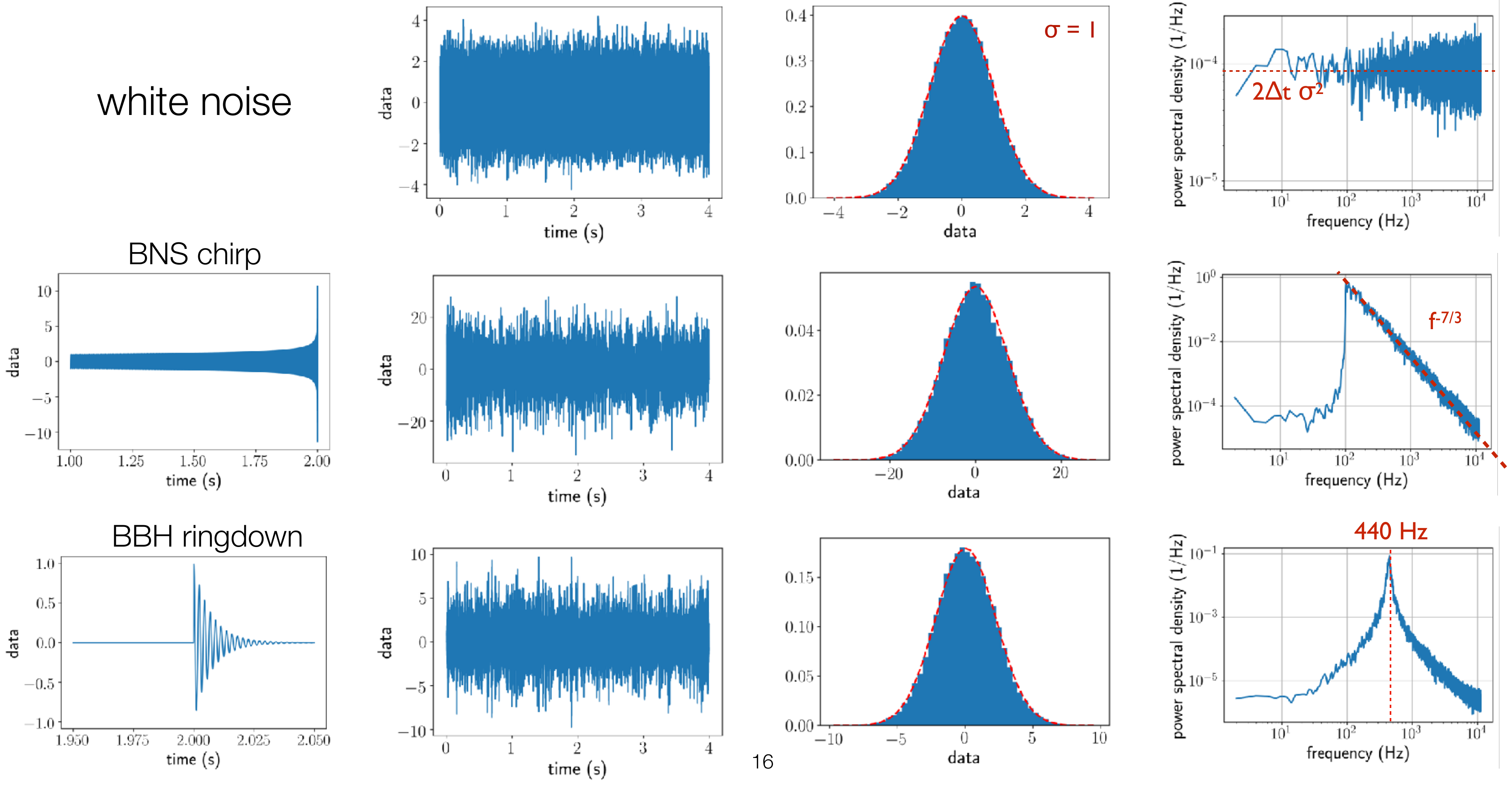}
\caption{Simulated time-domain data (including the signals for an
individual BNS merger and BBH ringdown), histograms, and power spectra
for three different types of Gaussian-stationary GWBs.}
\label{f:different_power_spectra}
\end{center}
\end{figure}

(iii) Stochastic backgrounds can also differ in their power 
spectra%
\footnote{If $x(t)$ is stationary time-domain data, then the 
power spectrum $P_x(f)$ is defined as the Fourier 
transform of the correlation function 
$C(t-t') \equiv \langle x(t)x(t')\rangle$, or,
equivalently, $\langle \tilde x(f)\tilde x^*(f')\rangle =
\frac{1}{2}P_x(f)\,\delta(f-f')$, where $\tilde x(f)$ is the Fourier
transform of $x(t)$.
The factor of $1/2$ is needed for a {\em one-sided} power spectrum;
see also \eqref{e:Gamma_def_freq}.}
as shown in Figure~\ref{f:different_power_spectra}.
Here we plot simulated time-domain data (including the signals for an
individual BNS merger and BBH ringdown%
\footnote{Our toy-model simulation for BBH ringdown is simply a 
damped sinusoid with frequency 440~Hz.
It has the correct qualitative behavior for a BBH ringdown, but 
is not meant to be astrophysically realistic.}), 
histograms, and power spectra
for three different types of GWBs.
For these toy-model simulations, we overlapped a sufficient number of 
individual BNS merger and BBH ringdown signals to produce 
Gaussian-stationary confusion-limited GWBs
(second and third columns).
The difference between these backgrounds shows up in their power 
spectra (fourth column).
The power spectra for the BNS merger and BBH ringdown backgrounds 
have the same shape as those for an individual BNS merger or 
BBH ringdown, scaled by the total number of sources contributing 
to the background.
%

\section{Mathematical characterization of a stochastic background}
\label{s:mathematical_characterization}

Since the individual signals comprising a GWB 
background are either too weak or too numerous to 
individually detect, the combined signal for the 
background is for all practical purposes 
{\em random}, similar to noise in an single detector.
Hence, we need to describe the GWB {\em statistically}, 
in terms of moments 
(i.e., ensemble averages) of the metric perturbations 
describing the GWB.

\subsection{Plane-wave expansion}
\label{s:plane_wave}

\begin{figure}[htbp!]
\begin{center}
\includegraphics[width=0.4\textwidth]{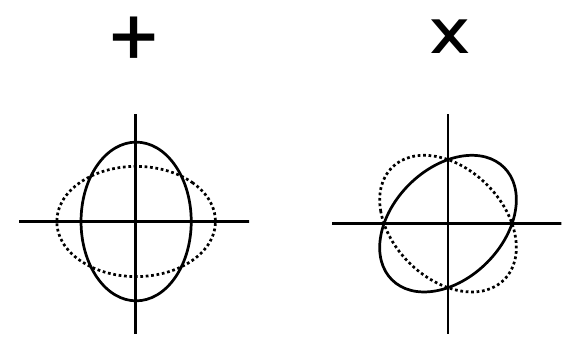}
\caption{The two orthogonal polarizations of a gravitational wave.
A circular ring of test particles in the plane orthogonal to 
the direction of propagation of the wave are alternately deformed
into ellipses, as space is ``squeezed" and ``stretched" by the 
passing of the wave.}
\label{f:polarizations}
\end{center}
\end{figure}

Recall that gravitational waves are time-varying 
perturbations to the geometry of space-time, 
which propagate away from the source at the speed 
of light~\cite{MTW:1973, Hartle:2002}.
In transverse-traceless coordinates
$(t,\vec x)\equiv (t,x^a)$, where $a=1,2,3$,
the metric perturbations corresponding to a 
plane wave (propagating in direction $\hat k\equiv-\hat n$) 
have two degrees of freedom, corresponding to the 
amplitudes of the 
plus ($+$) and cross ($\times$) polarizations of
the gravitational wave (Figure~\ref{f:polarizations}).
The metric perturbation for the most general GWB 
can thus be written as a superposition of such 
waves:
\be
h_{ab}(t,\vec x) =
\int_{-\infty}^\infty \D f\>
\int \D^2\Omega_{\hat k}\>
\sum_{A=+,\times}
h_A(f,\hat k)e^A_{ab}(\hat k) 
e^{i2\pi f(t-\hat k\cdot \vec x/c)}\,,
\label{e:planewave}
\ee
where $f$ denotes the frequency of the 
component waves, $\hat k$ their direction of
propagation, and $A=+,\times$ their polarization.
(The direction to a particular GW source is given 
by $\hat n=-\hat k$.)
\begin{figure}[htbp!]
\begin{center}
\includegraphics[width=0.4\textwidth]{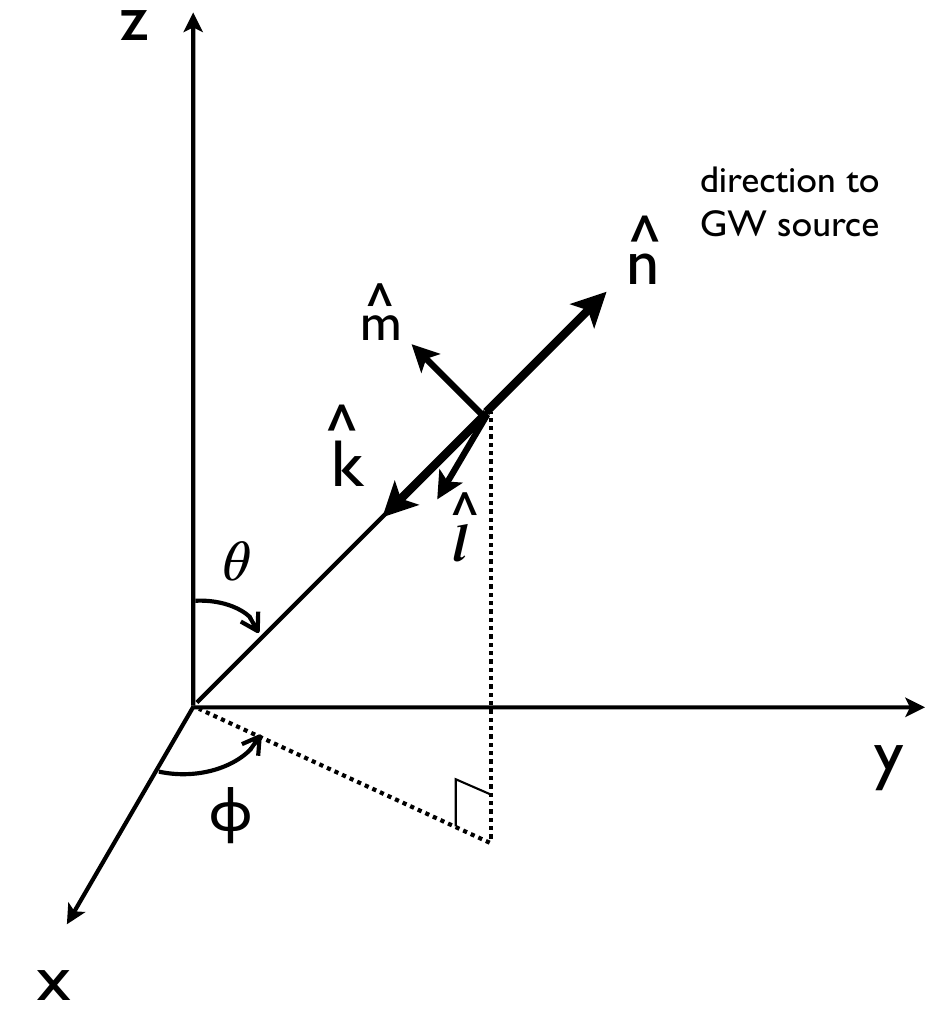}
\caption{Coordinate system and unit vectors used in the 
plane-wave expansion of a GWB.}
\label{f:plane_wave}
\end{center}
\end{figure}
The quantities $e^A_{ab}(\hat k)$ are polarization
tensors, given by
\be
\begin{aligned}
e_{ab}^+(\hat k)
&=\hat l_a\hat l_b-\hat m_a\hat m_b\,,
\\
e_{ab}^\times(\hat k)
&=\hat l_a\hat m_b+\hat m_a\hat l_b\,,
\end{aligned}
\ee
where $\hat l$, $\hat m$ are any two orthogonal unit 
vectors in the plane orthogonal to $\hat k$.
Typically, for stochastic background analyses, 
we take $\hat l$, $\hat m$ to be proportional to 
the standard angular unit vectors tangent to the sphere,
so that $\{\hat k, \hat l, \hat m\}$ is a right-handed
system (Figure~\ref{f:plane_wave}):
\be
\begin{aligned}
\hat k
&=-\sin\theta\cos\phi\,\hat x
-\sin\theta\sin\phi\,\hat y
-\cos\theta\,\hat z
= -\hat r\,,
\\
\hat l
&=+\sin\phi\,\hat x
-\cos\phi\,\hat y
= -\hat\phi\,,
\\
\hat m
&=-\cos\theta\cos\phi\,\hat x
-\cos\theta\sin\phi\,\hat y
+\sin\theta\,\hat z 
= -\hat\theta\,.
\end{aligned}
\label{e:klm_def}
\ee
For analyzing non-stochastic GW sources that have 
a symmetry axis (e.g., the angular momentum 
vector for binary inspiral), one takes 
$\hat l$ and $\hat m$ to be rotated relative 
to $-\hat\phi$ and $-\hat\theta$, 
where the rotation angle is the 
{\em polarization angle} of the source.

\subsection{Ensemble averages}
\label{s:ensemble_averages}

The quantities $h_A(f,\hat k)$ are the Fourier 
coefficients of the plane wave expansion.
Since the metric perturbations 
for a stochastic background are random variables, 
so too are the 
Fourier coefficients.
The probability distributions of the Fourier coefficients
thus define the statistical properties of the background.

Without loss of generality, we can assume that the 
expected value of the Fourier coefficients is 
zero,
\be
\langle h_A(f,\hat k)\rangle=0\,,
\ee
where angle brackets denote {\em ensemble average}
over different realizations of the background.
(The different realizations could be thought of 
as the different backgrounds observed by
different spatially-located observers in a homogeneous 
and isotropic universe.)
The second-order moments (i.e., quadratic expectation 
values) specify possible correlations between the 
Fourier coefficients.
For example, if the background is 
{\em unpolarized, stationary, and isotropic}, then
\be
\langle h_A(f,\hat k)h_{A'}^*(f',\hat k')\rangle
=\frac{1}{16\pi} S_h(f)\delta(f-f')\delta_{AA'}\delta^2(\hat k,\hat k')\,,
\label{e:quad_iso}
\ee
where $S_h(f)$ is the {\em strain power spectral 
density} of the background, 
having units of ${\rm strain}^2\,{\rm Hz}^{-1}$.
The fact that the RHS is proportional 
to $\delta(f-f')$ is a consequence of the assumption
of {\em stationarity}---i.e., that there is no 
preferred origin of time.
That the RHS depends on the polarization indices
only via  $\delta_{AA'}$ is a consequence of the 
background being unpolarized---i.e., that the 
$+$ and $\times$ polarization components are statistically
equivalent and uncorrelated with one another.
Similarly, the dependence on GW propagation directions 
only via $\delta^2(\hat k,\hat k')$ is a consequence of 
exact isotropy---i.e., that the power in the GWB has no preferred
direction, and that the GWs propagating in
different directions have uncorrelated phases.

If we drop the last assumption, allowing the background
to be either {\em anisotropic} or {\em statistically 
isotropic}, then the quadratic expectation values become
\be
\langle h_A(f,\hat k)h_{A'}^*(f',\hat k')\rangle
=\frac{1}{4} {\cal P}(f,\hat k)\delta(f-f')\delta_{AA'}
\delta^2(\hat k,\hat k')\,,
\label{e:quad_aniso}
\ee
where
\be
S_h(f) = \int \D^2\Omega_{\hat k}\,{\cal P}(f,\hat k)\,.
\label{e:Sh_aniso}
\ee
Here ${\cal P}(f,\hat k)$ is the strain power
spectral density per unit solid angle, with 
units ${\rm strain}^2\,{\rm Hz}^{-1}\,{\rm sr}^{-1}$.
For statistically isotropic backgrounds, the angular power 
spectrum is given by the coefficients $C_l$ of a Legendre 
series expansion (\ref{e:legendre_series})
of the two-point function 
$C(\theta)\equiv 
\langle {\cal P}(f,\hat k){\cal P}(f,\hat k')\rangle_{\rm sky\ avg}$, 
for all $\hat k$, $\hat k'$ having 
$\cos\theta = \hat k\cdot \hat k'$.

For {\em Gaussian} backgrounds, all cubic 
and higher-order moments are either identically zero 
or can be written in terms of the second-order moments.
Thus, the quadratic expectation values of the Fourier
coefficients completely characterize the statistical
properties of a Gaussian-distributed background.

\subsection{Energy density spectrum in gravitational waves}
\label{s:Omega_gw}

As mentioned above, $S_h(f)$ is the strain power 
spectral density of the GWB.
It can be related to the (normalized) 
{\em energy density spectrum}
\be
\Omega_{\rm gw}(f) 
\equiv \frac{1}{\rho_{\rm c}}\frac{\D\rho_{\rm gw}}{\D\ln f}
=\frac{f}{\rho_{\rm c}}\frac{\D\rho_{\rm gw}}{\D f}\,,
\label{e:Omega_gw}
\ee
where $\D\rho_{\rm gw}$ is the energy density in gravitational
waves contained in the frequency interval $f$ to $f+\D f$, and 
$\rho_{\rm c}\equiv {3 H_0^2 c^2}/{8\pi G}$
is the {\em critical} energy density (that needed to just 
close the universe today).
The result is~\cite{Allen-Romano:1999}
\be
S_h(f) = \frac{3H_0^2}{2\pi^2}\frac{\Omega_{\rm gw}(f)}{f^3}\,,
\label{e:S_h_and_Omega_gw}
\ee
which makes use of the relation
\be
\rho_{\rm gw} = \frac{c^2}{32\pi G}
\langle \dot h_{ab}(t,\vec x)\dot h^{ab}(t,\vec x)\rangle\,,
\label{e:rho_gw}
\ee
which gives the energy density in gravitational waves
in terms of the quadratic expectation values of the
metric perturbations.
You are asked in Exercise~\ref{exer:2} to derive
(\ref{e:S_h_and_Omega_gw}); to do so, 
you will also need to use the 
plane-wave expansion (\ref{e:planewave})
and the quadratic expectation values 
\eqref{e:quad_iso} or \eqref{e:quad_aniso}.

In addition to $S_h(f)$ and $\Omega_{\rm gw}(f)$, one
sometimes describes the strength of a GWB in terms of
the (dimensionless) {\em characteristic strain}
$h_c(f)$ defined by
\be
h_c(f) = \sqrt{f S_h(f)}\,.
\ee
For backgrounds described by a power-law dependence
on frequency,%
\footnote{There is no sum over $\alpha$ or $\beta$ in 
the following expressions.}
\be
h_c(f) = A_\alpha \left(\frac{f}{f_{\rm ref}}\right)^\alpha\,
\quad\Leftrightarrow\quad
\Omega_{\rm gw}(f) = \Omega_\beta\left(\frac{f}{f_{\rm ref}}\right)^\beta\,,
\label{e:powerlaw}
\ee
where $\alpha$ and $\beta$ are spectral indices,
and $A_\alpha$ and $\Omega_\beta$ are the amplitudes 
of the characteristic strain and energy density 
spectrum, respectively, at some reference frequency
$f=f_{\rm ref}$.
Using the above definitions and relationships between
$\Omega_{\rm gw}(f)$, $S_h(f)$, and $h_c(f)$, we have
\be
\Omega_\beta = \frac{2\pi^2}{3 H_0^2}f_{\rm ref}^2 A_\alpha^2\,,
\qquad
\beta = 2\alpha +2\,.
\ee
For standard inflationary backgrounds, $\Omega_{\rm gw}(f)={\rm const}$,
for which $\beta=0$ and $\alpha=-1$.
For GWBs associated with binary inspiral, 
$\Omega_{\rm gw}(f)\propto f^{2/3}$ (as we shall show below),
for which $\beta=2/3$ and $\alpha=-2/3$.
This last dependence is valid for both compact binary coalescences 
consisting of NSs and/or 
stellar-mass BHs (relevant for advanced LIGO, Virgo, etc.),
and also for SMBH binaries (relevant for pulsar timing searches).
 
\subsection{Calculating $\Omega_{\rm gw}(f)$ for an
astrophysically-generated background}
\label{s:Phinney_formula}

There is a relatively simple formula for calculating
the energy density spectrum $\Omega_{\rm gw}(f)$ 
produced by a collection of discrete astrophysical GW 
sources distributed throughout the universe~\cite{Phinney:2001}:
\be
\left.
\Omega_{\rm gw}(f) = \frac{1}{\rho_{\rm c}}\int_0^\infty \D z\>
n(z) \frac{1}{1+z}\left(f_{\rm s}\frac{\D E_{\rm gw}}{\D f_{\rm s}}
\right)\right|_{f_{\rm s}=f(1+z)}\,.
\label{e:phinney1}
\ee
We will call this the ``Phinney formula", since it was 
first written down by E.S.~Phinney in an unpublished
paper in 2001.
For this expression,
one needs only the comoving number density of
sources $n(z)$ as a function of the cosmological redshift $z$, 
and the energy spectrum of an individual source
$\D E_{\rm gw}/\D f_s$ as measured in its rest frame.
The source frame frequency $f_{\rm s}$ is related to the 
observed (present-day) frequency $f$ via $f_{\rm s}=f(1+z)$.
The factor of $1/(1+z)$ in the integrand is needed to 
redshift the energy measured in the source frame to that
measured today.
Note that the right-hand side of \eqref{e:phinney1} is
just the right-hand side of \eqref{e:Omega_gw} expanded in 
terms of its contribution from individual sources.

The above relationship can also be written in terms of 
the comoving rate density $R(z)$, which is related to the
comoving number density $n(z)$ via
\be
n(z)\,\D z = R(z)\,|\D t|_{t=t(z)}\,.
\ee
The final result is
\be
\left.
\Omega_{\rm gw}(f) = \frac{f}{\rho_{\rm c}H_0}\int_0^\infty \D z\>
R(z) \frac{1}{(1+z)E(z)}\left(\frac{\D E_{\rm gw}}{\D f_{\rm s}}
\right)\right|_{f_{\rm s}=f(1+z)}\,,
\label{e:phinney2}
\ee
where 
\be
E(z)\equiv \sqrt{\Omega_{\rm m}(1+z)^3 + \Omega_\Lambda}
\ee
is a cosmological factor that arises when evaluating 
$\D t/\D z$~\cite{Ryden:2003}.
$\Omega_{\rm m}$ and $\Omega_\Lambda$ are the 
fractional energy densities for matter
(ordinary baryonic matter plus dark matter) 
and dark energy, with numerical values roughly
equal to $0.30$ and $0.70$, respectively.
Exercise~\ref{exer:3} asks you to prove this 
``rate-version" of the Phinney formula, filling in some 
of the cosmology-related details.

\subsubsection{Example: $\Omega_{\rm gw}(f)$ for binary inspiral}
\label{s:binary_inspiral}

To illustrate the Phinney formula in action, we will 
verify the $\Omega_{\rm gw}(f)\propto f^{2/3}$
power-law dependence for binary inspiral, which we
stated without proof at the end of Section~\ref{s:Omega_gw}.
Since we are interested here only in the frequency 
dependence of $\Omega_{\rm gw}(f)$, we just need to
calculate the energy spectrum $\D E_{\rm gw}/\D f_{\rm s}$
for a single binary system.

\begin{figure}[htbp!]
\begin{center}
\includegraphics[width=0.45\textwidth]{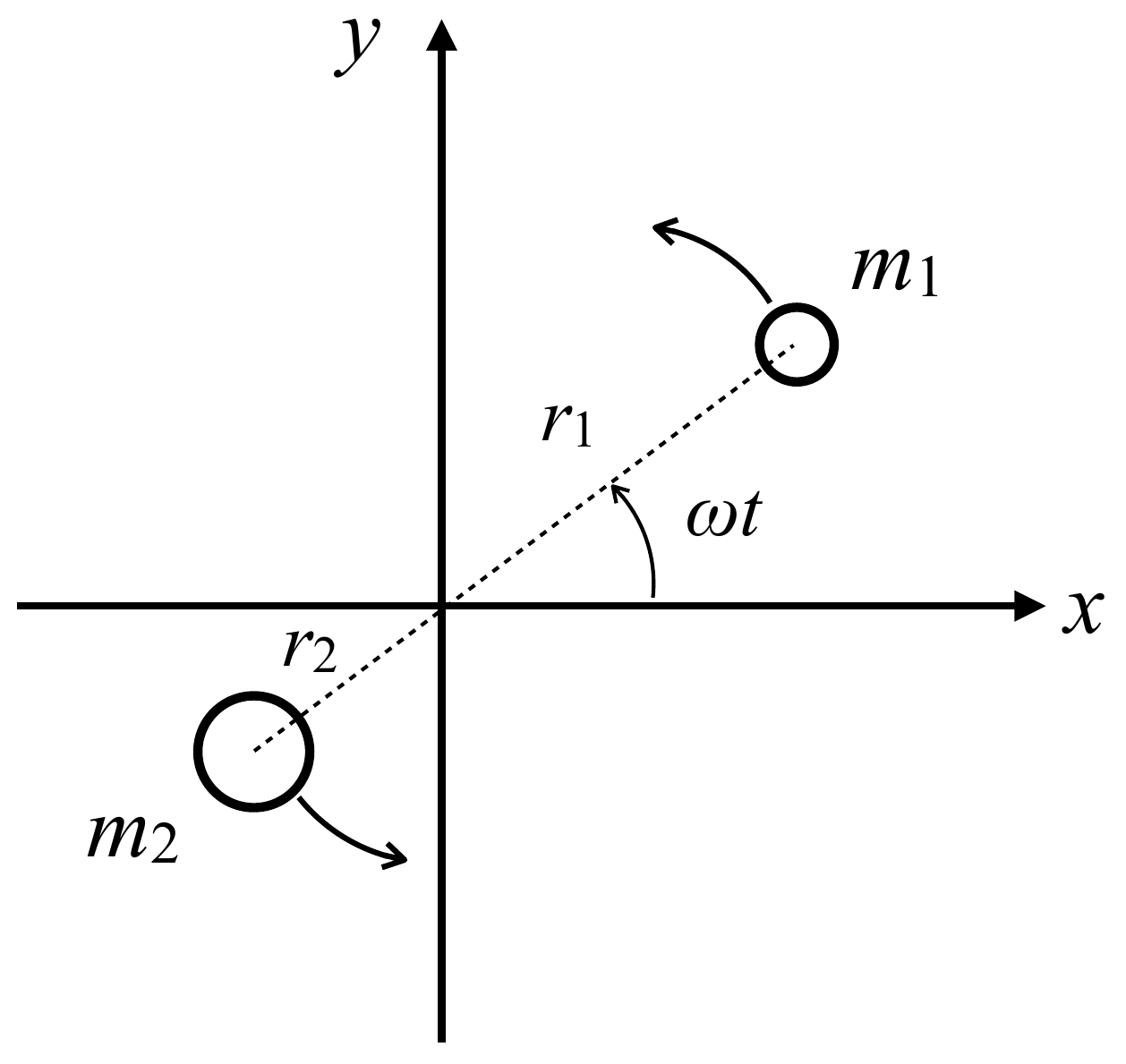}
\caption{Two masses $m_1$, $m_2$ in orbit around their common
of mass.}
\label{f:binary_inspiral}
\end{center}
\end{figure}

So let us consider two masses, $m_1$ and $m_2$, in circular 
orbits around their common center of mass (Figure~\ref{f:binary_inspiral}).
We make the standard definitions
\be
\begin{aligned}
r\equiv r_1+r_2\,,\quad
M\equiv m_1 + m_2\,,\quad
\mu\equiv \frac{m_1 m_2}{m_1 + m_2}
\end{aligned}
\ee
of the {\em relative separation}, {\em total mass}, 
and {\em reduced mass} of the system.
In terms of these quantities, Kepler's third law and 
the total orbital energy of the system can be written as
\be
\omega^2 r^3 = GM\,,\qquad
E_{\rm orb} = -\frac{GM\mu}{2r}\,,
\ee
where $\omega\equiv 2\pi f_{\rm orb}$ is the orbital 
angular frequency.
The power emitted in GWs comes from
the orbital energy
\be
\frac{\D E_{\rm gw}}{\D t} 
= -\frac{\D E_{\rm orb}}{\D t}\,,
\ee
which implies that the energy spectrum is given by
\be
\frac{\D E_{\rm gw}}{\D f_{\rm s}} 
= \frac{\D t}{\D f_{\rm s}}\frac{\D E_{\rm gw}}{\D t}
= -\frac{\D t}{\D f_{\rm s}}\frac{\D E_{\rm orb}}{\D t}\,.
\ee
It is now a relatively simple matter to evaluate the
RHS of the last expression, using Kepler's law to 
replace all occurences of $r$ and $\dot r$ with 
expressions involving $\omega$ and $\dot\omega$.
The final result is 
\be
\frac{\D E_{\rm gw}}{\D f_{\rm s}} 
\sim {\cal M}_{\rm c}^{5/3} f_{\rm s}^{-1/3}\,,
\qquad
{\cal M}_c^{5/3} \equiv M^{2/3}\mu\,,
\ee
where ${\cal M}_{\rm c}$ is the {\em chirp mass} of 
the system, and where we have ignored all numerical
factors.
Note that we also replaced the orbital angular frequency
$\omega$ by the GW frequency
$f_{\rm s} = 2 f_{\rm orb}$, with the factor of 2 
arising for quadrupolar radiation in general relativity.%
\footnote{For elliptical orbits, one should average
the radiated power, etc., over a period of the orbit.
There will also be contributions to the gravitational
radiation from harmonics other than just the 
quadrupole~\cite{Peters-Mathews:1963}.} 
Returning now to (\ref{e:phinney2}), we substitute
$f_s=(1+z)f$ and multiply by the factor of $f$ outside 
the integral to get
$\Omega_{\rm gw}(f) \propto f^{2/3}$ as claimed.

\section{Correlation methods}
\label{s:correlations}

As discussed above, a stochastic background of GWs 
is described by a {\em random} signal, which looks 
like noise in a single detector.
As such, standard search techniques like 
{\em matched filtering}~\cite{Wainstein-Zubakov:1971, Helstrom:1968}, 
which correlate the data 
against known, deterministic waveforms (e.g., BBH chirps) 
won't work when trying to detect a GWB.
Instead, we have to consider other possibilities:

(i) One possibility is to know the noise sources 
in our GW detector well enough (in both 
amplitude and spectral shape) that we can 
attribute any unexpected excess ``noise" to a GWB.
This was basically how Penzias and Wilson 
initially detected the CMB; they saw an excess 
noise temperature of $\sim\!3.5^\circ~{\rm K}$ in their
radio antenna that they could not attribute to 
any other noise source~\cite{Penzias-Wilson:1965}.
This is also what one hopes to do with LISA,
because the time-delay interferometry (TDI)~\cite{Estabrook:2000ef}
data combinations that one uses to remove the laser
frequency noise are orthogonal to one another~\cite{Prince:2002hp}.
Thus, one cannot cross-correlate these data 
streams to look for a GWB.
Instead, one must properly model the instrumental 
noise and astrophysical foreground (galactic WD 
binaries) in order to have a chance to detect a
cosmological GWB.
Studies by Adams and 
Cornish~\cite{Adams-Cornish:2010, Adams-Cornish:2014}
have shown that you can separate the detector
noise, astrophysical foreground, and cosmological
background using differences in their spectral
shapes and the modulation of the astrophysical 
background due to LISA's motion around the Sun 
(Figure~\ref{f:cyclostationary_data}).

(ii) Another possibility is to use data from 
multiple detectors.
Then we can look for evidence of a common 
disturbance in the multiple data streams that is
consistent with each detector's response to GWs.

Currently, (i) is not an option for ground-based 
interferometers since, even though the individual 
noise sources 
are understood pretty well, their amplitude is not 
known precisely enough to attribute any observed 
excess power to GWs.
One would need a really loud GWB relative to the
detector noise in order detect it in a way similar
to Penzias and Wilson's detection of the CMB.
But (ii) is an option as LIGO consists of two 
detectors, one in Hanford, WA, the other in Livingston, LA~\cite{LIGO:web}.
Virgo~\cite{Virgo:web}, in Italy, provides a third detector, 
and soon we will have two more large-scale 
interferometers: one in Japan, called Kagra~\cite{Kagra:web},
and the other in India, called IndIGO~\cite{Indigo:web}.
Cross-correlating data from multiple detectors works
for detecting a GWB since, even though the signal is 
random, it is the {\em same} signal in the different 
dectors (modulo the physical separation and relative
orientation of the detectors).
In effect, the random output of one detector is
used as a template for the data in another detector.
As we shall see below, the signal-to-noise ratio of 
the cross-correlation grows like the square-root of 
the observation time.
Thus, although the GWB might be weak 
relative to the noise, it can still be extracted from 
a cross-correlation measurement if it is observed
for a long enough period of time.

\subsection{Basic idea}
\label{s:basic_idea}

To illustrate the basic idea behind cross-correlation,
we will consider first the simplest possible scenario---i..e,
a single sample of data from two colocated and coaligned
detectors:
\be
\begin{aligned}
d_1 &= h + n_1\,,
\\
d_2 &= h + n_2\,.
\end{aligned}
\ee
Here $h$ denotes the common GW signal component, 
and $n_1$, $n_2$ denote the corresponding instrumental
noise components.
Cross-correlating the data for this case amounts to 
simply taking the product of the two data 
samples, $\hat C_{12}\equiv d_1 d_2$.
The expected value of the cross-correlation is
\be
\langle \hat C_{12}\rangle
=\langle d_1 d_2\rangle
= \langle h^2\rangle + \cancelto{0}{\langle h n_2\rangle} 
+ \cancelto{0}{\langle n_1 h\rangle}
+ \langle n_1 n_2\rangle\,,
\ee
where $\langle h n_2\rangle = 0 = \langle n_1 h\rangle$,
since the GW signal and instrumental noise are not correlated
with one another.
If we further assume that the noise in the two detectors
is {\em uncorrelated} (which is typically a valid assumption 
if the detectors are widely separated%
\footnote{Note that global magnetic fields, e.g., Schumann 
resonances, {\em can} produce environmental correlations in 
widely separated detectors~\cite{schumann1, schumann2, schumann3}.}), 
then $\langle n_1 n_2\rangle =0$, leaving
\be
\langle \hat C_{12}\rangle = \langle h^2\rangle\equiv S_h\,,
\ee
which is just the variance (i.e., power) in the GW signal.

\subsection{Extension to multiple data samples}
\label{s:multiple_samples}

The above analysis can be easily extended to the case of 
multiple samples:
\be
\begin{aligned}
d_{1i} &= h_i + n_{1i}\,,
\\
d_{2i} &= h_i + n_{2i}\,,
\end{aligned}
\ee
where $i=1,2,\cdots,N$.
As before, we will assume that the two detectors are
coincident and coaligned, and that the noise in the
two detectors are uncorrelated with the GW signal 
and with one another
\be
\langle n_{1i} h_j\rangle = 0\,,
\qquad
\langle n_{2i} h_j\rangle = 0\,,
\qquad
\langle n_{1i}n_{2j}\rangle=0\,.
\ee
We will also assume that the GWB 
and detector noise are both {\em white}, which means 
\be
\langle h_ih_j\rangle = S_h\,\delta_{ij}\,,
\qquad
\langle n_{1i}n_{1j}\rangle = S_{n_1}\,\delta_{ij}\,,
\qquad
\langle n_{2i}n_{2j}\rangle = S_{n_2}\,\delta_{ij}\,,
\ee
where $S_h$, $S_{n_1}$, $S_{n_2}$ are the variances
(i.e., power) in the GW signal and detector noise, respectively.%
\footnote{The assumption that both the GWB and detector noise
are white is made here just to simplify the analysis.
One can use cross-correlation methods for the more 
general case where the signal and noise power spectral
densities are non-trivial functions of 
frequency; see Section~\ref{s:optimal_filtering}.}
For this case, our cross-correlation statistic is the
average of the products of the individual data samples
\be
\hat S_h 
\equiv \hat C_{12} 
\equiv \frac{1}{N}\sum_{i=1}^N d_{1i} d_{2i}\,,
\label{e:Sh_ML}
\ee
which, as we shall see below, is again an estimator of the 
power in the GWB (hence the ``hat" over the 
$S_h$ on the LHS of this equation).

Using the above definitions and quadratic expectation 
values, it is easy to show that
\be
\mu\equiv \langle \hat C_{12}\rangle
= \frac{1}{N}\sum_{i=1}^N \langle d_{1i} d_{2i}\rangle
= \frac{1}{N}\sum_{i=1}^N \langle h_i^2\rangle
= S_h\,.
\ee
Thus, the cross-correlation statistic $\hat C_{12}$ is 
an (unbiased) estimator of the GW power $S_h$.
The variance in this estimator can be calculated via
\be
\sigma^2\equiv \langle \hat C_{12}^2\rangle-
\langle \hat C_{12}\rangle^2
=\left(\frac{1}{N}\right)^2
\sum_{i=1}^N \sum_{j=1}^N 
\left(\langle d_{1i} d_{2i} d_{1j} d_{2j}\rangle - 
\langle d_{1i} d_{2i}\rangle \langle d_{1j} d_{2j}\rangle\right)\,.
\label{e:sigma2_def}
\ee
To evaluate the RHS of the above equation, we make use of 
the identity
\be
\langle abcd\rangle =
\langle ab\rangle\langle cd\rangle + \langle ac\rangle \langle bd\rangle 
+\langle ad\rangle \langle bc\rangle\,,
\ee
which is valid for zero-mean Gaussian random variables.
Using this identity and the quadratic expectation values
between the signal and noise, we end up with
\be
\sigma^2 = \frac{1}{N}(S_1 S_2 + S_h^2)\,,
\label{e:sigma2_final}
\ee
where 
\be
S_1\equiv S_{n_1} + S_h\,,
\qquad
S_2\equiv S_{n_2} + S_h\,,
\ee
are the total power in the detector output (consisting of 
both signal and noise power).
Note that the factor of $1/N$ in \eqref{e:sigma2_final} 
comes from the double sum in \eqref{e:sigma2_def} having
non-zero contributions from only the diagonal terms $(i=j)$, 
which are all equal to one another.

Since the power in the GWB is expected to be weak compared
to the detector noise, the variance can be approximated
as $\sigma^2\simeq S_1 S_2/N$, for which the expected 
signal-to-noise ratio is given by
\be
\rho\equiv \frac{\mu}{\sigma}\simeq \frac{S_h}{\sqrt{S_1 S_2/N}}
\simeq \sqrt{N}\,\frac{S_h}{S_n}\,,
\label{e:rho_expected}
\ee
where $\sqrt{S_1 S_2}\simeq\sqrt{S_{n_1} S_{n_2}}\equiv S_n$.
This result verifies the statement made earlier that the signal-to-noise
ratio for a cross-correlation measurement grows like the square-root 
of the observation time (in this case, the total number of samples).

\subsection{Optimal filtering}
\label{s:optimal_filtering}

To handle the case of physically-separated and misaligned 
detectors, we need to include the non-trivial response of 
a GW detector to a GWB.  
We will do this in detail in 
Sections~\ref{s:nontrivial_response} and
\ref{s:nontrivial_correlations}.
Here, it suffices to simply define the {\em overlap function} 
(or overlap reduction function),
denoted $\Gamma_{12}(f)$, as the transfer function relating
the strain power in the GWB, $S_h(f)$, 
to the cross-correlated signal power 
in the two detectors~\cite{Flanagan:1993, Christensen:1997}:
\be
C_{12}(f) \equiv \Gamma_{12}(f) S_h(f)\,.
\ee
In terms of the quadratic expectation values of the GW 
signal in the two detectors, we have%
\footnote{The factor of $1/2$ is included on the RHS
so that the power spectrum is {\em one-sided}.
In other words, 
the total cross-correlated power in the GWB is
given by the integral of $\Gamma_{12}(f)S_h(f)$ over just
the {\em positive} frequencies.
The factor of $\delta(f-f')$ is a consequence of stationarity.}:
\be
\langle \tilde h_1(f) \tilde h_2^*(f')\rangle
=\frac{1}{2}\delta(f-f')\Gamma_{12}(f)S_h(f)\,,
\label{e:Gamma_def_freq}
\ee
where $\tilde h_1(f)$, $\tilde h_2(f)$ denote the 
Fourier transforms of the GW signal components 
$h_1(t)$, $h_2(t)$ in  the two detectors.
For comparison, the (auto-correlated) power spectra 
of the detector noise $P_{n_1}(f)$, $P_{n_2}(f)$ 
can be written in terms of the noise components 
$\tilde n_1(f)$, $\tilde n_2(f)$ via:
\be
\begin{aligned}
\label{e:noise_power_spectra}
\langle \tilde n_1(f) \tilde n_1^*(f')\rangle
&=\frac{1}{2}\delta(f-f')P_{n_1}(f)\,,
\\
\langle \tilde n_2(f) \tilde n_2^*(f')\rangle
&=\frac{1}{2}\delta(f-f')P_{n_2}(f)\,,
\end{aligned}
\ee
while the cross-correlated noise is assumed to be zero:
\be
\langle \tilde n_1(f) \tilde n_2^*(f')\rangle =0\,.
\ee
Plots of $\Gamma_{12}(f)$ for the 
LIGO Hanford-LIGO Livingston interferometer pair 
and for the LIGO Hanford-Virgo interferometer pair 
can be found in 
Section~\ref{s:examples-overlap}; other examples
of overlap functions are also given in that section.

Given the above definitions, we can now ask
the question: ``What is the optimal way to correlate 
data from two physically separated and possibly 
mis-aligned detectors to search for a GWB?"
To answer this question, we start by forming the 
generic cross-correlation
\be
\hat C_{12} = \int_{-T/2}^{T/2}\D t\>\int_{-T/2}^{T/2}\D t'\>
d_1(t)d_2(t) Q(t,t')\,,
\ee
where $Q(t,t')$ is an a~priori arbitrary filter 
function and $T$ is the observation time.
For stationary data, $Q(t,t')$ should depend only on
the difference between the two time arguments, 
$\Delta t\equiv t-t'$, 
so that $Q(t,t')\equiv Q(t-t')$.
In the Fourier domain, we can then write
\be
\hat C_{12} \simeq 
\int_{-\infty}^{\infty}\D f\>\int_{-\infty}^{\infty}\D f'\>
\delta_T(f,f')\tilde d_1(f)\tilde d^*_2(f') \tilde Q^*(f')\,,
\ee
where $\tilde Q(f)$ is the Fourier transform of 
$Q(\Delta t)$, and $\delta_T(f-f')$ is a finite-time
version of the Dirac delta function defined by 
$\delta_T(f-f')\equiv T\,{\rm sinc}\,[\pi(f-f')T]$, where
${\rm sinc}\,x \equiv \sin x/x$.

To proceed further we need to define what we mean by 
{\em optimal}.
A natural criterion in this context is to maximize the 
expected signal-to-noise ratio of $\hat C_{12}$ 
for a GWB with a fixed spectral shape $H(f)$.
(The expected signal-to-noise ratio is defined as 
in the previous section
$\rho\equiv \mu/\sigma$, where 
$\mu\equiv\langle \hat C_{12}\rangle$
and 
$\sigma^2\equiv \langle \hat C_{12}^2\rangle -\langle
\hat C_{12}\rangle^2$.) 
As you are asked to show in Exercise~\ref{exer:4},
this maximization condition  determines the form of the 
filter 
function $\tilde Q(f)$ up to an overall 
normalization~\cite{Allen:1997, Allen-Romano:1999}
\be
\tilde Q(f) \propto \frac{\Gamma_{12}(f) H(f)}
{P_1(f) P_2(f)}\,,
\ee
where $P_1(f)$, $P_2(f)$ are the total power in the two
detectors, 
\be
\begin{aligned}
P_1(f) \equiv P_{n_1}(f) + P_h(f)\,,
\qquad
P_2(f) \equiv P_{n_2}(f) + P_h(f)\,,
\end{aligned}
\ee
which are approximately equal to 
$P_{n_1}(f)$, $P_{n_2}(f)$ under the assumption that 
the GW signal is weak compared to the detector noise.
Note that the numerator of $\tilde Q(f)$ is proportional
to the expected value of the cross-correlated data in
the frequency domain, 
$\langle \tilde d_1(f)\tilde d_2^*(f)\rangle$,
while the denominator basically de-weights the 
correlation when the detector noise is large.
The dependence of $\tilde Q(f)$ on the spectral shape
$H(f)$ means that the optimal filter is tuned to a 
particular GWB.

The overall normalization of the optimal filter $\tilde Q(f)$ 
is not determined by the maximation condition, since
a constant multiplicative factor cancels out when
calculating  the signal-to-noise ratio $\rho=\mu/\sigma$.  
Typically, we use this freedom in the choice of 
normalization to set the expected value $\mu$ of the
cross-correlation equal to the overall amplitude of 
the background---i.e., $\mu = \Omega_{\rm gw}(f_{\rm ref})$.
In other words, for this choice of normalization, 
the measured value of the 
cross-correlation statistic, $\hat C_{12}$, is a 
{\em point estimate} of $\Omega_{\rm gw}(f_{\rm ref})$.

\section{Optimal filtering applied to some simple examples}
\label{s:simple_examples}

\begin{figure}[htbp!]
\begin{center}
\includegraphics[width=\textwidth]{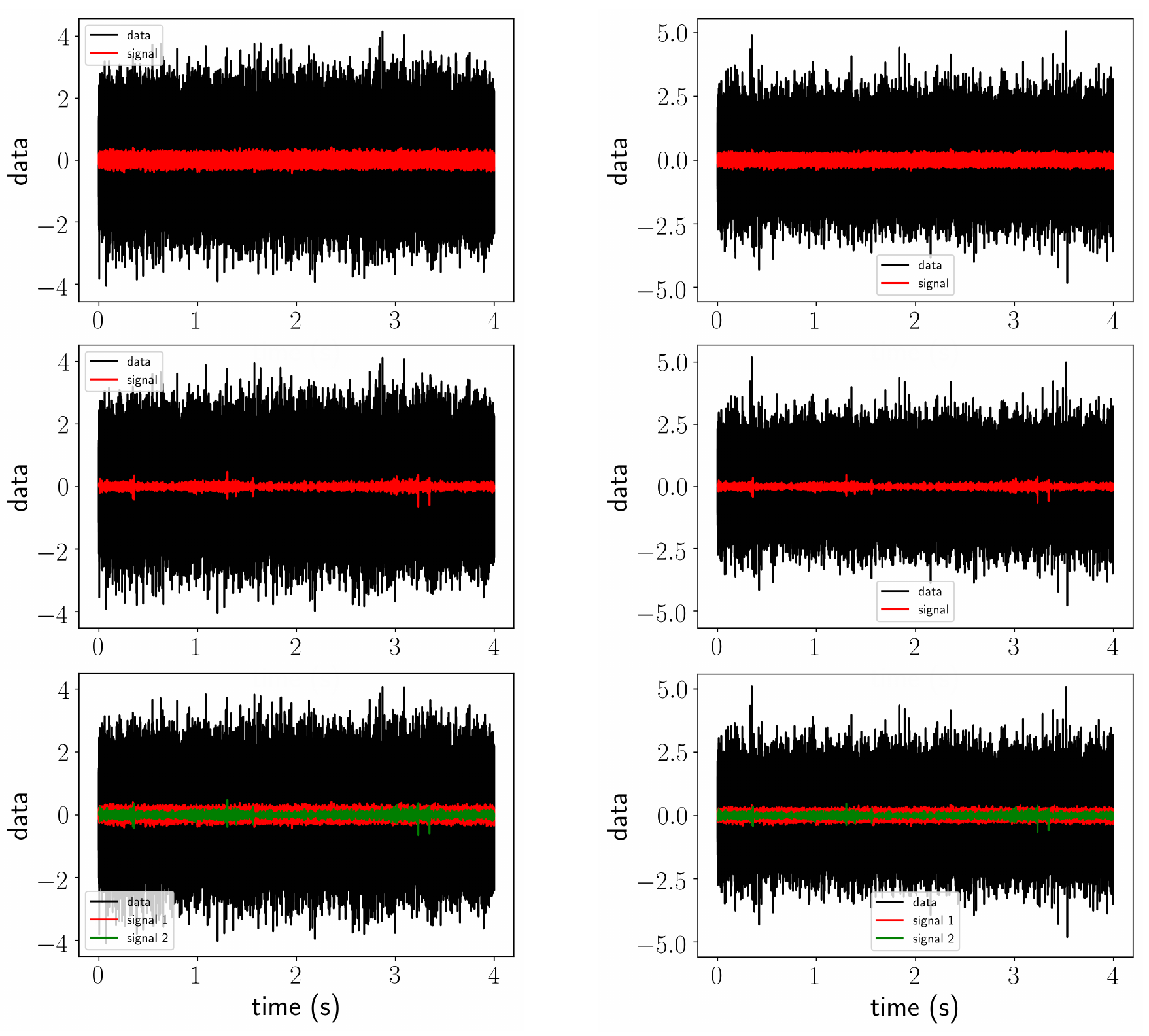}
\caption{Simulated time-domain data for the three different
cases discussed in the main text:
(top row) a white GWB in uncorrelated, white detector noise,
(middle row) a confusion-limited BNS background in uncorrelated, 
white detector noise,
(bottom row) a two-component background formed from the superposition 
of the GWBs from the top two rows in uncorrelated, white 
detector noise.
The two columns correspond to data in the two coincident
and coaligned detectors.
By eye one can see that signal components in the two detectors 
are identical, but the noise (and hence the data) in the two 
detectors are different.}
\label{f:simple_examples}
\end{center}
\end{figure}

We now apply the above correlation methods to analyze 
some simple examples involving simulated data.
(The simulations are solely meant to illustrate how 
optimal filtering works;
the amplitude and duration of the simulated data are 
not representative of real interferometer data.%
\footnote{The simulated data used for these examples 
can be found at \cite{github-code}.
Access to real GW data is available via the 
Gravitational-Wave Open Science Center (GWOSC)~\cite{gwosc}.})
We will consider three different GWBs injected into 
uncorrelated, white detector noise in two coincident
and coaligned detectors:
(i) a white GWB, 
(ii) a confusion-limited BNS background, 
and 
(iii) a two-component background, formed from the superposition
of the GWBs from (i) and (ii).
The simulated time-domain data for the three different 
cases are shown in Figure~\ref{f:simple_examples}.
Recall that a white GWB has a flat spectrum $H(f)=1$, 
while a confusion-limited background produced by BNS 
inspirals and mergers has spectral shape $H(f)=(f/f_{\rm ref})^{-7/3}$ 
(see Figure~\ref{f:different_power_spectra}).

\subsection{Single-component analyses}

We start by applying the single-component optimal-filter
analysis of the previous section.
For example (i), we find that the measured and injected
values of the amplitude of the GWB agree to 3.5\%, 
which is within 1-$\sigma$.
The corresponding optimally-filtered signal-to-noise
ratio is $\rho=2.9$.
For example (ii), the measured and injected values of
the amplitude of the GWB agree to 2.7\%, which again 
is within 1-$\sigma$.
The corresponding optimally-filtered signal-to-noise
ratio for this case is $\rho=12$.
Note that even though the overall amplitude of the
background is noticeably smaller for the confusion-limited
BNS background, the signal-to-noise ratio is considerably
larger (12 versus 2.9).
This is because the spectrum of the GW signal differs 
in this case from that of the detector noise,
which helps in distinguishing the signal and noise components.

Finally for example (iii), if we filter the data for 
the two components separately, 
we overestimate the amplitude of the white GWB component 
by 48\%, which is greater than 1-$\sigma$, and overestimate
the amplitude of the BNS background by 6.9\%, which is within 1-$\sigma$.
Basically, filtering the data for each GWB component 
separately typically leads to {\em overestimates}
of the amplitudes of the individual components, 
but {\em underestimates} of the error bars.
The overestimates arise since the other GWB component is
also contributing to the correlated signal.

\subsection{Multi-component analysis}

To better extract the amplitudes of the individual
components for example (iii), we need to go beyond 
single-component optimal-filtering, and consider 
a signal model that allows for a superpostion 
of multiple GWB components~\cite{Parida-et-al:2015}.
So instead of taking the cross-correlation to be a 
{\em single number}, $\hat C_{12}$, which is obtained 
by integrating the contributions from all 
frequencies, we will keep the frequency-dependence
explicit, defining
\be
\hat C_{12}(f)\equiv\frac{2}{T} \tilde d_1(f)\tilde d^*_2(f)
\,,
\ee
where $\tilde d_1(f)$, $\tilde d_2(f)$ are the Fourier
transforms of the time-domain data $d_1(t)$, $d_2(t)$ 
from the two detectors.
We will treat the values of $\hat C_{12}(f)$ 
for different frequencies $f$
as the `data points' from which to construct a 
{\em likelihood function}, which is the probability 
of the data given the parameters defining the signal
and noise models.%
\footnote{See Section~\ref{s:statistical_inference}
and John Veitch's lectures in this Volume 
for more details regarding likelihood functions and
statistical inference.}
For this case, the signal model is given by the expected 
value of the correlated data:
\be
\langle\hat C_{12}(f)\rangle
=\sum_\alpha \Gamma_{12}(f)A_\alpha H_\alpha(f)
\equiv \sum_\alpha M_\alpha(f) A_\alpha\,,
\ee
where $H_\alpha(f)$ are the different spectral
shapes having amplitudes $A_\alpha$.
(Abstractly, we can think of $M_\alpha(f) \equiv
\Gamma_{12}(f) H_\alpha(f)$ as a matrix with indices
$f$ and $\alpha$, where $f$ runs over different 
frequency bins and $\alpha$ runs over different 
spectral components.)
The noise model enters via the covariance matrix 
of the data:
\be
N_{12}(f,f') 
\equiv \langle \hat C_{12}(f)\hat C_{12}^*(f')\rangle -
\langle \hat C_{12}(f)\rangle \langle \hat C_{12}^*(f')\rangle
\simeq \delta_{ff'}\,P_1(f) P_2(f)\,,
\ee
which is the product of the noise power spectra
in the two detectors in the weak-signal approximation.
The likelihood function is then
\be
\begin{aligned}
p(\hat C|A,N)
&\propto\exp\left[-\frac{1}{2}
(\hat C-MA)^\dagger N^{-1}(\hat C-MA)\right]
\\
&\propto
\exp\left[-\frac{1}{2}\int_{-\infty}^\infty \D f
\frac{|\hat C_{12}(f)-\sum_\alpha M_\alpha(f)A_\alpha|^2}
{P_1(f)P_2(f)}\right]\,,
\end{aligned}
\ee
which is the probability of the cross-correlated data
$\hat C_{12}(f)$ 
given the amplitudes $A_\alpha$ of the GWB 
spectral components 
and the noise in the two detectors $N_{12}(f,f')$.
The advantage of using an 
index-free matrix notation, as we did in 
the first line of the above expression, is that 
we can use standard linear algebra calculations
to find the values of $A$ that maximize the 
likelihood.

Given the likelihood $p(\hat C|A,N)$, we can now obtain 
estimators of the amplitudes of the GWB components
by maximizing it with respect to the $A_\alpha$.
The final result (which you are asked to show in 
Exercise~\ref{exer:5}) is: 
\be
\hat A = F^{-1} X\,,
\ee
where
\be
F\equiv M^\dagger N^{-1} M\,,
\qquad 
X\equiv M^\dagger N^{-1} \hat C\,.
\ee
The quantity $F$ is called the 
{\em Fisher information matrix}.
In terms of its components,
\be
F_{\alpha\beta} = \int_{-\infty}^\infty \D f\> \frac{H_\alpha(f)\Gamma_{12}^2(f)H_\beta(f)}
{P_1(f)P_2(f)}\,.
\ee
Thus, we see that the Fisher matrix is a noise-weighted 
inner product 
of the spectral shapes $H_\alpha(f)$, $H_\beta(f)$ with one 
another.
Provided the spectral shapes are not degenerate (i.e., not 
propoportional to one another), then the Fisher matrix $F$
can be inverted and $\hat A$ calculated.
Otherwise, some form of regularization is needed to 
perform the matrix inversion.
The inverse of the Fisher matrix, $F^{-1}$, turns out to
equal the covariance matrix of the estimators $\hat A$.  

Using the above multi-component formalism, we are now able 
to extract the ampltitude of the white GWB component to 7.3\%, 
corresponding to a signal-to-noise ratio of 1.4, 
and to extract
the amplitude of the BNS background component to 3.8\%,
corresponding to a signal-to-noise ratio of 6.0.
In essence, the {\em joint} multi-component analysis 
properly takes into account the {\em covariance}
between the spectral shapes of the two components, 
allowing for unbiased, minimal variance estimates of 
the amplitudes $A_\alpha$.

\part{Details / Examples}
\label{p:details}

In the second part of these lecture notes, 
we describe the non-trivial response of a beam
detector to gravitational waves, calculate the overlap function
between a pair of detectors, and introduce a Bayesian method that
can optimally search for the astrophysical background 
produced by stellar-mass binary BHs throughout the universe.

\section{Non-trivial detector response}
\label{s:nontrivial_response}

To understand stochastic background searches on a 
more quantitative level, we need to describe the 
non-trivial response of a GW detector to a passing GW.
In Section~\ref{s:optimal_filtering}, we defined the 
overlap function $\Gamma_{12}(f)$
for a pair of detectors, but we didn't specify how 
to calculate it, or how its form differs for different
GW detectors.
In this and the following section, we will develop
the tools that we need to do these calculations.

\subsection{Beam detectors and different types of 
detector response}

\begin{figure}[htbp!]
\begin{center}
\includegraphics[width=\textwidth]{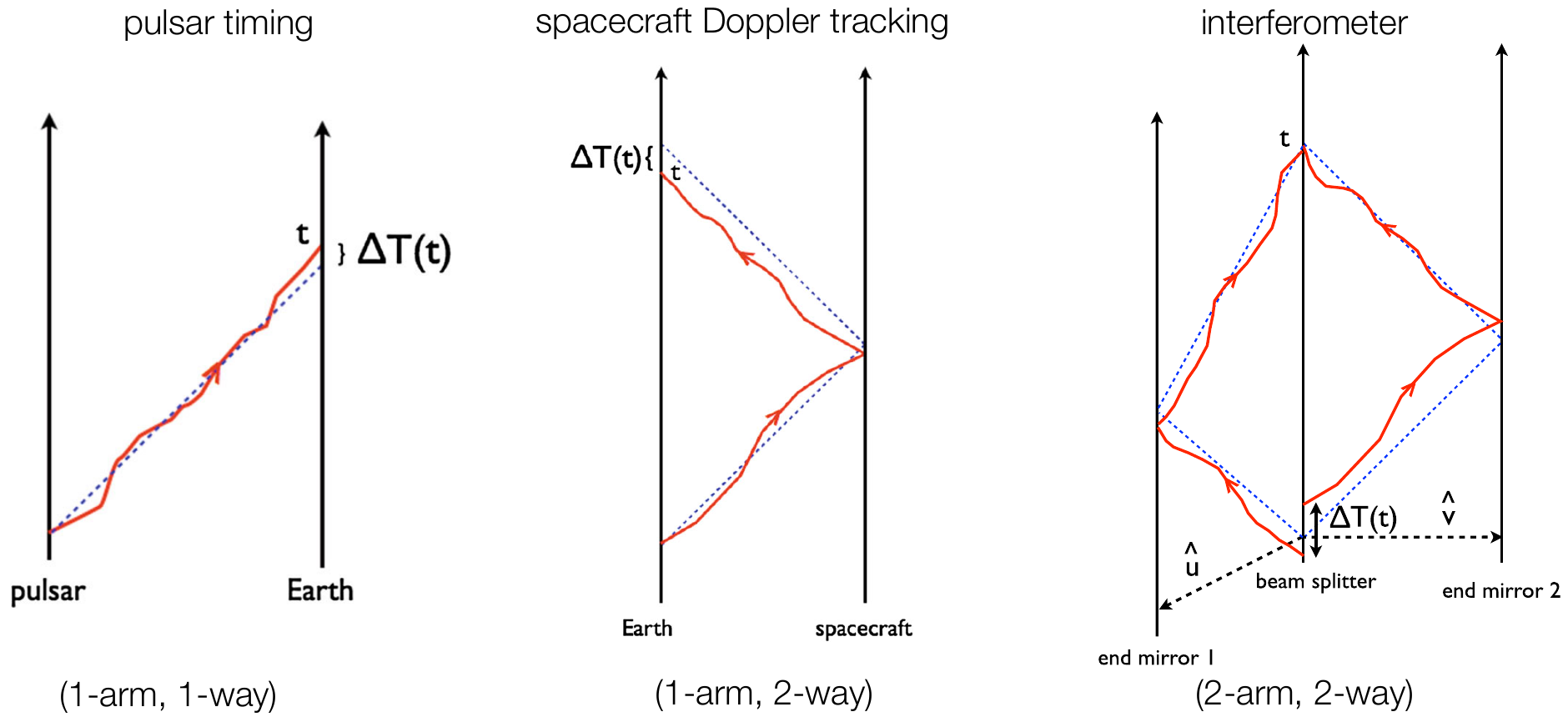}
\caption{Spacetime diagram showing the response of beam
detectors to a passing GW.  
Left: pulsar timing; middle: spacecraft Doppler
tracking; right: interferometer (ground or space-based).
A passing GW perturbs the path of the photon (red trajectory) 
relative to its nominal path in the absence of the wave 
(blue dotted line), leading to a 
difference in the expected arrival time of the photon.
(Figure adapted from \cite{Romano-Cornish:2017}.)}
\label{f:beam_detectors}
\end{center}
\end{figure}

For simplicity, we will restrict our attention to 
{\em beam detectors}, which use electromagnetic radiation
to monitor the separation of two or more test masses.
Laser interferometers (both ground-based and space-based),
spacecraft Doppler tracking, and pulsar timing arrays
are all examples of beam detectors.
(A resonant-bar detector, like that first used by
Joseph Weber, is a much different type of detector.
Roughly speaking, a resonant bar detector responds
like a giant tuning fork to a passing GW, provided
the GW has frequencies equal to the resonant frequencies 
of the bar~\cite{MTW:1973}.)
The response of a beam detector to a passing GW is 
the change in the light-travel time $\Delta T(t)$ 
between the two 
masses relative to the nominal light-travel time.
This is illustrated schematically in Figure~\ref{f:beam_detectors}.

In the literature, one might see the detector response
written in terms of strain $\Delta L(t)/L$, 
fractional Doppler frequency $\Delta v(t)/\nu_0$, or 
phase $\Delta\Phi(t)$, instead of the timing residual
$\Delta T(t)$.
Despite the apparent differences in the responses, 
they are all derivable from the change in light travel
time $\Delta T(t)$ via the relations:
\be
\begin{aligned}
&h(t)\equiv \Delta T(t)\quad &({\rm pulsar\ timing})\\
&h(t)\equiv \frac{\Delta L(t)}{L} = \frac{\Delta T(t)}{T}
\quad&({\rm LIGO,\ Virgo,\ }\cdots) \\
&h(t)\equiv \frac{\Delta\nu(t)}{\nu_0}=\frac{\D \Delta T(t)}{\D t}
\quad &({\rm spacecraft Doppler\ tracking})\\
&h(t)\equiv \Delta\Phi(t) = 2\pi \nu_0\,\Delta T(t)
\quad &({\rm LISA})\,.
\end{aligned}
\ee
Hence, once we know how to calculate the timing residual
response $\Delta T(t)$, we can easily calculate all the
other quantities listed above.

\subsection{Detector response functions}
\label{e:det_response}

Gravitational waves are weak.
As such, a GW detector acts 
like a {\em linear} system,%
\footnote{It is a linear system 
because second and higher-order terms in the 
metric perturbations can be safely ignored.}
converting metric perturbations $h_{ab}(t,\vec x)$ 
to the detector output.
Mathematically, this is represented by the 
{\em convolution} of the metric perturbations with the 
{\em response function} of the detector:
\begin{equation}
h(t) = (\mb{R}*\mb{h})(t,\vec x)\equiv
\int_{-\infty}^\infty {\rm d}\tau
\int {\rm d}^3 y\>
R^{ab}(\tau,\vec y)h_{ab}(t-\tau, \vec x-\vec y)\,.
\end{equation}
Here $h(t)$ is the output of the detector at time $t$.
The vector $\vec x$ is the location of detector, and 
$R^{ab}(\tau,\vec y)$ is the {\em impulse repsonse}
of the detector.
Expanding $h_{ab}(t-\tau,\vec x-\vec y)$ as a sum of
plane waves \eqref{e:planewave}, and substituting 
this into the right-hand side of the above expression, 
we find that the 
Fourier transform $\tilde h(f)$ of $h(t)$ can be written as
\be
\tilde h(f)=\int {\rm d}^2\Omega_{\hat n}
\sum_{A=+,\times} R^A(f,\hat k)\,h_A(f,\hat k)\,,
\ee
where
\be
R^A(f,\hat k) \equiv R^{ab}(f,\hat k)e^A_{ab}(\hat k)
\label{e:R^A_def}
\ee
and
\be
R^{ab}(f,\hat k) \equiv e^{-i 2\pi f\hat k\cdot\vec x/c}\,
\int_{-\infty}^\infty {\rm d}\tau \int {\rm d}^3 y\>
R^{ab}(\tau,\vec y)\,e^{-i2\pi f(\tau-\hat k\cdot\vec y/c)}\,.
\label{e:R^ab_def}
\ee
Note that $R^A(f,\hat k)$ is the 
detector response for a plane-wave
with frequency $f$, propagation direction $\hat k$, and polarization $A$.

\subsection{Examples}

We now calculate the detector response functions for a couple of
examples.

\subsubsection{Detector response for a one-arm, one-way detector}

\begin{figure}[htbp!]
\begin{center}
\includegraphics[width=0.2\textwidth]{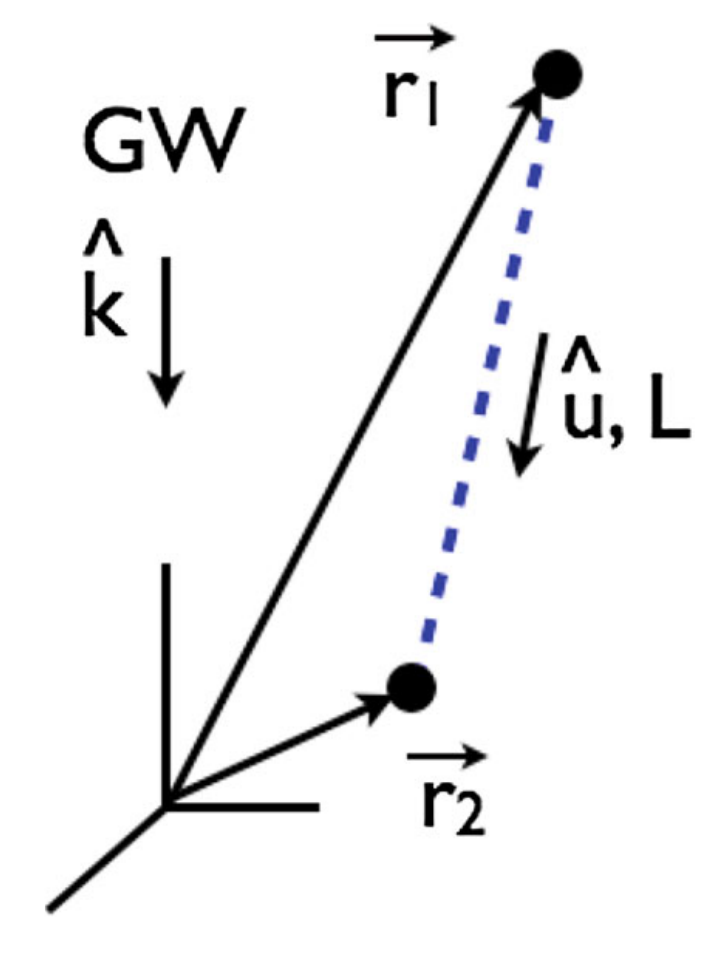}
\caption{Geometry for a one-arm, one-way beam detector, relevant for 
a pulsar timing residual measurement.
The GW propagates in the $\hat k$ direction; the electromagnetic wave
(e.g., a radio pulse from a pulsar) propagates in the $\hat u$ direction
(opposite of the direction to the pulsar, $\hat p=-\hat u$).
(Figure taken from \cite{Romano-Cornish:2017}.)}
\label{f:one_arm_one_way}
\end{center}
\end{figure}

For our first example, we will consider the timing response of a one-arm,
one-way beam detector, which is relevant for pulsar timing observations.
The geometry of the situation is shown in Figure~\ref{f:one_arm_one_way}.
The timing residual response is then given by
\be
h(t)\equiv
\Delta T(t) = \frac{1}{2c} u^a u^b\int_0^L{\rm d}s\> h_{ab}(t(s),\vec x(s))\,,
\ee
where
\be
t(s) = (t-L/c) + s/c\,, \qquad
\vec x(s) = \vec r_1 + s\hat u
\ee
is a parametric representation of the photon path from the 
source ($s=0$) to the detector ($s=L$).
Note that we do not need to include any corrections to
the straight-line path for the photon given above, as the 
metric perturbations are already first-order and we 
can ignore all second and higher-order terms in the calculation.

To do the integral, we first substitute $t(s)$ and 
$\vec x(s)$ for $t$ and $\vec x$ in the plane-wave 
expansion for $h_{ab}(t,\vec x)$.
The $s$ dependence shows up only in the exponential:
\begin{equation}
\begin{aligned}
e^{i2\pi f(t(s)-\hat k\cdot\vec x(s)/c)}
&= e^{i2\pi f(t-L/c + s/c -\hat k\cdot(\vec r_1 + s\hat u)/c)}
\\
&= e^{i2\pi f(t-L/c -\hat k\cdot\vec r_1/c)}
e^{i2\pi f(1 -\hat k\cdot\hat u)s/c}\,,
\end{aligned}
\end{equation}
and the integral over $s$ is easy to do:
\be
\int_0^L {\rm d}s\>
e^{i2\pi f(1-\hat k\cdot\hat u)s/c}=
\frac{c}{i2\pi f}\frac{1}{1-\hat k\cdot\hat u}\left[e^{\frac{i2\pi f L}{c}
(1-\hat k\cdot\hat u)}-1\right]\,.
\ee
Then including all the other factors and rearranging terms, 
you should find (Exercise~\ref{exer:6}):
\be
R^A(f,\hat k) = \frac{1}{i2\pi f}\,
\frac{1}{2}u^a u^b e^A_{ab}(\hat k)\frac{1}{1-\hat k\cdot \hat u}\,
\left[ 1-e^{-\frac{i 2\pi fL}{c}(1-\hat k\cdot\hat u)}\right]
\,e^{-i2\pi f\hat k\cdot\vec r_2/c}\,.
\label{e:response_one_arm_one_way}
\ee
In the context of pulsar timing, the two terms in square brackets 
are called the {\em Earth term} and {\em pulsar term}, respectively.
The pulsar term encodes information about the phase of the GW at
the location of the pulsar, at the time the radio pulse was emitted.
The pulsar term is usually ignored for stochastic background searches, as 
this term for different pulsars will not be correlated with one other
(since the spatial distance between two pulsars, of order kpc, 
is much greater than the wavelengths of the GWs that pulsar timing arrays are 
sensitive to, of order 10~light-years).

\begin{figure}[htbp!]
\begin{center}
\includegraphics[width=0.5\textwidth]{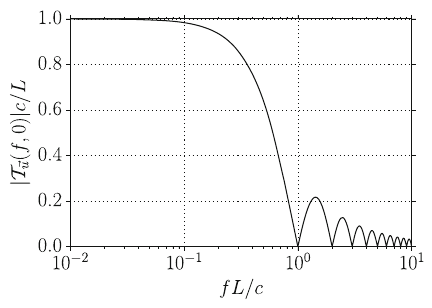}
\caption{Plot of the absolute value of the timing transfer 
function $|{\cal T}_{\vec u}(f,0)|$ for normal incidence.
Note the nulls in the response when $L$ equals an integer 
number of GW wavelengths $\lambda \equiv c/f$.
(Figure taken from \cite{Romano-Cornish:2017}.)}
\label{f:timing_transfer}
\end{center}
\end{figure}

Both terms {\em are} important for LISA data analysis, however,
as the wavelengths of the GWs that LISA will be sensitive to
are of the same order of magnitude as the lengths of LISA's arms 
(i.e., the separation between the spacecraft).
For this case, one defines a {\em timing transfer function} for 
one-way photon propagation as
\be
\begin{aligned}
{\cal T}_{\vec u}(f,\hat k\cdot \hat u) 
&\equiv\frac{1}{i2\pi f}\frac{1}{1-\hat k\cdot\hat u}
\left[1-e^{-\frac{i2\pi f L}{c}(1-\hat k\cdot\hat u)}\right]
\\
&= \frac{L}{c}\,e^{-\frac{i\pi fL}{c}(1-\hat k\cdot\hat u)}
{\rm sinc}\left(\frac{\pi fL}{c}[1-\hat k\cdot\hat u]\right)\,,
\end{aligned}
\ee
where ${\rm sinc}\,x\equiv \sin x/x$.
Note that for normal incidence (i.e., $\hat k\cdot\hat u=0$), 
the timing transfer function has zeroes when $L$ is equal to 
an integer number of GW wavelengths $\lambda\equiv c/f$---i.e.,
when $fL/c$ equals an integer
(Figure~\ref{f:timing_transfer}).
This is because a photon's trajectory undergoes an integer
number of cycles of contraction and dilation produced by the
GW when $fL/c=1,2,\cdots$, thereby giving a net zero effect. 

\begin{figure}[htbp!]
\begin{center}
\includegraphics[width=0.4\textwidth]{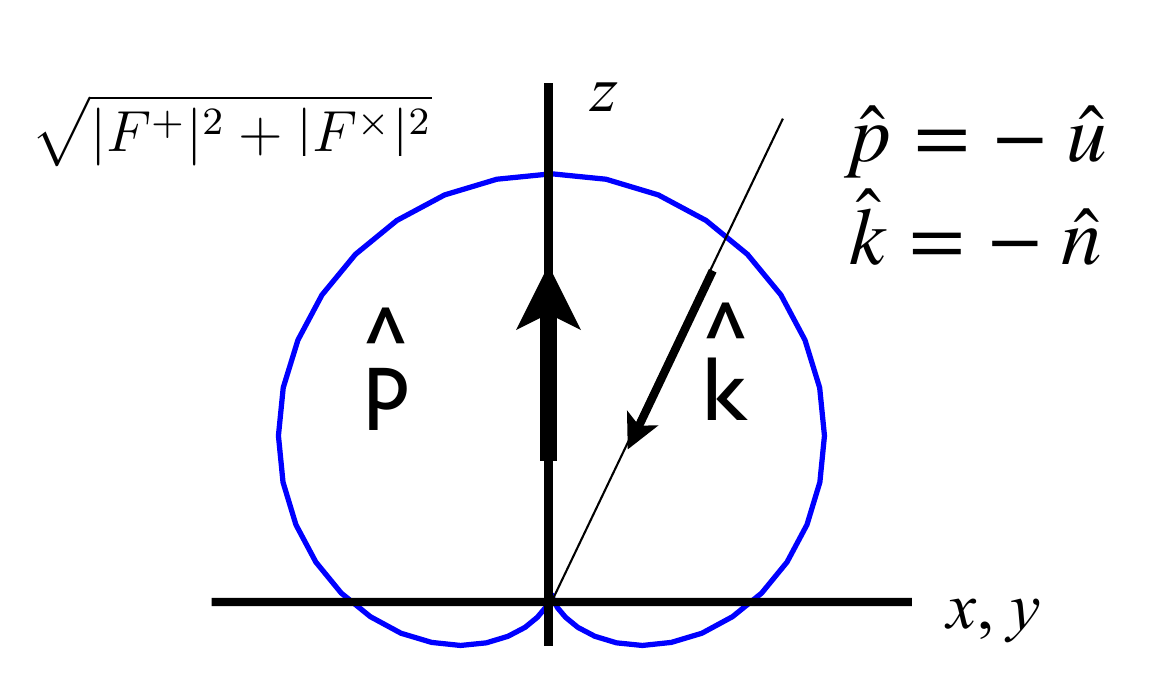}
\caption{Polarization-averaged Doppler frequency response
for pulsar timing, where we have ignored the pulsar term.
The response is axially symmetric around the $z$-axis, which
we've chosen to be in the direction to the pulsar $\hat p=-\hat u$.}
\label{f:one_arm_one_way_peanut}
\end{center}
\end{figure}

Returning to \eqref{e:response_one_arm_one_way} and its 
application to pulsar timing analyses, 
note that the factor $1/(i2\pi f)$ goes away for the Doppler 
frequency response, $\Delta\nu(t)/\nu_0$, and that the
phase term $e^{-i2\pi f\hat k\cdot\vec r_2/c}$ equals one
if we take the $\vec r_2$ to be the origin of coordinates, e.g.,
at the solar system barycenter.
Thus, ignoring the pulsar term, the Doppler frequency 
response is given simply by
\be
F^A(\hat k) 
=\frac{1}{2}\frac{u^a u^b}{1-\hat k\cdot\hat u}\,e_{ab}^A(\hat k)\,.
\label{e:F^A(k)}
\ee
A plot of the root-summed-squared response (summed over the 
two polarizations)
is shown in Figure~\ref{f:one_arm_one_way_peanut} for the 
case $\hat u=-\hat z$ (or, equivalently, $\hat p=\hat z$ where 
$\hat p=-\hat u$ is the direction to the pulsar).
The response is maximum when the GW and radio pulse propagate 
in the same direction---i.e., when $\hat k=\hat u$.
It is zero when they propagate in opposite directions.
These results follow from 
\be
\begin{aligned}
&u^a u^b e^+_{ab}(\hat k) 
= -\sin^2\theta = -(1-\cos\theta)(1+\cos\theta)\,,
\\
&u^a u^b e^\times_{ab}(\hat k)=0\,,
\\
& 1-\hat k\cdot\hat u = 1-\cos\theta\,,
\end{aligned}
\ee
for which
\be
F^+(\hat k) = -\frac{1}{2}(1+\cos\theta)\,,
\qquad
F^\times(\hat k) = 0\,.
\label{e:FA_Earth_z}
\ee
Here $\theta$ is the angle between $\hat k$ and $\hat u$
(which is the usual polar angle measured from the $z$-axis).

\begin{figure}[htbp!]
\begin{center}
\includegraphics[width=0.4\textwidth]{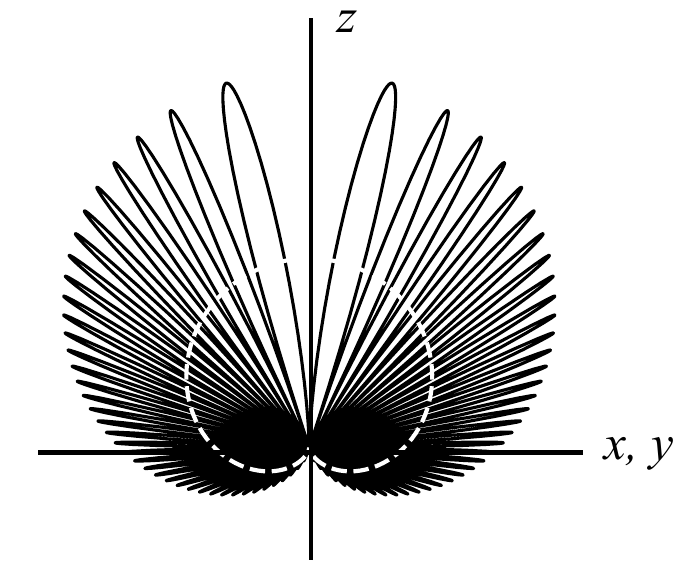}
\caption{Same as Figure~\ref{f:one_arm_one_way_peanut} but 
including the pulsar term and taking $fL/c=20$.
The response without the pulsar term is shown as a 
dashed-white curved for comparison.}
\label{f:one_arm_one_way_peanut_full}
\end{center}
\end{figure}

Note that if we include the pulsar term in the response,
then $F^A(\hat k)$ in \eqref{e:FA_Earth_z} 
should be multiplied 
by a term proportional to $\sin(\pi f L[1-\cos\theta]/c)$.
This introduces a null at $\theta=0$ and at other values 
of $\theta$ satisfying
\be
\cos\theta = 1-\frac{nc}{fL}\,,
\qquad n=0,1,\cdots\,,{\rm int}[2fL/c]\,.
\ee
In Figure \ref{f:one_arm_one_way_peanut_full} we show
the full root-summed-squared response 
including the pulsar term,
taking $fL/c=20$ for illustration purposes.
(For most pulsars, $fL/c$ will be of order 100 
or more, as the distance to typical pulsars is 
of order a kpc or more.)
The response without the pulsar term 
(Figure~\ref{f:one_arm_one_way_peanut}) is also
shown for comparison.
%

\subsubsection{Detector response for a laser interferometer 
in the short-antenna limit}

\begin{figure}[htbp!]
\begin{center}
\includegraphics[width=0.6\textwidth]{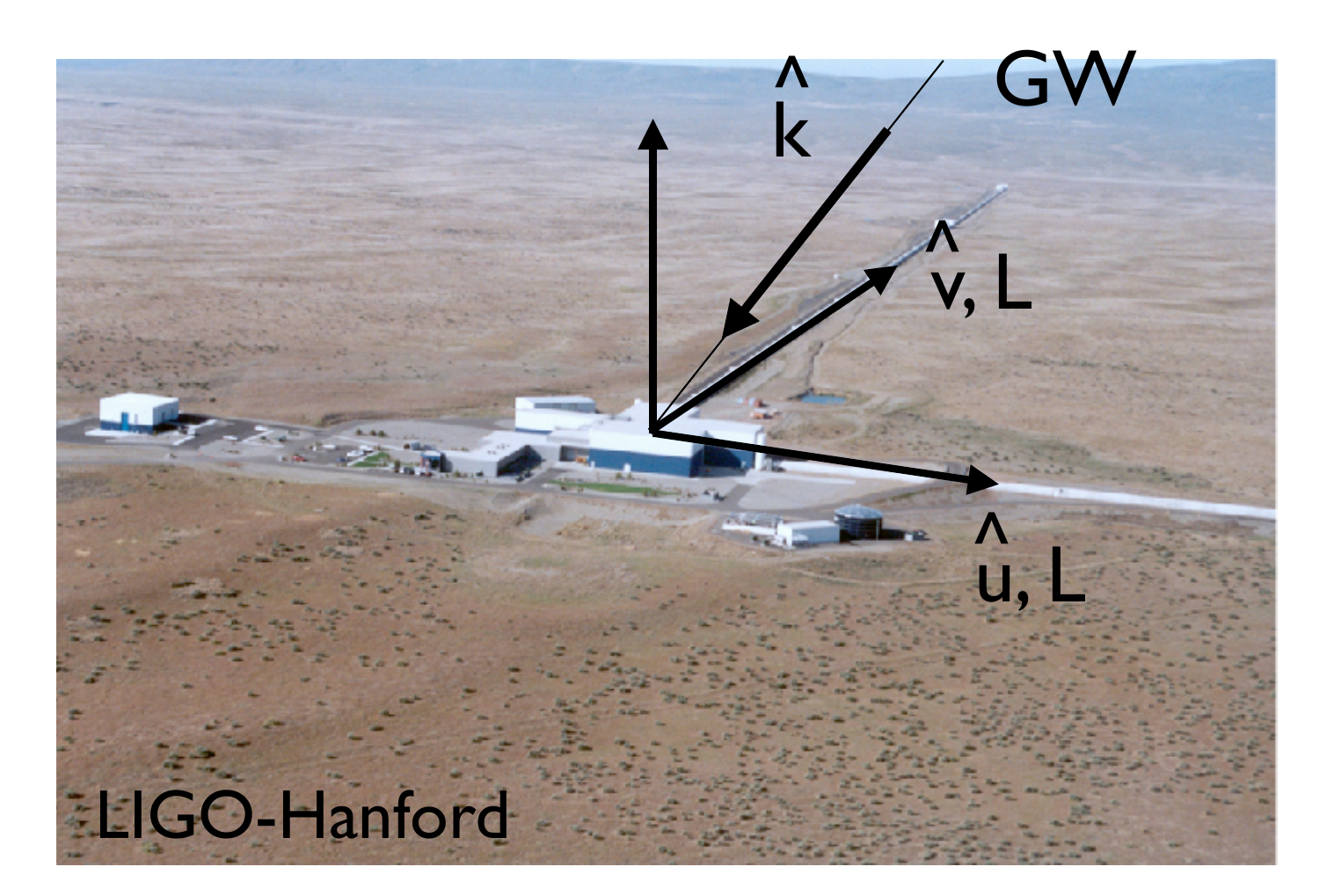}
\caption{Geometry for a ground-based interferometer GW response
calculation.
(Shown here is the LIGO Hanford Observatory (LHO), 
in Hanford, WA.)
The GW propagates in the $\hat k$ direction;
$\hat u$, $\hat v$ are unit vectors that point along the 
two arms of the interferometer.
In the short-antenna approximation, the length $L$ of the 
arms does not enter the expression for the strain response.}
\label{f:LHO_geometry}
\end{center}
\end{figure}

\begin{figure}[htbp!]
\begin{center}
\includegraphics[width=\textwidth]{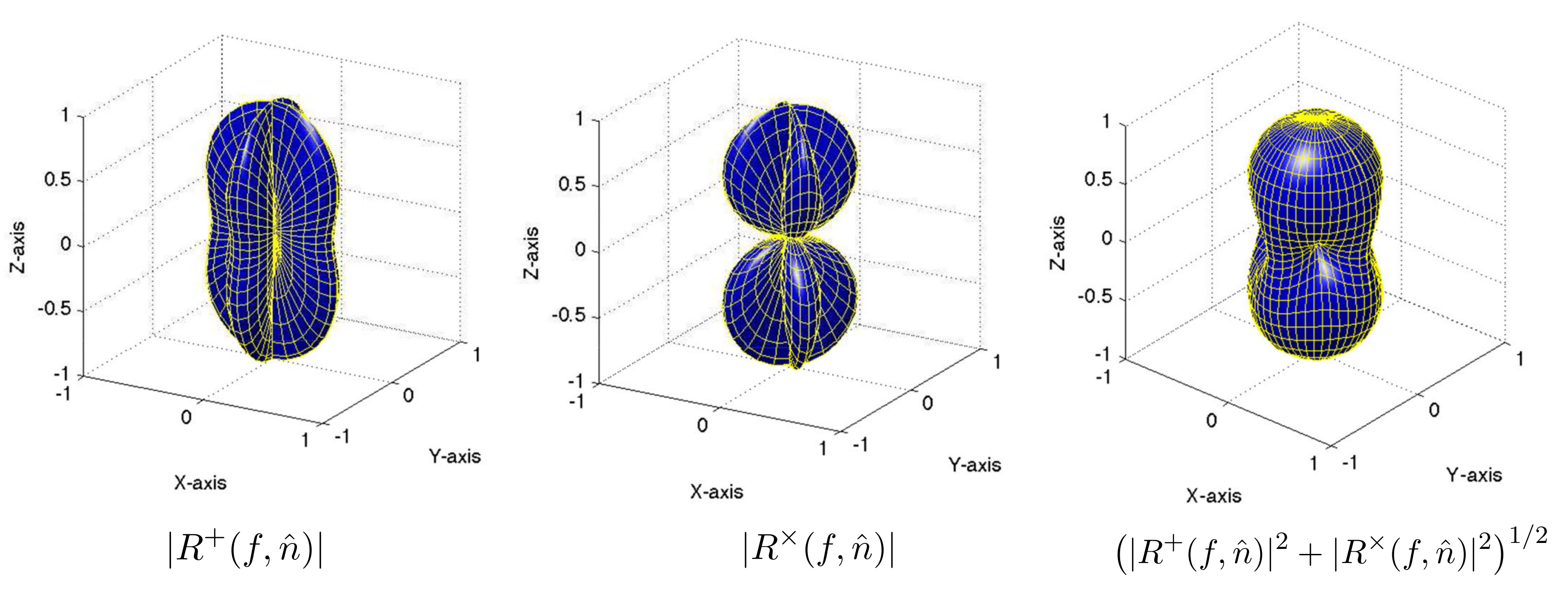}
\caption{Beam pattern functions for a ground-based interferometer
like LIGO in the short-antenna approximation---i.e., $f\lesssim {\rm few\ kHz}$.
The vertex of the interferometer is at the origin of coordinates,
and the interferometer arms are assumed to be orthogonal, pointing along
the $x$ and $y$ directions.}
\label{f:LIGO_beam_patterns}
\end{center}
\end{figure}

Another simple example of a detector response function is
for a equal-arm laser interferometer, like LIGO, in the 
{\em short-antenna} (or long-wavelength) 
approximation (Figure~\ref{f:LHO_geometry}).
This approximation is valid when the wavelength of the GW is 
much larger than the dimensions of the detector.  
The GW phase is then effectively constant as a photon
travels down and back an interferometer arm.
The integral over the photon path is simply proportional
to the nominal round-trip propagation time $2L/c$.
Defining the strain response of the interferometer as
\be
h(t) \equiv \frac{1}{2}\left(
\frac{\Delta T_{\vec u,\, {\rm roundtrip}}(t)}{T}
-
\frac{\Delta T_{\vec v,\, {\rm roundtrip}}(t)}{T}
\right)\,,
\ee
one can show that
\be
R^A(f,\hat k)\simeq\frac{1}{2}\left(u^au^b-v^av^b\right)\,e^A_{ab}(\hat k).
\ee
The quantity multiplying $e^A_{ab}(\hat k)$ in the 
expression for the reponse function above is called the 
{\em detector tensor}
\be
D^{ab}\equiv\frac{1}{2}\left(u^a u^b - v^a v^b\right)\,.
\ee
Plots of the {\em beam pattern functions}
$|R^+(f,\hat k)|$ and $|R^\times(f,\hat k)|$ for the 
two polarizations individually, and the 
root-summed-squared response (summed over both polarization)
are shown in Figure~\ref{f:LIGO_beam_patterns}.
The last plot showing the root-summed-squared response 
is sometimes called the LIGO ``peanut".
It illustrates that a laser-interferometer in the 
short-antenna approximation is a very blunt instrument,
being senstive to a very large portion of the sky. 
The only nulls are in the plane spanned by the arms, 
in the directions of the perpendicular bisectors of 
the arms.
%

\section{Non-trivial correlations}
\label{s:nontrivial_correlations}

In this section, we describe how to correlate the outputs
of two detectors, taking into account their non-trivial
response to GWs.

\subsection{Overlap function}

Detectors in different locations and with different 
orientations respond differently to a passing GW.
The {\em overlap function} encodes the reduction in 
sensitivity of a cross-correlation analysis due to 
the separation and misalignment of the detectors.

Let $I$ and $J$ label two detectors, and let 
$h_I(t)$ and $h_J(t)$ denote the corresponding response
of these detectors to an unpolarized and isotropic
GW background.
The expected correlation of the two detector outputs
can then be written as
\be
\langle h_I(t)h_J(t')\rangle 
= \frac{1}{2}\int_{-\infty}^\infty{\rm d}f\>
e^{i2\pi f(t-t')}\Gamma_{IJ}(f)S_h(f)\,,
\label{e:Gamma-time}
\ee
where $S_h(f)$ is the (1-sided) strain power spectral density
of the GWB, cf.~\eqref{e:quad_iso} and \eqref{e:S_h_and_Omega_gw}, 
and $\Gamma_{IJ}(f)$ is the overlap function: 
\be
\Gamma_{IJ}(f) 
= \frac{1}{8\pi}\int {\rm d}^2\Omega_{\hat k}\> 
\sum_A R_I^A(f,\hat k) R_J^{A}{}^{*}(f,\hat k)\,.
\label{e:Gamma}
\ee
Recall from \eqref{e:R^A_def}, \eqref{e:R^ab_def}
that the location of the detector is already included
in the response functions $R^A(f,\hat k)$ via
the phase factor $e^{-i2\pi \hat k\cdot \vec x/c}$.
If we explicitly display this dependence by writing
$R^A(f,\hat k) \equiv \bar R^A(f,\hat k)e^{-i2\pi f\hat k\cdot\vec x/c}$,
then
\be
\Gamma_{IJ}(f) 
= \frac{1}{8\pi}\int {\rm d}^2\Omega_{\hat k}\> 
\sum_A \bar R_I^A(f,\hat k) \bar R_J^{A}{}^{*}(f,\hat k)\,
e^{-i2\pi f\hat k\cdot(\vec x_I-\vec x_J)/c}\,.
\ee
One often sees this alternative expression for $\Gamma_{IJ}(f)$
in the literature, e.g., \cite{Flanagan:1993, Christensen:1997, Allen-Romano:1999}.

The interpretation of $\Gamma_{IJ}(f)$ as encoding 
the reduction in sensitivity of a cross-correlation
analysis due to the physical separation and relative
orientation of the detectors
is most easily seen in the frequency domain, where
\eqref{e:Gamma-time} becomes
\be
\langle \tilde h_I(f)\tilde h_J^*(f')\rangle 
= \frac{1}{2}\delta(f-f')\Gamma_{IJ}(f)S_h(f)\,.
\label{e:Gamma-freq}
\ee
From this expression, we see that $\Gamma_{IJ}(f)$ 
is a transfer function between the strain power 
$S_h(f)$ in the GWB and the detector cross-power
$C_{IJ}(f) = \Gamma_{IJ}(f)S_h(f)$.

For statistically anisotropic backgrounds, it 
turns out that the integrand of 
$\Gamma_{IJ}(f)$ is the most important quantity for
describing the cross-correlation.
This is because for this case
\be
\langle \tilde h_I(f)\tilde h_J^*(f')\rangle 
= \frac{1}{4}\delta(f-f')\int {\rm d}^2\Omega_{\hat k}\> 
\sum_A R_I^A(f,\hat k) R_J^{A}{}^{*}(f,\hat k) {\cal P}(f,\hat k)\,,
\label{e:aniso-corr}
\ee
where ${\cal P}(f,\hat k)$ is the GW power on the 
sky, coming from direction $\hat n=-\hat k$; see \eqref{e:quad_aniso}
and \eqref{e:Sh_aniso}.
One typically expands 
${\cal P}(f,\hat k)$ in terms of spherical harmonics 
\be
{\cal P}(f,\hat k) = \sum_{l=0}^\infty\sum_{m=-l}^l
{\cal P}_{lm}(f)Y_{lm}(\hat k)\,,
\ee
for which \eqref{e:aniso-corr} becomes
\be
\langle \tilde h_I(f)\tilde h_J^*(f')\rangle 
= \frac{1}{2}\delta(f-f')\sum_{l=0}^\infty\sum_{m=-l}^l
\Gamma_{IJ,lm}(f){\cal P}_{lm}(f)\,.
\label{e:Gamma-freq-aniso}
\ee
with~\cite{Allen-Ottewill:1997, Thrane-et-al:2009}
\be
\Gamma_{IJ,lm}(f) \equiv 
\frac{1}{2}
\int{\rm d}^2\Omega_{\hat k}\>
\sum_A R_I^A(f,\hat k) R_J^{A}{}^{*}(f,\hat k) 
Y_{lm}(\hat k)\,.
\ee
So up to a factor of $1/4\pi$, the spherical harmonic 
components of the integrand of the overlap function
\eqref{e:Gamma} encode the reduction in sensitivity
when doing a cross-correlation for anisotropic backgrounds.
Interested readers can find much more discussion
about anisotropic backgrounds in Section~7 of
\cite{Romano-Cornish:2017}, and references to the
original work cited therein.

\subsection{Examples}
\label{s:examples-overlap}

Given \eqref{e:Gamma} for $\Gamma_{IJ}(f)$ and 
explicit expressions for the response functions 
$R^A_I(f,\hat k)$ for different detectors, we can 
now calculate the overlap function for different 
detector pairs.

\subsubsection{Overlap function for a pair of laser
interferometers in the short-antenna limit}

\begin{figure}[htbp!]
\begin{center}
\includegraphics[width=0.49\textwidth]{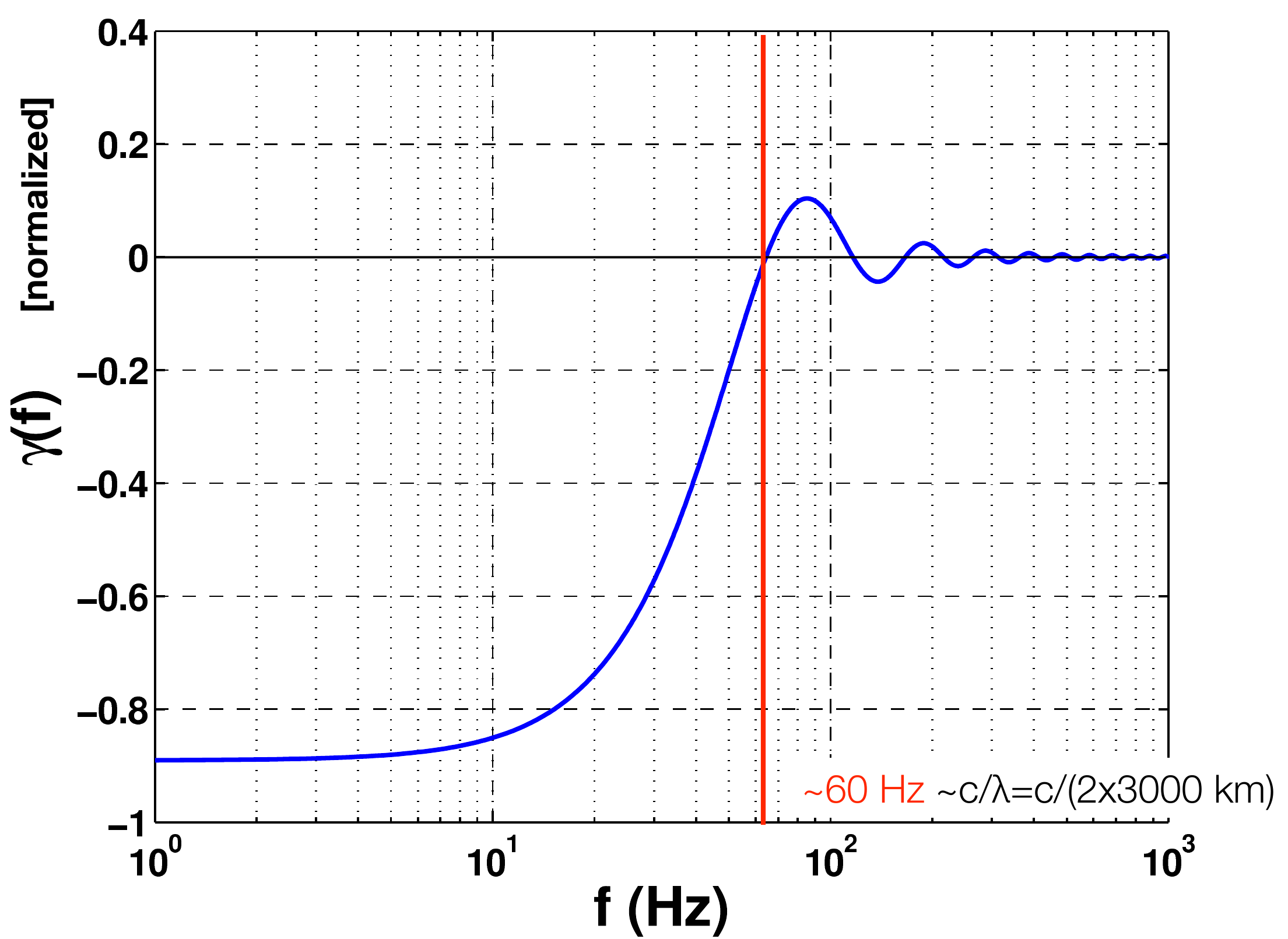}
\includegraphics[width=0.49\textwidth]{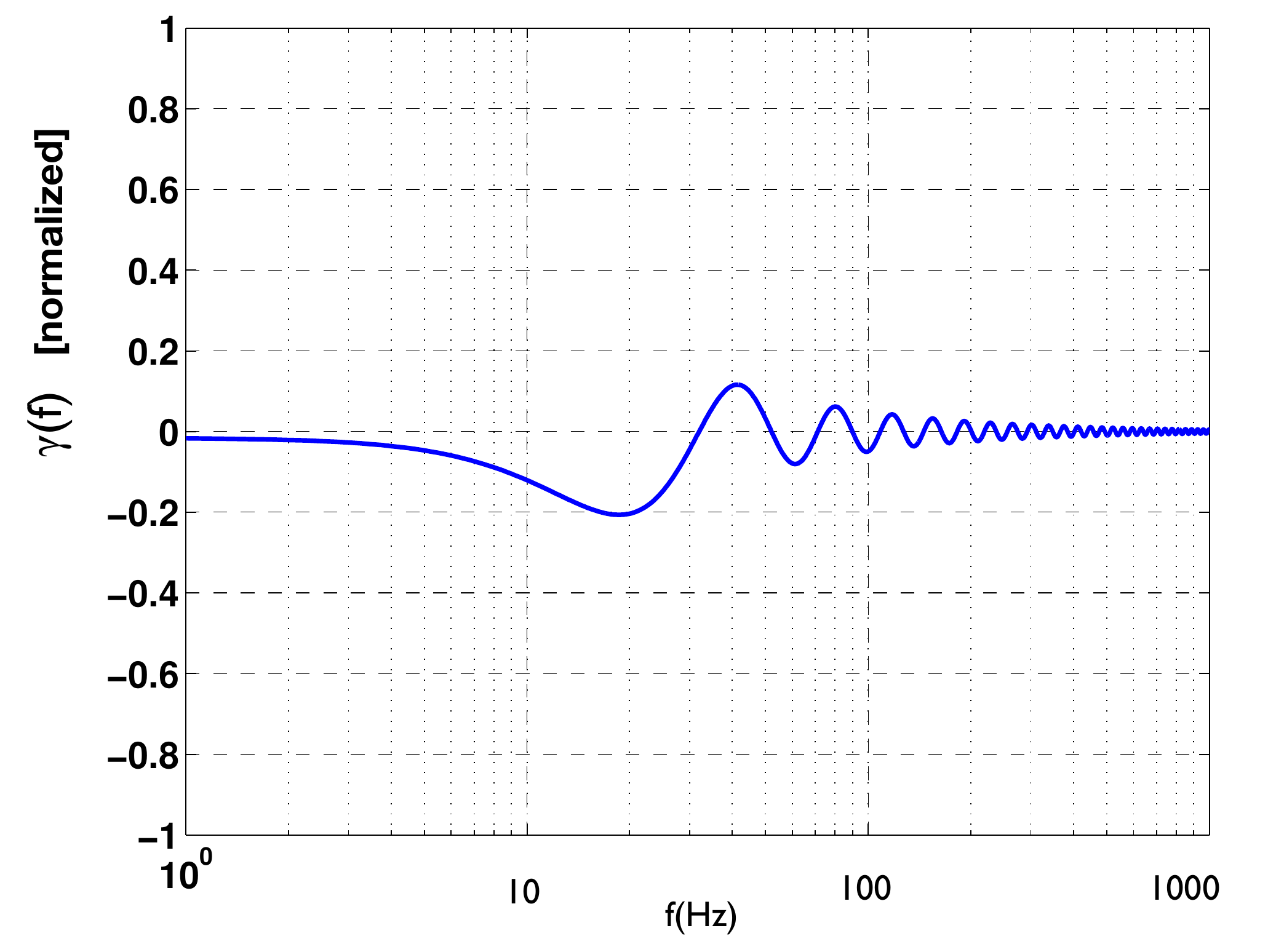}
\caption{Normalized overlap function for ground-based
interferometers calculated in the short-antenna approximation.
Left panel: LHO-LLO overlap funtion.
Right panel: LHO-Virgo overlap function.}
\label{f:orfs}
\end{center}
\end{figure}

Our first example will be the overlap function 
for pairs of laser interferometers in the short-antenna
approximation.
For concreteness, we will consider the 
LIGO Hanford-LIGO Livingston detector pair (which we will
denote LHO-LLO)
and the LIGO Hanford-Virgo (LHO-Virgo) detector pair.
Plots of these overlap functions, normalized to unity
for coincident and coaligned detectors 
(denoted $\gamma(f)$) are shown in Figure~\ref{f:orfs}.

For the LHO-LLO detector pair,
note that as $f\rightarrow 0$, $\gamma(f)\rightarrow -0.89$.
The minus sign indicates that the two interferometers
are rotated by $90^\circ$ relative to one another.
The fact that the absolute value $|\gamma(0)|=0.89$ is 
not exactly equal to 1, even though the overlap function
is normalized, indicates that the two interferometers
aren't exactly (anti) aligned.
The two interferometers are separated by $27.2^\circ$ as 
seen from the center of the Earth, so the tangent planes 
of the interferometers are tilted relative to one another due 
to the curvature of the Earth.
In addition, the first zero crossing of the overlap function
occurs at approximately 60~Hz, which corresponds (roughly)
to the frequency (50~Hz) of a GW having a wavelength 
equal to 
twice the separation ($2\times 3000$~km) between the two observatories.
For lower frequencies, the two interferometers are driven
(on average) by the same positive (or negative) part of a passing GW;
while for slightly larger frequencies, the two interferometers 
start to be driven by parts of the GW having opposite signs.  
The zero crossings correspond to the transitions between
these in-phase and out-of-phase excitations of the 
interferometers.

For the LHO-Virgo detector pair, note 
that in the limit $f\rightarrow 0$, $\gamma(f)\simeq 0$.
This is because the two interferometers effectively respond
to the two orthogonal polarizations of a GW, corresponding 
to a rotation of the interferometer arms by $45^\circ$.
This $45^\circ$ misalignment is also the reason for the 
(overall) reduced amplitude of the LHO-Virgo overlap 
function relative to that for LHO-LLO.  
The fact that the first zero crossing for the LHO-Virgo 
overlap function is just over $30~{\rm Hz}$ (almost half 
that for LHO-LLO) is due the larger separation between the
LHO and Virgo interferometers, compared to LHO 
and LLO.

\subsubsection{Overlap function for pulsar timing arrays}

\begin{figure}[htbp!]
\begin{center}
\includegraphics[width=0.7\textwidth]{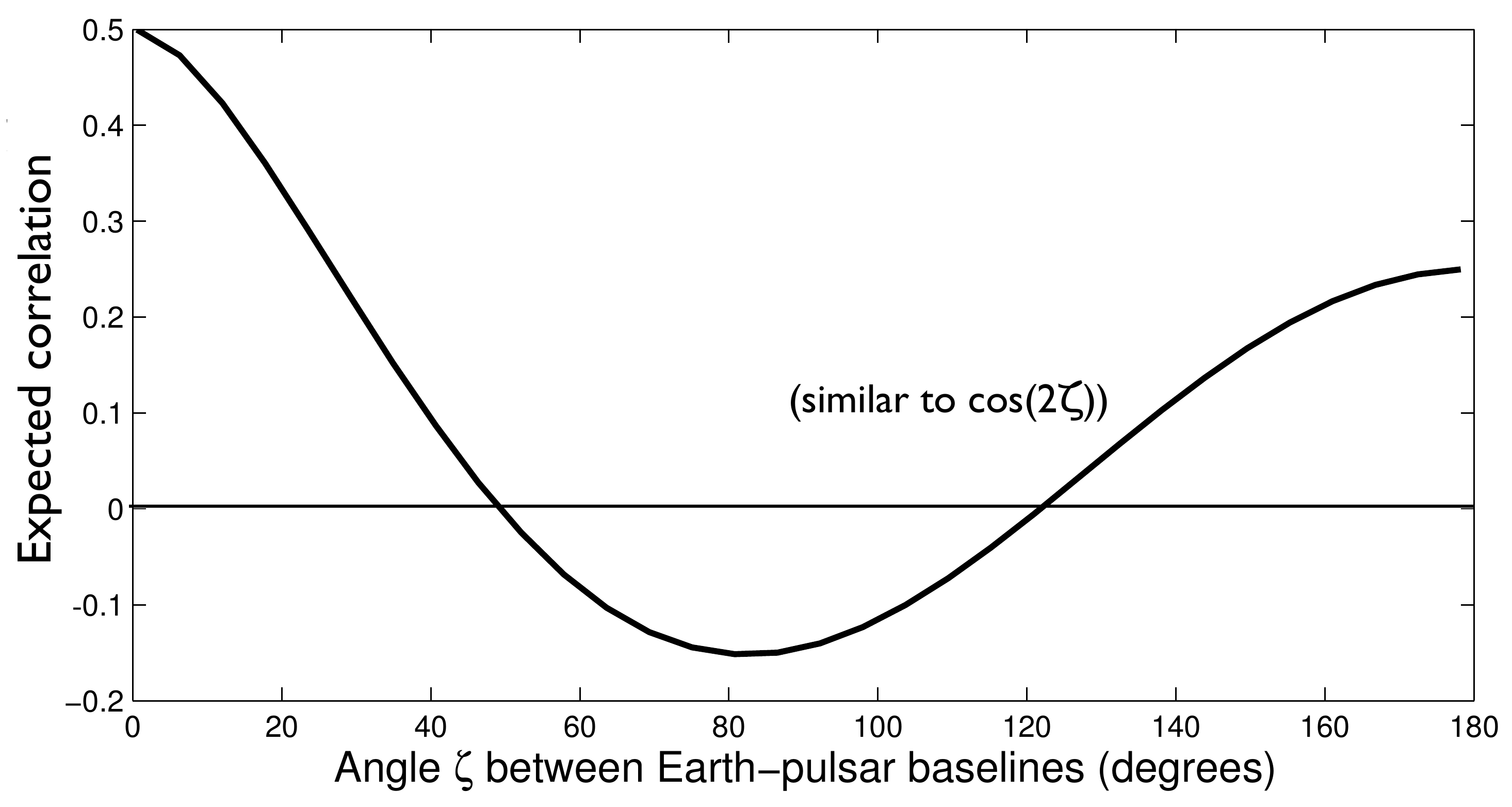}
\caption{Hellings and Downs curve.
Plotted are the values of the expected correlation for 
an unpolarized,
isotropic GWB as a function of the angle $\zeta$ between
two Earth-pulsar baselines.}
\label{f:HD_curve}
\end{center}
\end{figure}

If one uses \eqref{e:F^A(k)} for the Doppler frequency
repsonse of a pulsar timing measurement, then the correlation
between two Earth-pulsar baselines is just a single number as 
the response functions $F^A_{I,J}(\hat k)$ are independent of
frequency.
This number, which can be interpreted as the expected
correlation between the two pulsar timing measurements, 
depends on the angular separation between the two 
Earth-pulsar baselines~\cite{Hellings-Downs:1983}:
\be
\chi(\zeta_{IJ})\equiv
\frac{1}{2} + \frac{3}{2}\left(\frac{1-\cos\zeta_{IJ}}{2}\right)
\left[\ln\left(\frac{1-\cos\zeta_{IJ}}{2}\right) - \frac{1}{6}\right]
+\frac{1}{2}\,\delta_{IJ}\,,
\label{e:HD}
\ee
where $\zeta_{IJ} = \cos^{-1}(\hat p_I\cdot\hat p_J)$, with 
$\hat p_{I,J}$ being unit vectors pointing in the directions
to the pulsars ($\hat p=-\hat u$ in the notation 
of \eqref{e:F^A(k)}).
A plot of this expected correlation as a function of the 
angular separation between the Earth-pulsar baselines is
shown in Figure~\ref{f:HD_curve}.
This is called the {\em Hellings and Downs curve}, originally 
calculated in 1983 by Hellings and Downs~\cite{Hellings-Downs:1983}.
This calculation assumes that the GWB is unpolarized
and isotropic.
Generalizations of the Hellings-Downs curve allowing for 
anisotropy and non-general-relativity polarization modes
can be found in e.g., \cite{Mingarelli-et-al:2013, 
Gair-et-al:2014, Lee-et-al:2007, Chamberlin-Siemens:2012, Gair-et-al:2015}.
The quadrupolar nature of GWs in general relativity is
apparent in the Hellings-Down curve, with an angular dependence
that is qualitatively similar to $\cos(2\zeta)$, where 
$\zeta$ is the angle between two Earth-pulsar baselines.

\begin{figure}[htbp!]
\begin{center}
\includegraphics[width=0.8\textwidth]{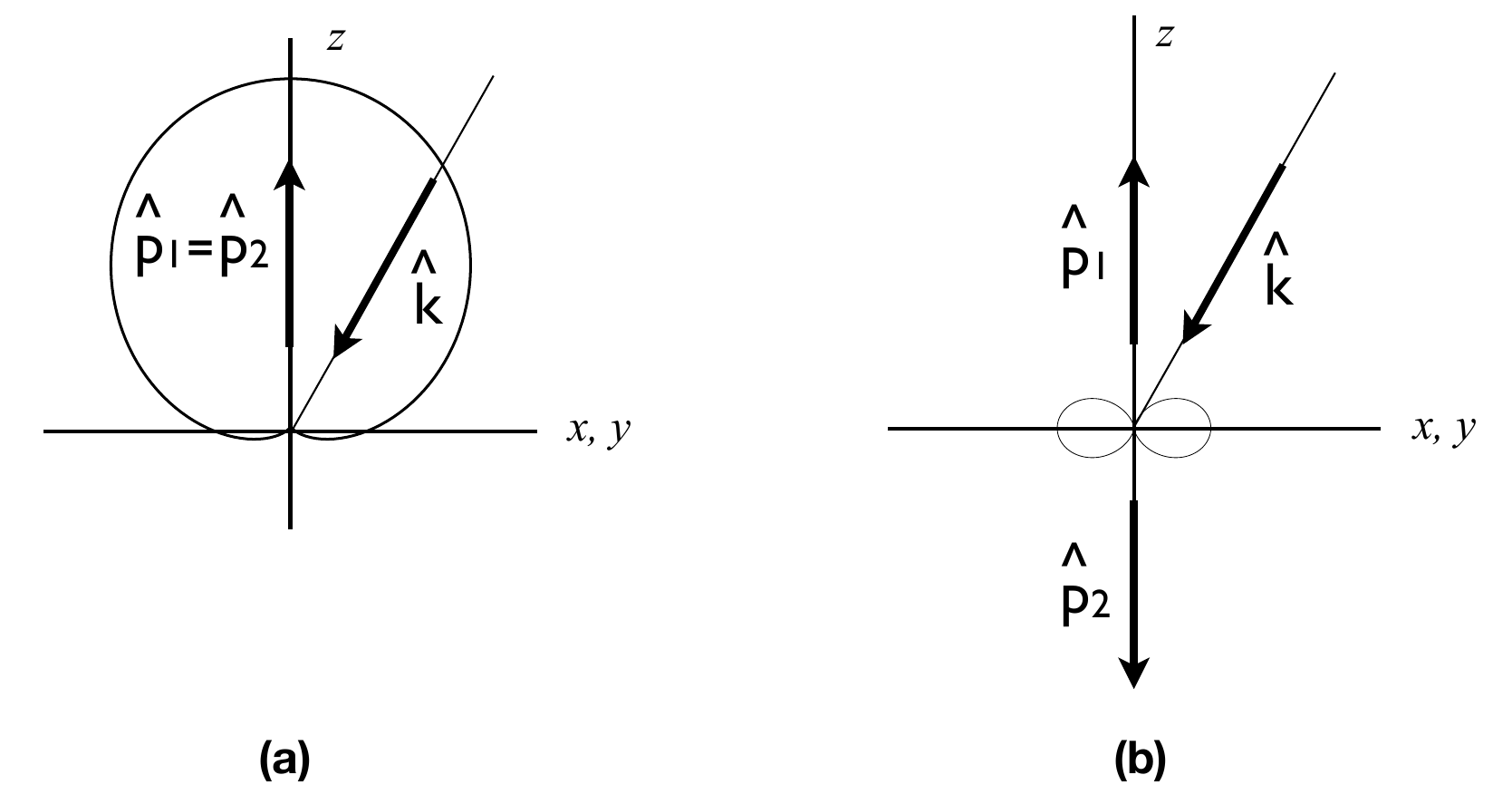}
\caption{Graphical representation of the integrand of the 
(Earth-only) overlap function for pulsar timing Doppler frequency
measurements.
Panel (a): integrand for two Earth-pulsar baselines having
$\zeta = 0^\circ$ ($\hat p_1=\hat p_2=\hat z)$;
Panel (b): integrand for two Earth-pulsar baselines having
$\zeta = 180^\circ$ ($\hat p_1=-\hat p_2 =\hat z)$.
These functions are axially symmetric around the $z$-axis,
which we've chosen to be in the direction to pulsar 1.}
\label{f:pulsar_overlap}
\end{center}
\end{figure}

The fact that $\chi(0^\circ)$ is twice as large as
$\chi(180^\circ)$ can easily be demonstrated by using 
\eqref{e:FA_Earth_z} for the relevant response functions.
Taking the two pulsars to point in the same direction
($\hat p_1=\hat p_2 = \hat z$), we have
\be
\sum_A F_1^A(\hat k) F_2^A(\hat k) = \frac{1}{4}(1+\cos\theta)^2\,,
\ee
while having them point in opposite directions
($\hat p_1 = -\hat p_2=\hat z$) leads to
\be
\sum_A F_1^A(\hat k) F_2^A(\hat k) = \frac{1}{4}(1+\cos\theta)(1-\cos\theta)
= \frac{1}{4}\sin^2\theta\,.
\ee
These functions are plotted in Figure~\ref{f:pulsar_overlap},
where $\theta$ is the usual polar angle measured with respect
to the $z$-axis.
Multiplying by $1/8\pi$ and integrating 
over the sphere, we find:%
\footnote{The factor of 3 difference between 
these two values for $\Gamma_{12}$ and $\chi(0^\circ)=1/2$
and $\chi(180^\circ)=1/4$ is due to a normalization factor that is conventionally
applied to relate $\Gamma_{IJ}$ to $\chi(\zeta_{IJ})$.}
\be
\Gamma_{12} = \frac{1}{6}\quad({\rm for}\ \hat p_1=\hat p_2=\hat z)\,,
\qquad
\Gamma_{12} = \frac{1}{12}\quad({\rm for}\ \hat p_1=-\hat p_2=\hat z)\,.
\ee
From Figure~\ref{f:pulsar_overlap}, we see that when the 
two pulsars both point in the $\hat z$ direction, 
the majority of support for the overlap function comes from 
sky directions $\hat n=-\hat k$ having $z>0$.
When the two pulsars point in opposite directions, 
$\hat p_1=-\hat p_2=\hat z$,
the majority of support for the overlap function comes 
from sky directions in the $xy$-plane, which is
a smaller contribution.

\subsubsection{Overlap function for a pair of electric dipole 
antennae}

\begin{figure}[htbp!]
\begin{center}
\includegraphics[width=0.25\textwidth]{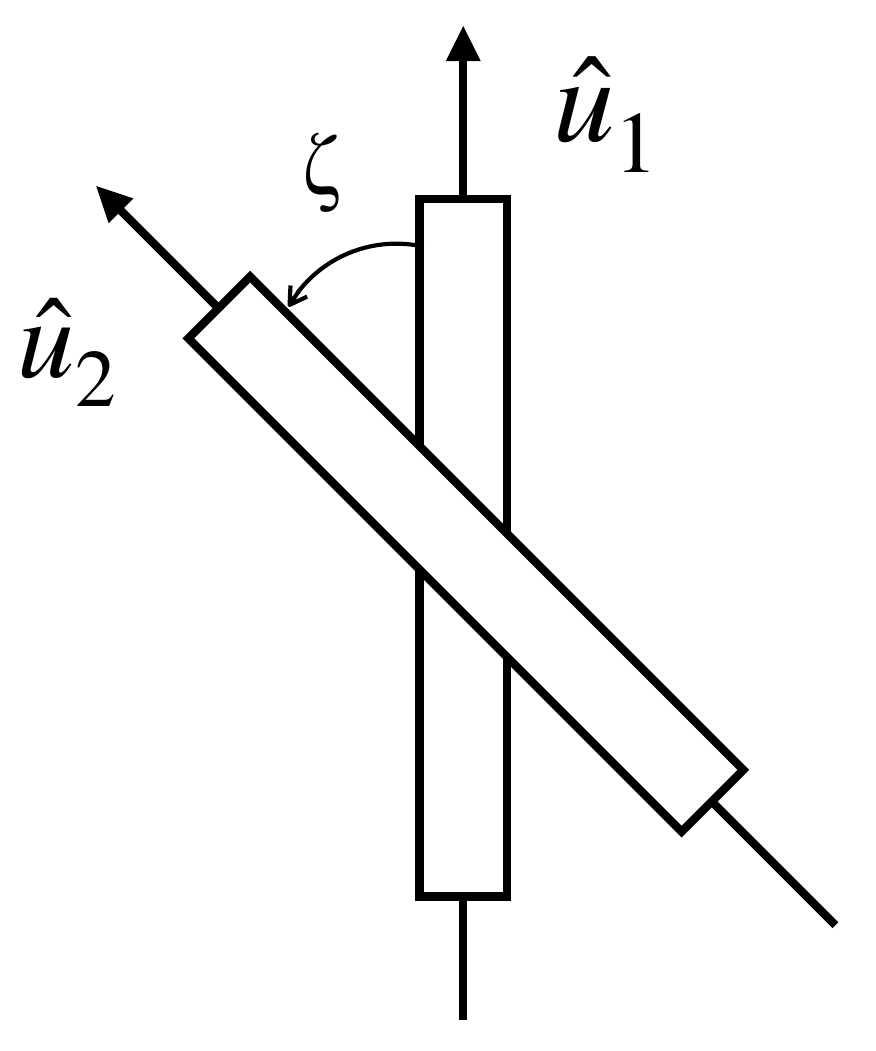}
\caption{Geometry for calculating the overlap function
for a pair of short, colocated electric dipole antennae, 
for an unpolarized and isotropic electric field
(Exercise \ref{exer:7}).}
\label{f:dipole-orf}
\end{center}
\end{figure}

For the final example, you are asked in Exercise~\ref{exer:7} to 
calculate the overlap function for a pair of short, colocated
electric dipole antennae in the presence of an unpolarized
and isotropic electric field $\vec E(t,\vec x)$; see also~\cite{Jenet-Romano:2015}.
The two dipole antennae point in different directions
separated by an angle $\zeta$ as shown in Figure~\ref{f:dipole-orf}.
To do the calculation, you should use the fact that the 
response of dipole antenna $I$ to the electric field is 
\be
r_I(t) = \hat u_I\cdot\vec E(t,\vec x_0)\,.
\ee
The electric field can be expanded in a manner
similar to that for an unpolarized, isotropic GWB:
\be
\vec E(t,\vec x) = \int_{-\infty}^\infty df\>\int d^2\Omega_{\hat k}\> 
\sum_{\alpha =1}^2 \tilde E_\alpha(f,\hat k)\hat\epsilon_\alpha(\hat k) 
e^{i2\pi f(t-\hat k\cdot \vec x/c)}\,,
\ee
where the polarization vectors are given by
\be
\hat\epsilon_1(\hat k) = \hat \theta\,,
\quad
\hat\epsilon_2(\hat k) = \hat \phi\,.
\ee
In addition, the Fourier components $\tilde E_\alpha(f,\hat k)$
satify the quadratic expectation values, cf.~\eqref{e:quad_iso}:
\be
\langle \tilde E_\alpha(f,\vec k) \tilde E_{\alpha'}^*(f',\hat k')\rangle
=\frac{1}{16\pi} S_E(f)\delta(f-f')\delta_{\alpha\alpha'}
\delta^2(\hat k,\hat k')\,.
\ee
With these definitions, it is then just a matter of 
``turning the crank" to calculate
the expectation value $\langle r_1(t) r_2(t')\rangle$, and from
that you can read off the overlap function $\Gamma_{12}(f)$
according to~\eqref{e:Gamma-time}.
You should find
\be
\Gamma_{12}(f) =\frac{1}{3}\,\hat u_1\cdot\hat u_2 
=\frac{1}{3}\cos\zeta\,.
\ee
The dipole nature of the antennae shows up in the $\cos\zeta$ 
dependence of the overlap function.

\section{Statistical inference}
\label{s:statistical_inference}

In order to discuss our final example (Section~\ref{s:nonstationary}),
which is an optimal search for the popcorn-like background
produced by stellar-mass BBH mergers throughout the 
universe, we need to go beyond the frequentist 
statistics that we have used so far 
(Section~\ref{s:correlations}), 
and introduce some concepts from the field of Bayesian inference.
So here we briefly introduce Bayesian inference by comparing 
it to frequentist statistics, focusing mainly on those topics 
needed for the stochastic search that we shall describe in
Section~\ref{s:nonstationary}.
Readers who are interested in more details should 
consult John Veitch's contribution to this Volume,
Section~3 of~\cite{Romano-Cornish:2017}, and e.g.,
\cite{Gregory:2005}.

\subsection{Comparing frequentist statistics and Bayesian inference}

We start by listing the key ingredients of these two formulations 
of statistical inference.
\medskip

\noindent{Frequentist statistics:}

\bi

\i probabilities are long-run relative occurrence of 
outcomes of repeatable experiments (i.e., random variables);
probabilities cannot be assigned to hypotheses or parameters, 
which have fixed but unknown values.

\i one usually starts by writing down a likelihood 
function $p(d|H)$, which is the probability distribution 
for the measured data $d$, assuming the truth of a particular
hypothesis $H$.

\i to estimate the value of parameters and/or to decide
between different hypotheses, one constructs 
{\em statistics}, which are particular functions of the data.

\i to make probabilistic statements about 
parameter estimates or hypothesis testing, 
one needs to calculate
the probability distributions of the statistics;
this can be done either analytically or numerically 
(e.g., using time slides to produce different realistic
noise realizations of the data).

\i given the probability distributions of the statistics,
one can then construct confidence intervals and 
$p$-values (the probability of obtaining a detection
statistic value as large or larger than what was measured) 
for parameter estimation and hypothesis testing.

\ei

\noindent{Bayesian inference:}

\bi

\i probability is degree of belief (or confidence)
in any proposition, and hence can be assigned to 
hypotheses and parameters.
(This is a more general definition of probability than 
that used in frequentist statististics.)

\i like a frequentist, one usually starts by writing 
down a likelihood function $p(d|H)$.

\i in addition to the likelihood function, one needs to
specify prior probability distributions for the various 
parameters and hypotheses that one is considering.

\i one uses Bayes' theorem to update the prior degree 
of belief in a parameter value or hypothesis in light 
of new data.

\i one constructs posterior distributions and odds 
ratios (or {\em Bayes factors}, see Section~\ref{s:bayes_factors}) 
for parameter estimation and hypothesis testing (also 
called model selection).

\ei
\noindent
In a nutshell, the main different between Bayesian and 
frequentist statistics is the definition of probability.
As such, certain probabilistic statements that you can
make as a Bayesian are not valid from a frequentist perspective.
Hence, Bayesian and frequentist statistics often ask 
(and subsequently answer) {\em different} questions 
about the data.
Nonetheless, despite this fundamental difference in approach,
if the data are sufficiently informative (i.e., if the 
likelihood is peaked relative to the prior distributions 
for the parameters, see Figure~\ref{f:informative_data}), 
then both Bayesian and frequentist analyses give more or 
less consistent results.

\subsubsection{Likelihood functions}
\label{s:likelihood}

As mentioned above, the starting point for most 
frequentist and Bayesian analyses is a likelihood function,
which we can write schematically as
\be
{\rm likelihood} = p({\rm data}| {\rm parameters}, {\rm model})\,.
\ee
For example, for Gaussian-distributed detector noise 
and a Gaussian-distributed GWB, the likelihood 
function for the noise-only model ${\cal M}_0$ and 
signal+noise model ${\cal M}_1$ are given by
\begin{align}
&p(d|S_{n_1}, S_{n_2}, {\cal M}_0) 
= \frac{1}{\sqrt{{\rm det}(2\pi C_n)}}\, 
\exp\left[-\frac{1}{2} d^T C_n^{-1} d\right]\,,
\label{e:noise_likelihood}
\\
&p(d|S_{n_1}, S_{n_2}, S_h, {\cal M}_1) 
= \frac{1}{\sqrt{{\rm det}(2\pi C)}}\, 
\exp\left[-\frac{1}{2} d^T C^{-1} d\right]\,,
\label{e:stochastic_likelihood}
\end{align}
where $C_n$ and $C$ are the covariance matrices
for the noise-only and signal+noise models, respectively.
For $N$ samples of white noise and white GWB in
two colocated and coaligned detectors:
\be
C_n = \begin{bmatrix}
S_{n_1}\,{\mathsf 1}_{N\times N} & {\mathsf 0}_{N\times N}\\
{\mathsf 0}_{N\times N} & S_{n_2}\,{\mathsf 1}_{N\times N}
\end{bmatrix}\,,
\qquad
C = \begin{bmatrix}
(S_{n_1} + S_h)\,{\mathsf 1}_{N\times N} & S_h\,{\mathsf 1}_{N\times N}\\
S_h\,{\mathsf 1}_{N\times N} & (S_{n_2}+S_h)\,{\mathsf 1}_{N\times N}
\end{bmatrix}\,,
\label{e:covariance_matrices}
\ee
where ${\mathsf 1}_{N\times N}$ and ${\mathsf 0}_{N\times N}$
denote the unit matrix and zero matrix, respectively, in $N$ dimensions.
For this simple case, there are only three relevant 
parameters: $S_{n_1}$, $S_{n_2}$ for the detector noise, 
and $S_h$ for the GWB.
Also, by assuming that the detectors are colocated and 
coaligned, we don't have to worry about including an 
overlap function in the
off-diagonal blocks of the signal+noise covariance matrix $C$.

\subsection{Frequentist analyses}
\label{s:frequentist}

Starting from the likelihood functions for the noise-only
and signal+noise models, we can construct the 
{\em maximum-likelihood ratio} statistic:
\be
\Lambda_{\rm ML}(d)\equiv\frac{
{\rm max}_{S_{n_1},S_{n_2},S_h}\,p(d|S_{n_1} S_{n_2},S_h,{\cal M}_1)}
{{\rm max}_{S_{n_1},S_{n_2}}\,p(d|S_{n_1}, S_{n_2},{\cal M}_0)}\,.
\label{e:Lambda_ML}
\ee
The values of the parameters that maximize the likelihood
for the signal+noise model can be used as 
frequentist estimators of the true values of the parameters 
$S_{n_1}$, $S_{n_2}$, $S_h$.
In Exercise~\ref{exer:8} you are asked to show that the 
data combinations
\be
\hat C_{11} \equiv 
\frac{1}{N}\sum_{i=1}^N d_{1i}^2\,,
\qquad
\hat C_{22} \equiv 
\frac{1}{N}\sum_{i=1}^N d_{2i}^2\,,
\qquad
\hat C_{12} \equiv 
\frac{1}{N}\sum_{i=1}^N d_{1i}d_{2i}\,,
\ee
are maximum-likelihood estimators of 
\be
S_1\equiv S_{n_1}+S_h\,,\quad
S_2\equiv S_{n_2}+S_h\,,\quad
S_h\,.
\ee
Note that the maximum-likelihood estimator 
$\hat S_h\equiv \hat C_{12}$ of $S_h$ is just the standard 
cross-correlation statistic introduced in \eqref{e:Sh_ML}.
The maximum-likelihood estimators of the detector
noise $S_{n_1}$, $S_{n_2}$ are then
\be
\hat S_{n_1}\equiv 
\hat S_1-\hat S_h=
\hat C_{11}-\hat C_{12}\,,
\qquad
\hat S_{n_2}\equiv
\hat S_2-\hat S_h=
\hat C_{22}-\hat C_{12}\,.
\ee
In addition, in Exercise~\ref{exer:9}, you are asked to 
show that 
\be
\Lambda(d)\equiv 2\ln(\Lambda_{\rm ML}(d))
\simeq\frac{\hat C_{12}^2}{\hat C_{11}\hat C_{22}/N}\,,
\label{e:Lambda}
\ee
which holds in the weak-signal approximation, 
where $S_h\ll S_{n_1}, S_{n_2}$.
The quantity $\Lambda(d)$ can be used as a frequentist detection 
statistic, comparing its value for the given data $d$ 
to a threshold $\Lambda_*$.
If $\Lambda(d)\ge \Lambda_*$, we reject the null 
hypothesis (the noise-only model) and claim detection 
of a GW signal; if $\Lambda(d)<\Lambda_*$, we accept
the null-hypothesis and reject the signal+noise hypothesis.
Note that the right-hand-side of \eqref{e:Lambda} is the 
square of the 
(power) signal-to-noise ratio, cf.~\eqref{e:rho_expected},
which illustrates a 
useful general relation between signal-to-noise ratios 
and the maximum-likelihood statisitic.

\subsection{Bayesian analyses}
\label{s:bayesian}

\begin{figure}[htbp!]
\begin{center}
\includegraphics[width=0.45\textwidth]{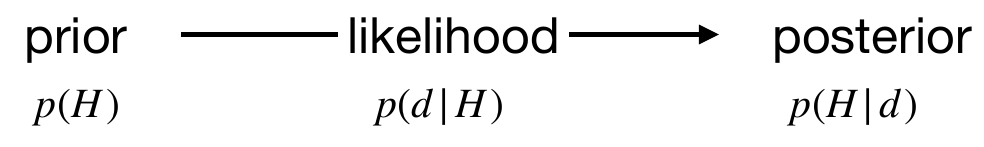}
\caption{Schematic representation of Bayes' theorem.}
\label{f:bayes_theorem}
\end{center}
\end{figure}

Not suprisingly, Bayesian analyses make use of Bayes'
theorem:
\be
p(H|d) = \frac{p(d|H) p(H)}{p(d)}\,,
\label{e:bayes_theorem}
\ee
which converts probabilities about the data $d$ given 
a hypothesis $H$ (the likelihood 
$p(d|H)$) to probabilites about the hypothesis given
the data (the posterior distribution $p(H|d)$).%
\footnote{Conditional probabilities 
$p(A|B)$ and $p(B|A)$ are not equal in general.
Paraphrasing an example from Louis Lyons:
the probability that a person is pregnant ($A$) 
given that that person is a woman  ($B$) is about 
about 30\%; while the probability that a person is 
a woman ($B$) given that that person is
pregnant ($A$) is 100\%.}
Here, $p(H)$ is the prior probability distribution 
for $H$, and $p(d)$ is the {\em evidence} or 
{\em marginalized likelihood}:
\be
p(d) \equiv \int {\rm d}H\>
p(d|H)p(H)\,.
\ee
Note that the evidence is simply the normalization 
factor needed to insure $\int {\rm d}H\>p(H|d)=1$.
The importance of Bayes' theorem is that it updates 
our degree of belief in a hypothesis in light
of new data.
It maps the prior probability $p(H)$ to the posterior 
probability $p(H|d)$ via the likelihood function 
$p(d|H)$
(Figure~\ref{f:bayes_theorem}).

In the context of searches for stochastic GW backgrounds,
Bayes' theorem has the form:
\be
p(S_{n_1}, S_{n_2}, S_h|d,{\cal M}_1) 
= \frac{p(d|S_{n_1}, S_{n_2}, S_h, {\cal M}_1)
p(S_{n_1}, S_{n_2}, S_h|{\cal M}_1)}{p(d|{\cal M}_1)}\,,
\ee
where $p(d|S_{n_1}, S_{n_2}, S_h,{\cal M}_1)$ is the likelihood
function \eqref{e:stochastic_likelihood}.
Here ${\cal M}_1$ is our signal+noise model and
$S_{n_1}$, $S_{n_2}$, $S_h$ are the parameters 
describing this model.
The quantity 
$p(S_{n_1}, S_{n_2}, S_h|d,{\cal M}_1)$ is the 
{\em joint} posterior probability distribution for
the parameters $S_{n_1}$, $S_{n_2}$, $S_h$ of 
model ${\cal M}_1$ given the data $d$.
The posterior distribution for $S_h$ alone is 
given by integrating over $S_{n_1}$, $S_{n_2}$:
\be
p(S_h|d,{\cal M}_1) 
= \int {\rm d}S_{n_1}\>\int{\rm d}S_{n_2}\>p(S_{n_1}, S_{n_2}, S_h|d,{\cal M}_1)\,.
\ee

\subsubsection{Bayes factors and model selection}
\label{s:bayes_factors}

To assess which of two models 
${\cal M}_0$, ${\cal M}_1$ is more consistent with
the observed data $d$, we form 
the ratio of the posterior distributions 
$p({\cal M}_0|d)$, $p({\cal M}_1|d)$.
Using Bayes' theorem, we obtain
\be
\frac{p({\cal M}_1|d)}{p({\cal M}_0|d)} =
\frac{p(d|{\cal M}_1)\,p({\cal M}_1)}
{p(d|{\cal M}_0)\,p({\cal M}_0)}\,,
\ee
where the common evidence term $p(d)$ in
\eqref{e:bayes_theorem} has canceled out 
when taking the ratio of the two posteriors.
Thus, we see that the posterior odds ratio
$O_{10}(d)\equiv p({\cal M}_1|d)/p({\cal M}_0|d)$
is equal to the prior odds ratio
$O_{10}\equiv p({\cal M}_1)/p({\cal M}_0)$
times the {\em Bayes factor}
\be
{\cal B}_{10}(d)\equiv \frac{p(d|{\cal M}_1)}{p(d|{\cal M}_0)}\,.
\label{e:bayes_factor}
\ee
The numerator and denominator in the Bayes factor 
are the marginalized likelihoods obtained by 
marginalizing the full likelihood functions
$p(d|\theta_\alpha,{\cal M}_\alpha)$ over the 
model parameters $\theta_\alpha$:
\be
p(d|{\cal M}_\alpha) \equiv
\int {\rm d}\theta_\alpha\>
p(d|\theta_\alpha, {\cal M}_\alpha)p(\theta_\alpha|{\cal M}_\alpha)\,,
\ee
where $\alpha=0,1$ labels the two models.
If there is no {\em a~priori} reason to prefer one model over 
the other (i.e., if the prior odds ratio $O_{10}=1$), then the 
posterior odds ratio for the two models is equal to the 
Bayes' factor, $O_{10}(d) = {\cal B}_{10}(d)$.

\begin{figure}[htbp!]
\begin{center}
\includegraphics[width=0.5\textwidth]{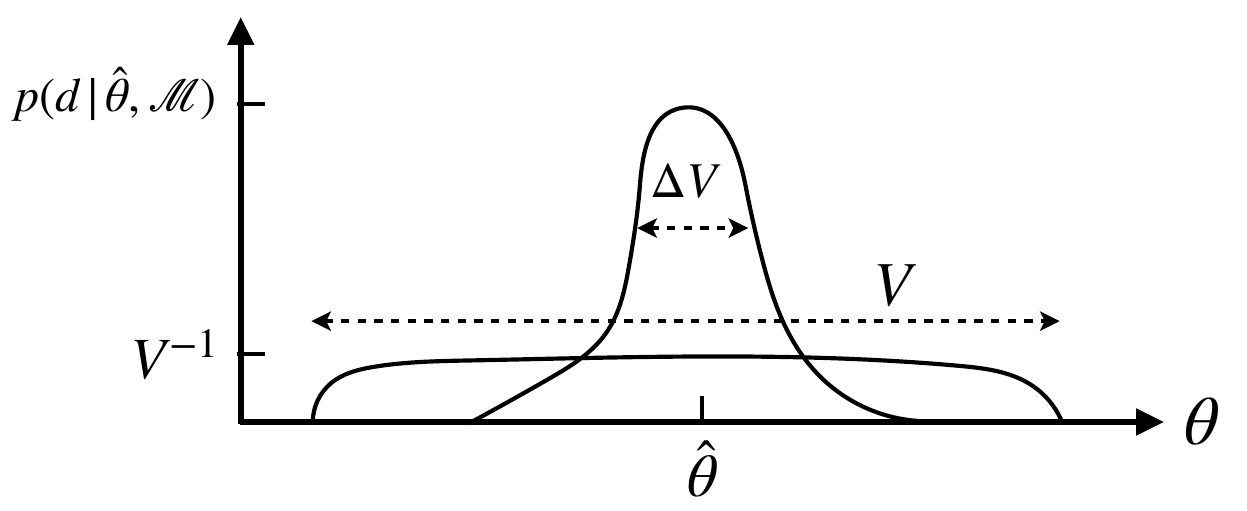}
\caption{Schematic representation of the likelihood function 
$p(d|\theta,{\cal M})$ and prior probability distribution 
$p(\theta|{\cal M})$ for the model parameters $\theta$,
when the data $d$ are informative.
In this case, the likelihood function is peaked relative to the 
prior probability distribution, with maximum at 
$\theta=\hat\theta$ and characteristic width $\Delta V$.
The full parameter space volume is denoted by $V$.}
\label{f:informative_data}
\end{center}
\end{figure}

Using the above definitions, we are now in a position to 
relate Bayesian and frequentist inference, at least in the 
case where the data are informative.
By this we mean that the likelihood function for a given
model is peaked relative to the prior probability 
distribution for its model parameters 
(Figure~\ref{f:informative_data}).
For this case, the marginalized likelihoods functions 
have the approximate form 
\be
p(d|{\cal M}_\alpha) 
\simeq
p(d|\hat\theta_\alpha, {\cal M}_\alpha)
\,{\Delta V_\alpha}/{V_\alpha}\,,
\ee
where $\hat\theta_\alpha$ denote the parameter values that
maximize the likelihood, $\Delta V_\alpha$ is the range of
parameter values over which the likelihood is peaked, and 
$V_\alpha$ denotes the full parameter volume.
This approximation is called the {\em Laplace approximation}.
Substituting these expressions into \eqref{e:bayes_factor} we find
\be
{\cal B}_{10}(d) 
\equiv\frac{p(d|{\cal M}_1)}{p(d|{\cal M}_0)}
\simeq \frac{p(d|\hat\theta_1,{\cal M}_1)}{p(d|\hat\theta_0,{\cal M}_0)}
\,\frac{\Delta V_1/V_1}{\Delta V_0/V_0}
\simeq\Lambda_{\rm ML}(d)
\,\frac{\Delta V_1/V_1}{\Delta V_0/V_0}\,,
\ee
where $\Lambda_{\rm ML}(d)$ is just the maximum-likelihood
ratio for the two models.
This last expression can also be written as
\be
2\ln({\cal B}_{10}(d)) \simeq \Lambda(d) + 
2\ln\left(\frac{\Delta V_1/V_1}{\Delta V_0/V_0}\right)
\label{e:BF-SNR}
\ee
where $\Lambda(d)\equiv 2\ln(\Lambda_{\rm ML}(d))$
plays the role of a frequentist detection statistic, 
and the last term is an Occam's factor that penalizes
models that use more parameter space volume than needed
to fit the data.
As shown in~\eqref{e:Lambda}, $\Lambda(d)$ is effectively
a squared signal-to-noise ratio.
The key observation here is that the ratio of marginalized
likelihoods, i.e., the Bayes factor, 
is well approximated by a maximum-likelihood 
ratio times an Occam's penalty factor when the data are informative.

\subsubsection{Bayesian signal priors}
\label{s:signal_priors}

The final piece of information that we will need 
for discussing the Bayesian search in 
Section~\ref{s:nonstationary} is the choice 
of signal prior.
We shall see in that section that by choosing the 
signal prior appropriately, we can properly model 
the popcorn-like nature of a GWB produced by
stellar-mass BBH mergers throughout the universe.

Here we illustrate the effect of chosing different
priors for two simple cases: 
(i) a deterministic GW signal (a sinusoid), and 
(ii) a Gaussian-stationary stochastic background in 
Gaussian-distributed noise.
For both of these cases, the difference between the 
observed data $d$ and signal model $h$ is the noise $n$.
So we can write down a generic likelihood function for 
the data $d$ by equating it to the 
Gaussian-distributed noise likelihood for the residuals $d-h$:
\begin{equation}
p(d|C_n,h) \equiv p_n(d-h|C_n) 
= \frac{1}{\sqrt{{\rm det}(2\pi C_n)}}\,
\exp\left[-\frac{1}{2}(d-h)^T C_n^{-1}(d-h)\right]\,,
\label{e:likelihood_generic}
\ee
where $C_n$ is the noise covariance matrix.
To proceed further we need to specify the form of the 
signal $h$.

(i) For a deterministic GW signal, we expect the signal
samples to have a precise form, e.g., a sine wave
parametrized by its amplitude, frequency, and initial phase 
(Figure~\ref{f:det_signal_prior}).
For this case the signal prior is a Dirac delta function that
sets the signal samples to the model waveform,
\begin{equation}
p(h|A, f_0, \phi_0)
=\delta\left(h(t)- A\sin(2\pi f_0 t + \phi_0)\right)\,.
\label{e:prior_deterministic}
\ee
Multiplying the likelihood \eqref{e:likelihood_generic}
by this prior and then (trivially) marginalizing over 
the signal samples $h$ yields
\be
\begin{aligned}
&p(d|C_n,A,f_0,\phi_0)
\equiv \int {\rm d}h\>p_n(d-h|C_n)p(h|A, f_0, \phi_0) 
\\
&\quad
=\frac{1}{\sqrt{{\rm det}(2\pi C_n)}}\,
\exp\left[-\frac{1}{2}\sum_{i,j}
(d_i-A\sin(2\pi f_0 t_i + \phi_0)) 
[C_n^{-1}]_{ij}(d_j-A\sin(2\pi f_0 t_j + \phi_0))\right]\,.
\label{e:likelihood_deterministic}
\end{aligned}
\ee
This marginalized likelihood function with priors on the range of
parameter values for $A$, $f_0$, $\phi_0$ (for the signal) 
and the covariance matrix $C_n$ (for the noise) then completely defines 
the deterministic signal+noise model.
\begin{figure}[htbp!]
\begin{center}
\subfigure[]{\includegraphics[width=0.25\textwidth]{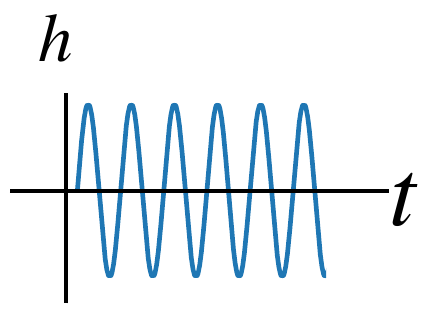}
\label{f:det_signal_prior}}
\hspace{1 in}
\subfigure[]{\includegraphics[width=0.25\textwidth]{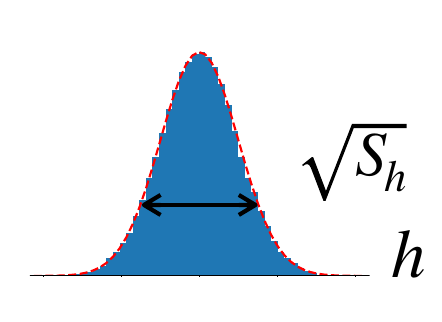}
\label{f:stoch_signal_prior}}
\caption{Different signal priors for $h(t)$.
Panel (a): Determinsitic (sinusoid) signal prior.
Panel (b): Stochastic signal prior.
For the stochastic signal prior, $h(t)$ values are drawn from a
Gaussian distribution with variance $S_h$.}
\label{f:det_stoch_signal_priors}
\end{center}
\end{figure}

(ii) For a stochastic GW signal, we cannot predict with
certainty what the
individual samples $h$ will be; we can only say that they 
are drawn from some probability distribution.  
Taking that distribution to be a Gaussian with zero mean
and variance $S_h$ (Figure~\ref{f:stoch_signal_prior}),
we have
\begin{equation}
p(h|S_h) 
= \frac{1}{\sqrt{2\pi S_h}}\,\exp\left[-\frac{1}{2}\frac{h^2}{S_h}\right]\,.
\label{e:stoch-prior}
\ee
So multiplying the likelihood \eqref{e:likelihood_generic}
by this prior and marginalizing over the signal samples
$h$ yields for this case
\begin{equation}
p(d|C_n,S_h) 
\equiv\int {\rm d}h\>p_n(d-h|C_n)p(h|S_h) 
= \frac{1}{\sqrt{{\rm det}(2\pi C)}}\,\exp\left[-\frac{1}{2}d^T C^{-1} d\right]\,,
\ee
where $C \equiv C_n + S_h$.
This marginalized likelihood with a prior on the 
variance $S_h$ for the stochastic signal samples 
and covariance matrix $C_n$ for the noise then
completely defines the Gaussian-stationary stochastic
signal+noise model.

In Exercise~\ref{exer:10} you are asked to extend the
above analysis for a stochastic background to the 
case of two coincident and coaligned detectors with 
uncorrelated detector noise.
You should start with the generic two-detector likelihood
function
\begin{equation}
p(d|C_n,h) \equiv p_n(d-h|C_n) 
= \frac{1}{\sqrt{{\rm det}(2\pi C_n)}}\,\exp\left[-\frac{1}{2}(d-h)^T C_n^{-1}(d-h)\right]\,,
\ee
where
\begin{equation}
C_n = \begin{bmatrix}
S_{n_1} & 0\\
0 & S_{n_2}
\end{bmatrix}\,,
\ee
and then marginalize over $h$ using \eqref{e:stoch-prior}.
The final result should be
\be
p(d|C_n, S_h)
= \frac{1}{\sqrt{{\rm det}(2\pi C)}}\,\exp\left[-\frac{1}{2}d^T C^{-1} d\right],
\ee
where
\begin{equation}
C = \begin{bmatrix}
S_{n_1} + S_h & S_h\\
S_h & S_{n_2} + S_h
\end{bmatrix}\,.
\ee
Note that the stochastic background contributes to both 
the diagonal and off-diagonal components of the covariance matrix.
(The overlap function doesn't appear since we have assumed
coincident and coaligned detectors.)
This marginalized likelihood is usually taken as the starting
point for all stochastic cross-correlation searches using
multiple detectors; see~\eqref{e:stochastic_likelihood}.

\section{Searching for the background of binary black-hole mergers}
\label{s:nonstationary}

As discussed in Section~\ref{s:BBH-BNS-LIGO}, the non-continuous 
popcorn-like background from BBH mergers is a potential signal 
for the advanced LIGO and Virgo network of detectors.
The recent detections of several large signal-to-noise ratio
BBH and BNS mergers imply the existence of a stochastic GW 
background composed of the more distant, weaker events.
In 2018, Smith \& Thrane~\cite{Smith-Thrane:2018} 
proposed an alternative to the standard cross-correlation 
method (Section~\ref{s:correlations})
to search for the BBH component, optimally suited for the 
popcorn-nature of the signal.
This was done by describing the BBH 
background with a ``mixture" signal prior consisting of
a BBH chirp in a certain fraction $\xi$ of the analyzed 
segments, and just noise for the remaining segments.
Also, as the individual signals will not be resolvable, they 
choose to marginalize over the BBH chirp parameters 
leaving only the probability parameter $\xi$ (which is 
simply related to the rate of BBH merger signals) to estimate.
Although in principle they can do the analysis with a
single detector, they use two detectors to help 
discriminate against glitches.

\subsection{Analysis details}

The key components of their analysis are as follows:
\smallskip

\noindent
1) Begin by splitting the data into short (e.g., 4 sec) 
segments, $i=1,2,\cdots,N_{\rm seg}$, which 
should contain at most one BBH merger signal.

\smallskip
\noindent
2) Choose a {\em mixture} signal prior for the signal model:
\begin{equation}
p(h|\xi, \vec\lambda) 
= \xi\,\delta\left(h-{\rm chirp}(\vec\lambda)\right) +
(1-\xi)\,\delta(h)\,,
\label{e:mixture-signal-prior}
\ee
which consists of a BBH chirp signal with probability
$\xi$ and just noise ($h=0$) with probability $(1-\xi)$
(Figure~\ref{f:mixture_signal_priors}).
This mixture signal prior captures the non-continuous
popcorn-like nature of the BBH mergers.
\begin{figure}[htbp!]
\begin{center}
\subfigure[]{\includegraphics[width=0.25\textwidth]{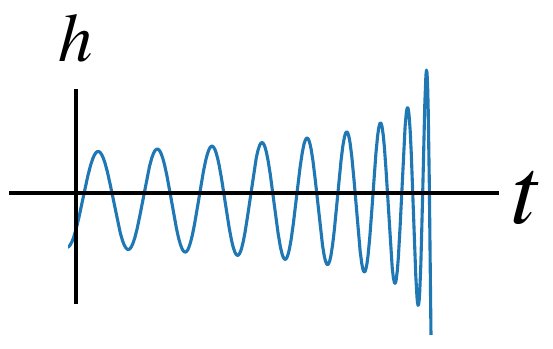}}
\hspace{1 in}
\subfigure[]{\includegraphics[width=0.25\textwidth]{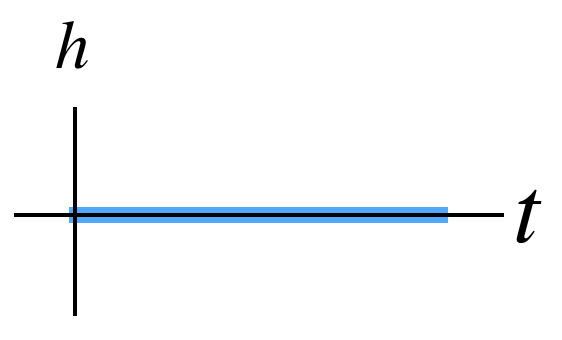}}
\caption{The two components of the mixture signal prior for the 
Bayesian BBH merger search.
Panel (a): With probability $\xi$, the signal prior for $h(t)$ is a 
chirp waveform.
Panel (b): With probability $(1-\xi)$, the signal prior for $h(t)$ is
just noise, i.e., $h(t)=0$.}
\label{f:mixture_signal_priors}
\end{center}
\end{figure}

\smallskip
\noindent
3) For each segment of data $d_i$, marginalize
the generic likelihood function
\begin{equation}
p(d_i|C_n,h)\equiv p_n(d_i-h|C_n)
\ee
over the signal samples $h$ using the signal prior 
\eqref{e:mixture-signal-prior}:
\be
\begin{aligned}
p(d_i|\xi, \vec\lambda, C_n)
&=\int {\rm d}h\>p(d_i|C_n,h)p(h|\xi,\vec\lambda)
\\
&= \xi\,p_n(d_i-{\rm chirp}(\vec\lambda)|C_n) + (1-\xi)\,p_n(d_i|C_n)\,.
\end{aligned}
\ee
Note that this is a mixture-Gaussian likelihood function.

\smallskip
\noindent
4) Further marginalize over the BBH signal
parameters $\vec\lambda$, and use an estimate of 
the detector noise thus fixing $C_n\equiv \bar C_n$:%
\begin{equation}
p(d_i|\xi)=\int {\rm d}\vec\lambda\>
p(d_i|\xi,\vec\lambda, \bar C_n)\,p(\vec\lambda)
\equiv (S_i-N_i)\xi + N_i\,,
\label{e:p(d_i|xi)}
\ee
where $N_i \equiv p(d_i|\bar C_n)$ and 
$S_i\equiv \int {\rm d}\vec\lambda\> p(d_i|\xi,\vec\lambda,\bar C_n)\,p(\vec \lambda)$.

\smallskip
\noindent
5) Calculate the posterior for each segment
using Bayes' theorem
\be
p(\xi|d_i)
=\frac{p(d_i|\xi)p(\xi)}{p(d_i)}\,,
\ee
taking $p(\xi)={\rm const}$.

\smallskip
\noindent
6) Finally, combine segments
by multiplying the individual marginalized likelihoods
\be
p(d|\xi) = \prod_{i} p(d_i|\xi)\,.
\ee
The final posterior distribution $p(\xi|d)$ is
proportional to the product of the individual segment 
posteriors $p(\xi|d_i)$ since $p(\xi)={\rm const}$.

\subsection{Illustrating the analysis method on simulated data}

We now illustrate the method on some simulated toy-model
data.%
\footnote{The simulated data and analysis routines are
publicly available at \cite{github-code}.}
The simulated time-series for the two detectors are each 
only 10~s long, and our simulated BBH chirps are less than 
0.25~sec in duration.
\begin{figure}[htbp!]
\begin{center}
\includegraphics[width=\textwidth]{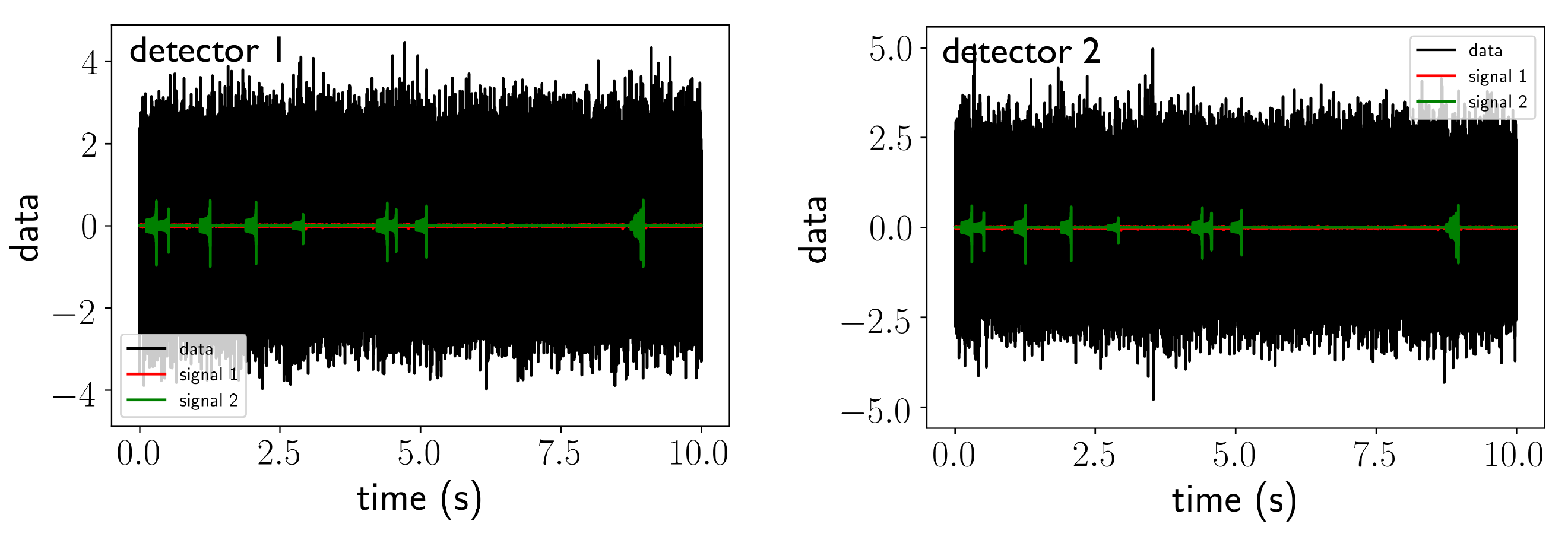}
\caption{Simulated BBH and BNS data in two coincident and coaligned
detectors.
The confusion-limited BNS background is shown in orange;
the popcorn-like BBH background is shown in green.
The black trace is the data consisting of the BBH and BNS signals
plus white Gaussian-stationary noise, uncorrelated in the two
detectors.}
\label{f:BBH-BNS-simulated-data}
\end{center}
\end{figure}
(Since this is only a toy-model, I didn't worry about 
making it astrophysically realistic.)
We divided the simulated data into 40 segments 
(each of duration 0.25~s), and we injected 10
signals into uncorrelated white Gaussian noise in two 
coincident and coaligned detectors, corresponding to 
a injected value of the probability parameter $\xi = 0.25$ 
(Figure~\ref{f:BBH-BNS-simulated-data}).
The signal parameters $\vec\lambda$ that we 
marginalized over were just the amplitude and time of 
arrival of a BBH chirp in each segment.  
We assumed that we knew the shape and duration of the signal.

The final result of the analysis is the posterior distribution
for $\xi$, shown in Figure~\ref{f:posterior_xi}.
\begin{figure}[htbp!]
\begin{center}
\includegraphics[width=0.4\textwidth]{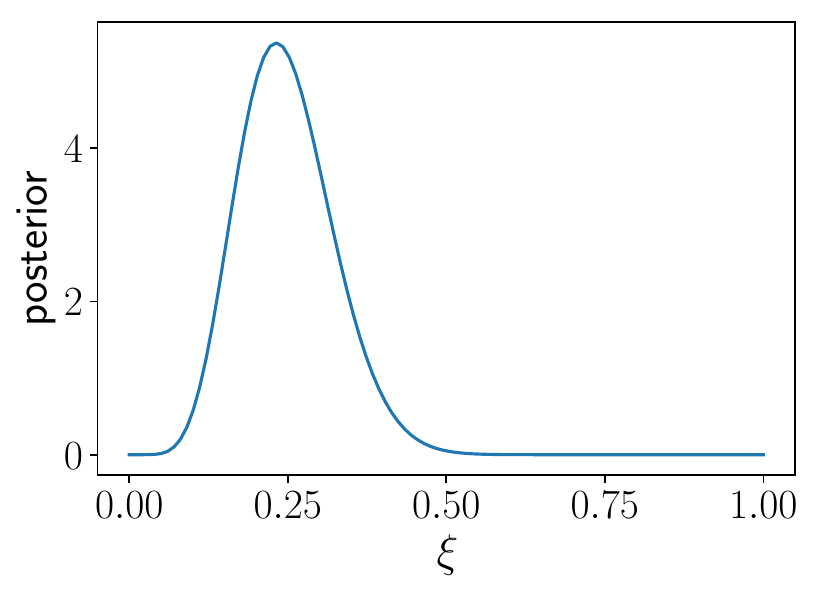}
\caption{The cumulative posterior distribution for $\xi$ after 
combining all 40~segments of data.}
\label{f:posterior_xi}
\end{center}
\end{figure}
One sees that it is peaked around the value of $\xi$ used for 
the injections, $\xi=0.25$.
Figure~\ref{f:posteriors_xi_seg} shows the posterior distributions
for $\xi$ for the first 16 segments (first 4~sec) of data.
Note that these distributions are all linear in $\xi$, as to be
expected from \eqref{e:p(d_i|xi)} for the individual-segment 
likelihood functions.
In addition,
the cumulative posterior distributions for $\xi$, obtained by 
combining the likelihood functions for the first $n$ segments of
data are shown in Figure~\ref{f:posteriors_xi_cum} for 
$n=1$, 2, 3, 4, 10, 20, 30, and 40 segments.
As $n$ increases, the product of the individual linear functions 
of $\xi$,
some with positive slope (when there is evidence in favor of the 
presence of a signal) and some with negative slope (when there is
evidence in favor of the absence of a signal), give rise to a 
distribution that gets more and more peaked.
The bottom-rightmost plot is just the final posterior distribution
shown in Figure~\ref{f:posterior_xi}.
\begin{figure}[htbp!]
\begin{center}
\includegraphics[width=0.24\textwidth]{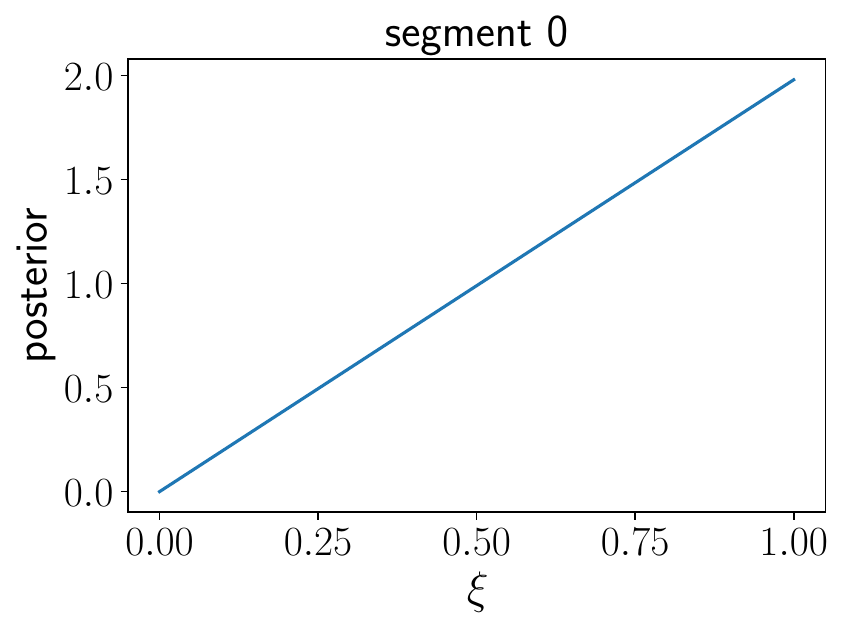}
\includegraphics[width=0.24\textwidth]{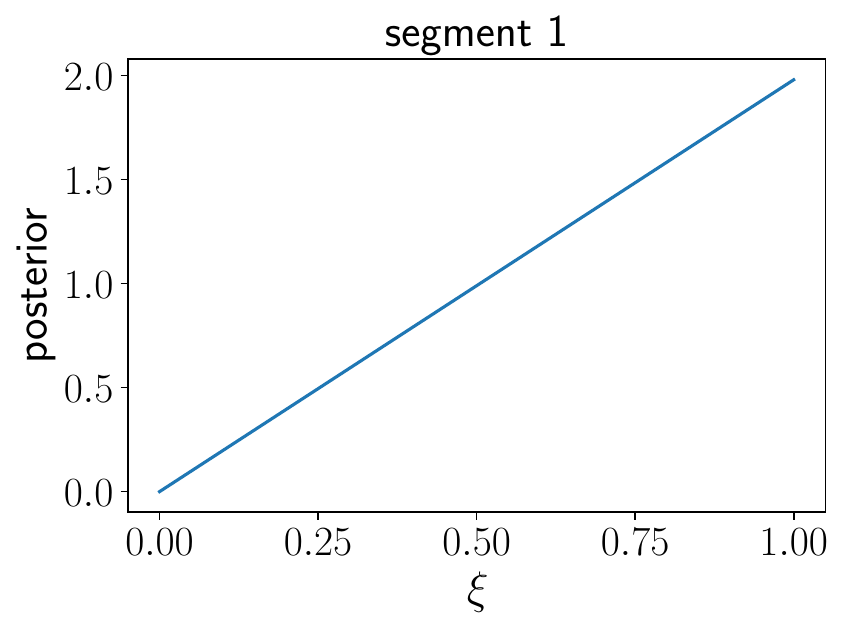}
\includegraphics[width=0.24\textwidth]{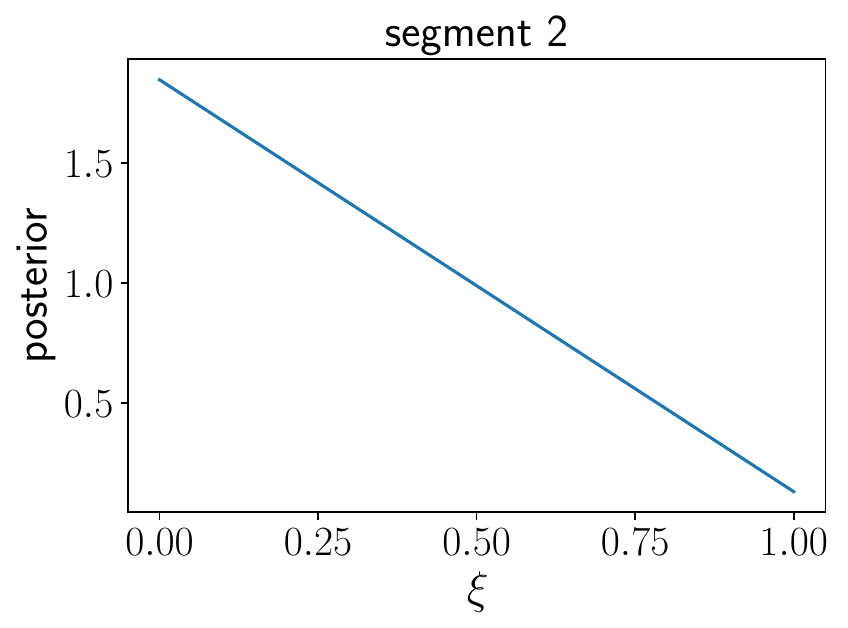}
\includegraphics[width=0.24\textwidth]{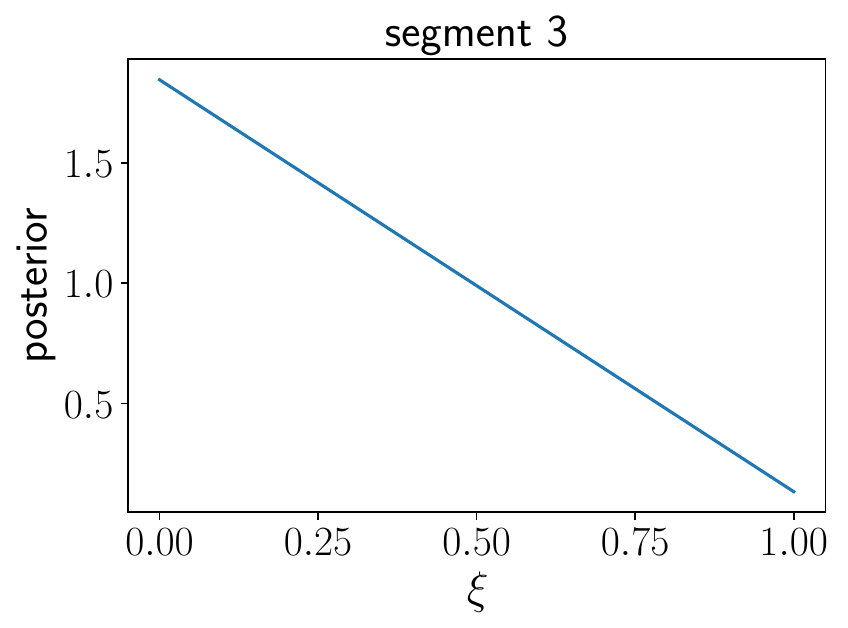}
\includegraphics[width=0.24\textwidth]{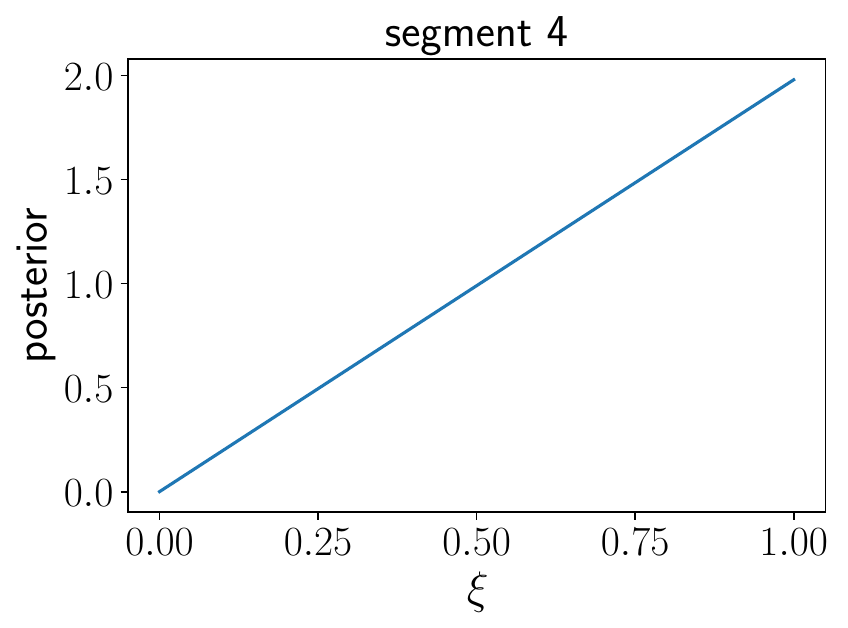}
\includegraphics[width=0.24\textwidth]{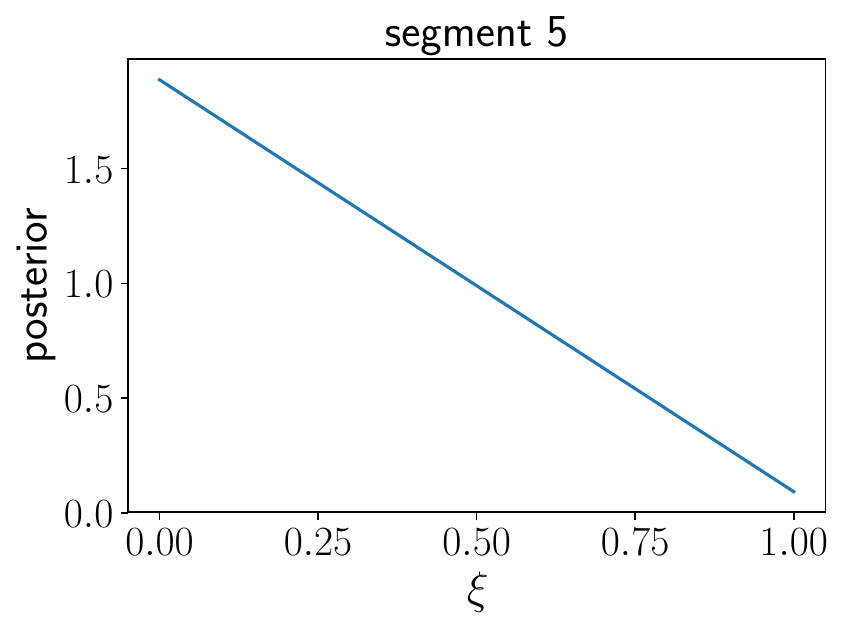}
\includegraphics[width=0.24\textwidth]{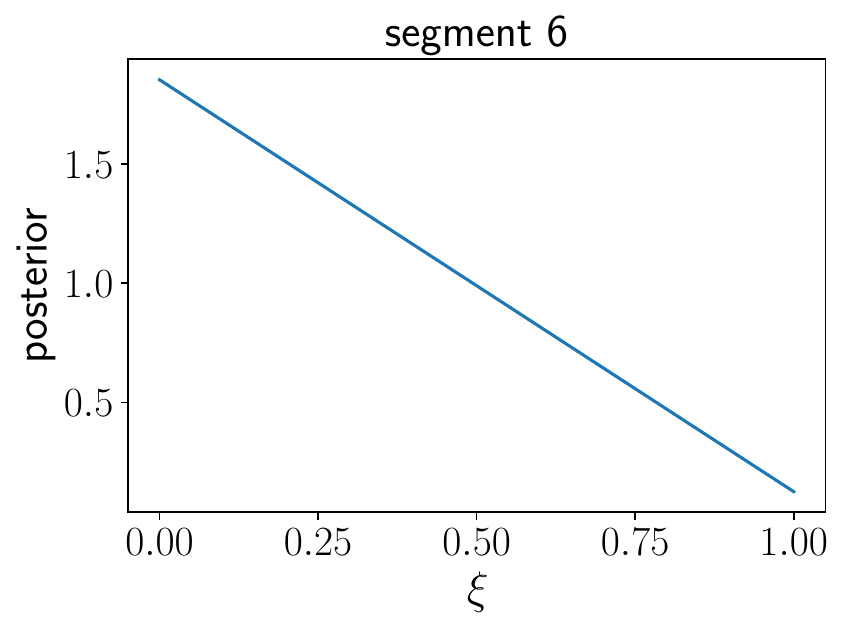}
\includegraphics[width=0.24\textwidth]{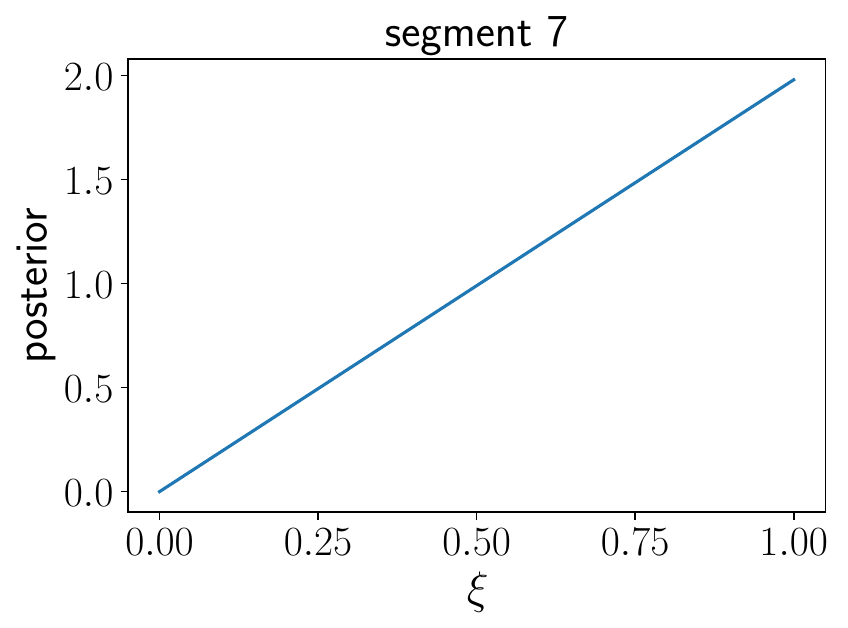}
\includegraphics[width=0.24\textwidth]{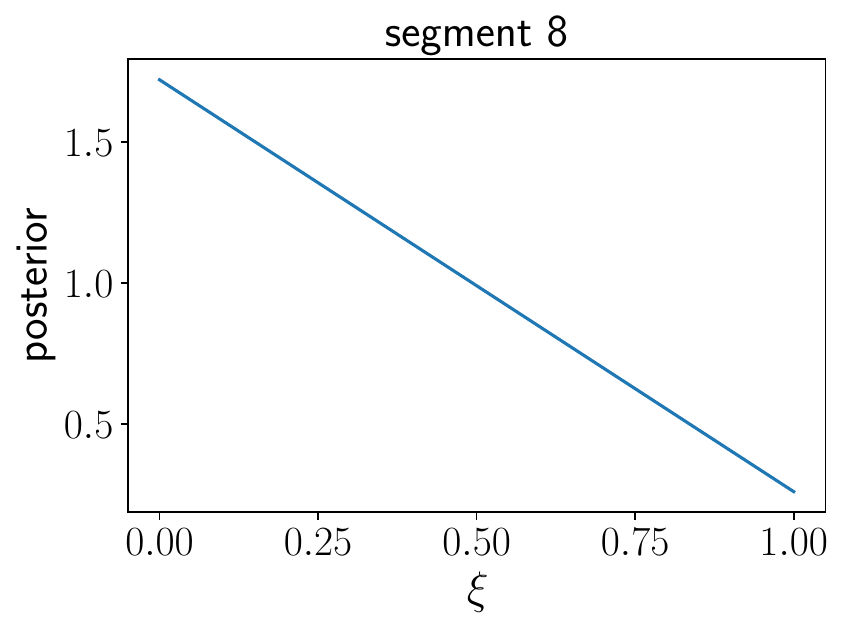}
\includegraphics[width=0.24\textwidth]{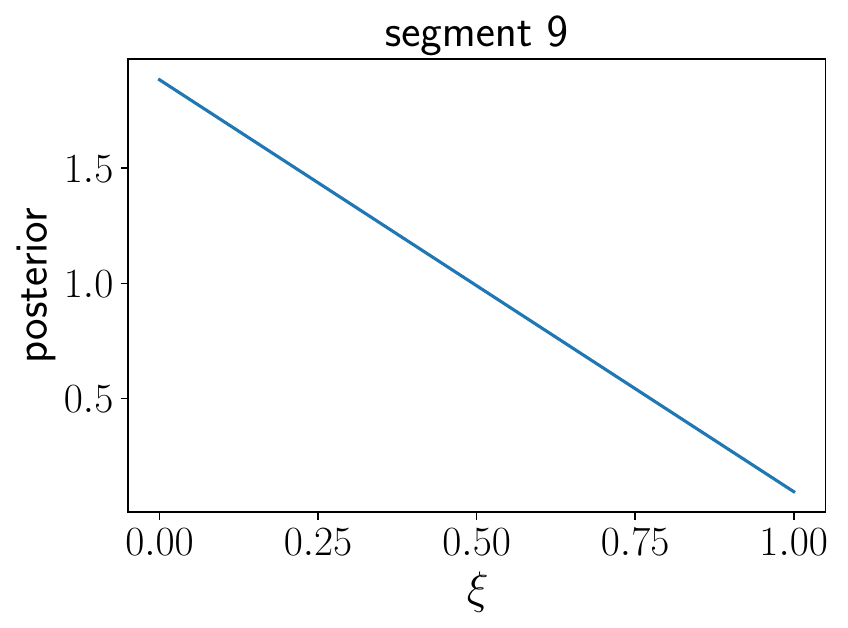}
\includegraphics[width=0.24\textwidth]{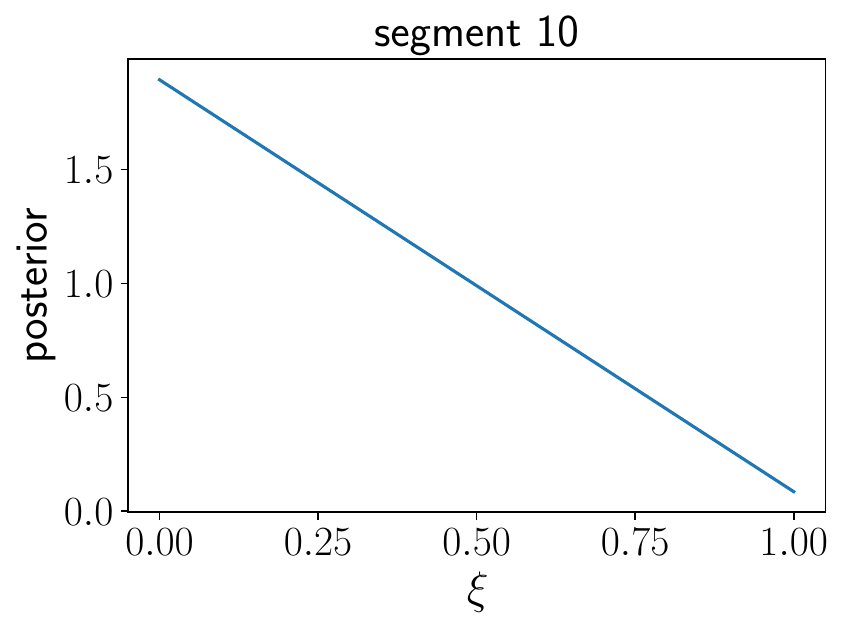}
\includegraphics[width=0.24\textwidth]{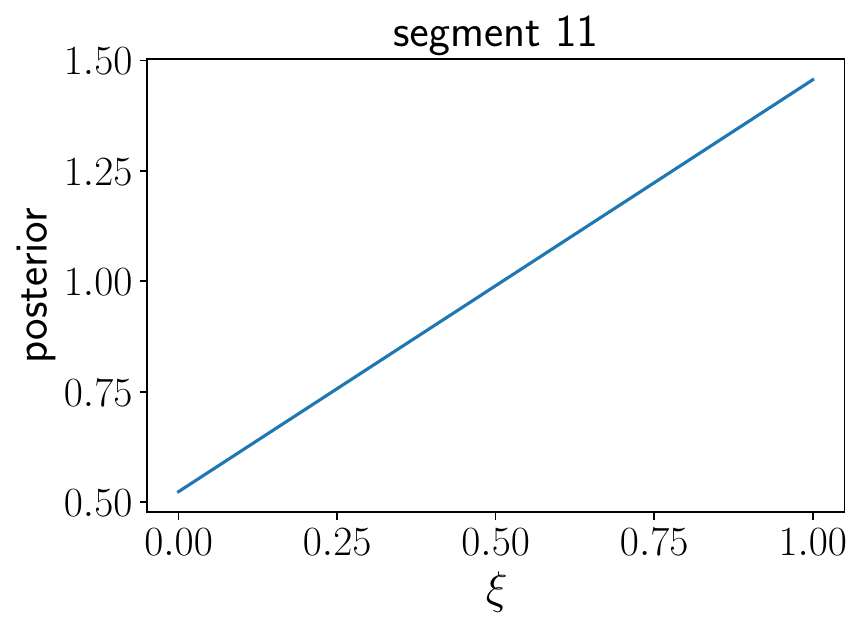}
\includegraphics[width=0.24\textwidth]{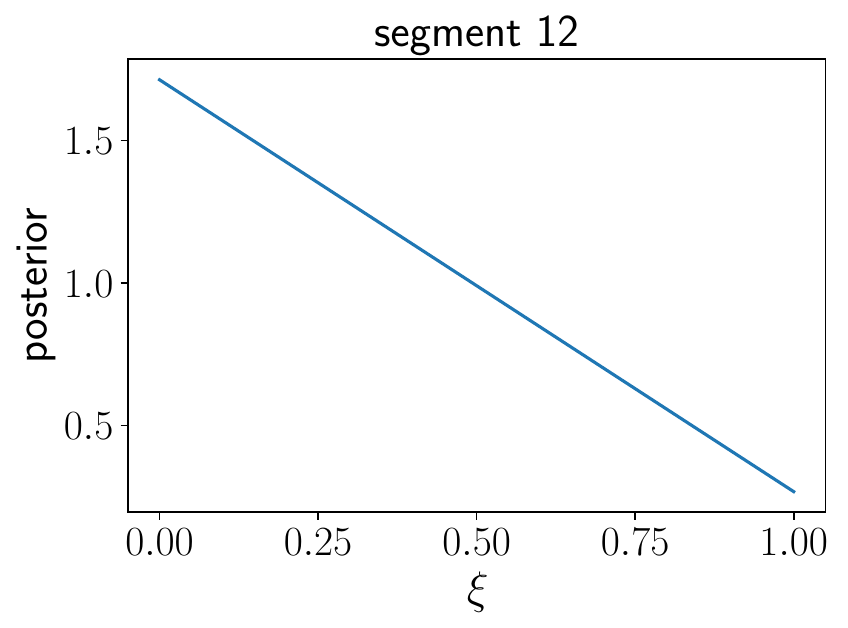}
\includegraphics[width=0.24\textwidth]{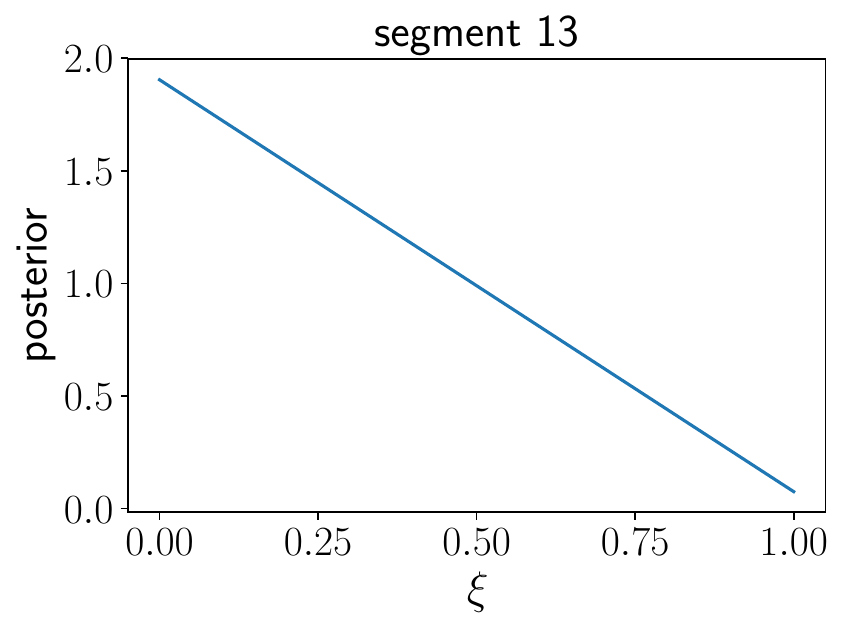}
\includegraphics[width=0.24\textwidth]{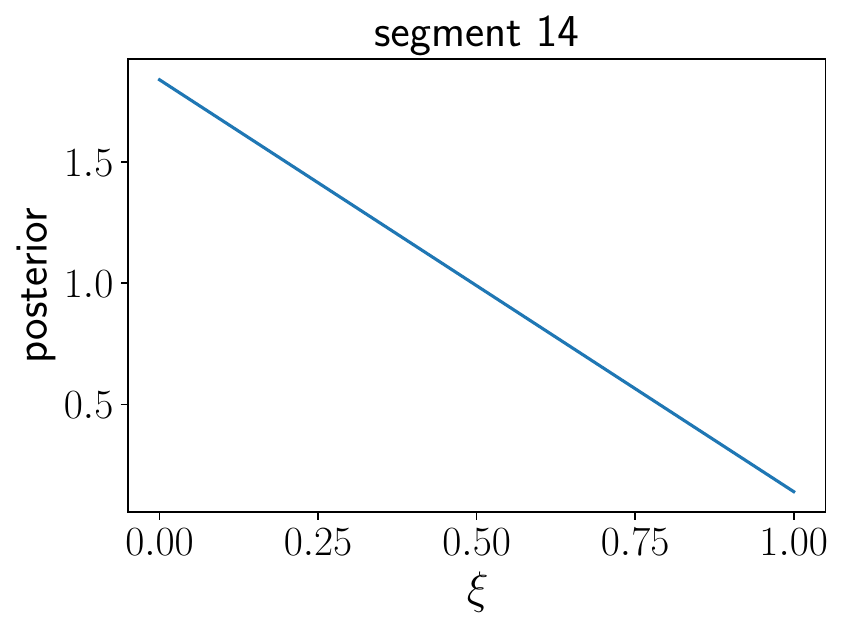}
\includegraphics[width=0.24\textwidth]{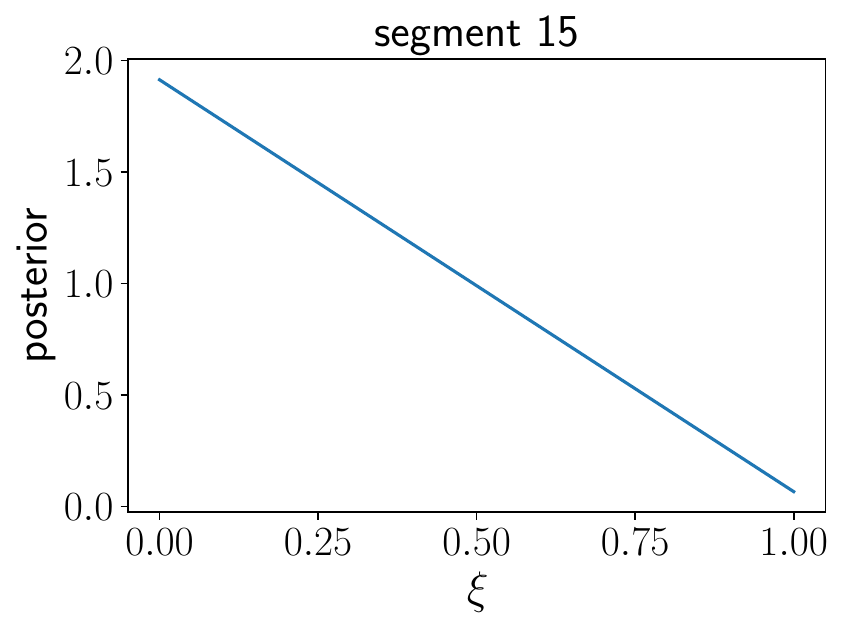}
\caption{Posterior distributions for $\xi$ for the first 16 segments
(first 4~sec) of data.
Since the injected signals were relatively large, the posteriors
having positive (negative) slope correspond to the segments 
having (not having) an injected BBH chirp signal.}
\label{f:posteriors_xi_seg}
\end{center}
\end{figure}
\begin{figure}[htbp!]
\begin{center}
\includegraphics[width=0.24\textwidth]{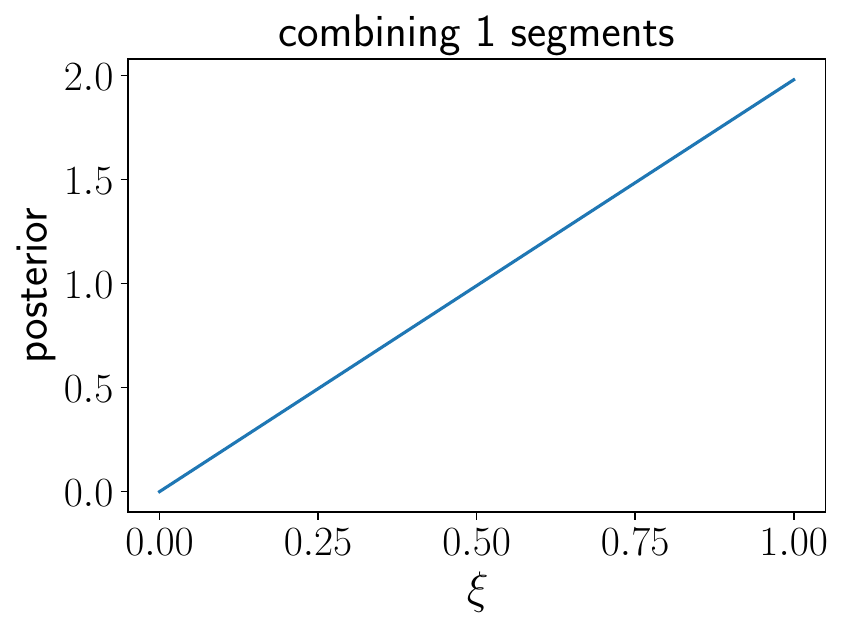}
\includegraphics[width=0.24\textwidth]{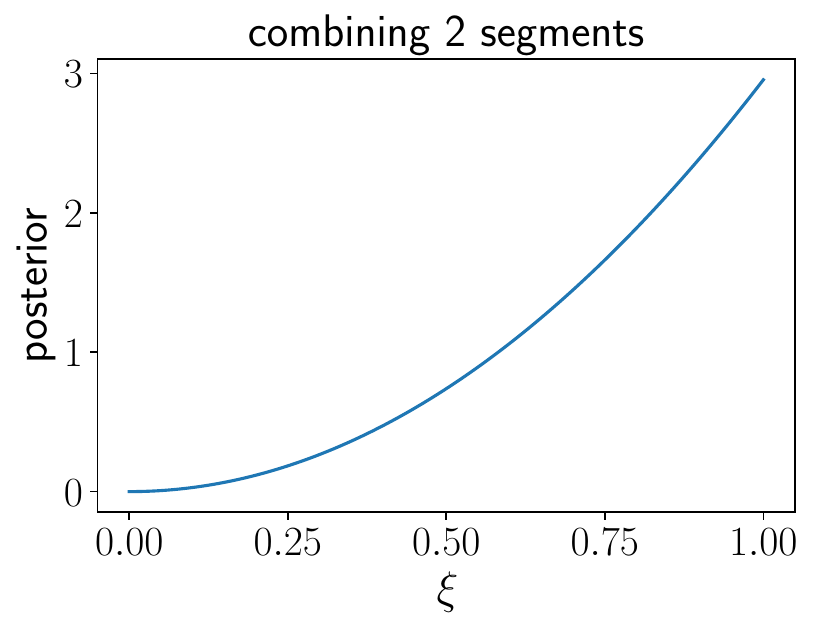}
\includegraphics[width=0.24\textwidth]{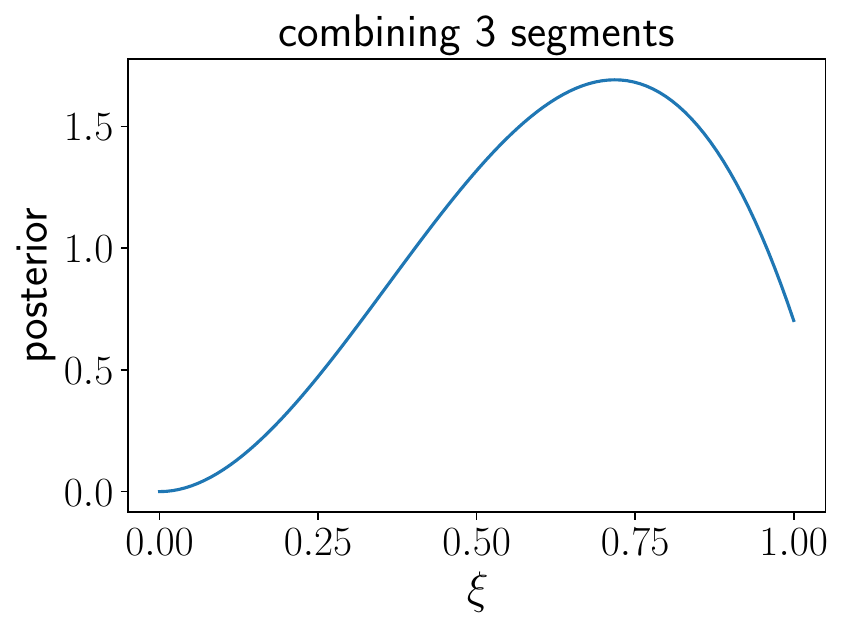}
\includegraphics[width=0.24\textwidth]{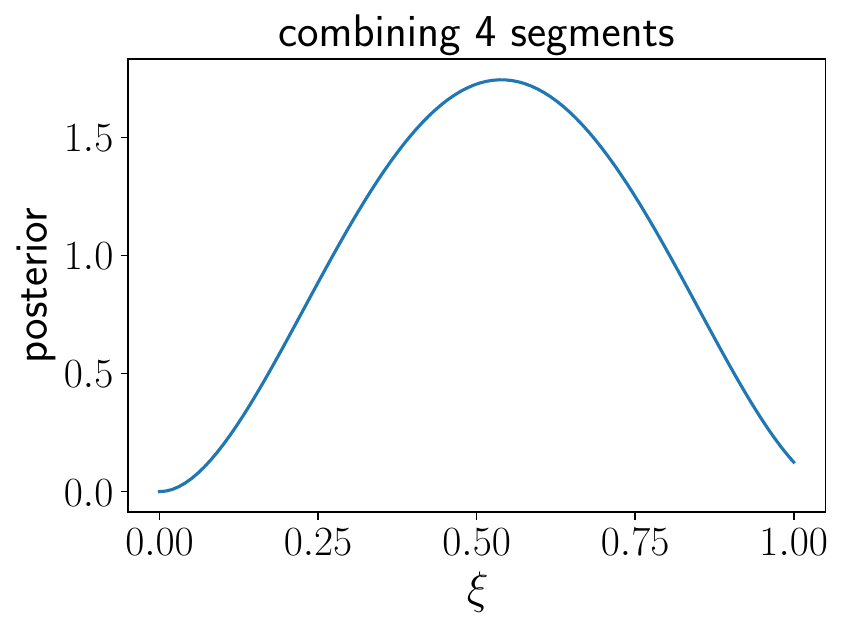}
\includegraphics[width=0.24\textwidth]{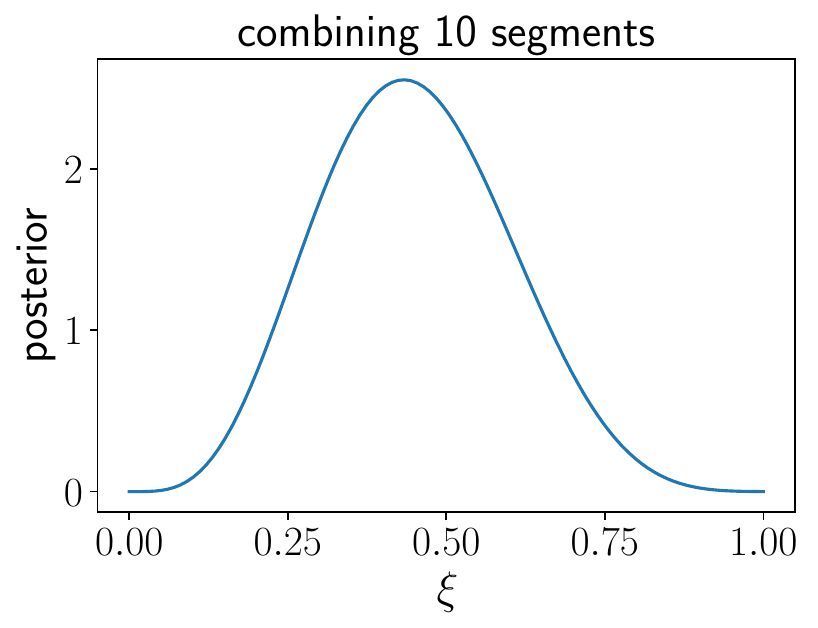}
\includegraphics[width=0.24\textwidth]{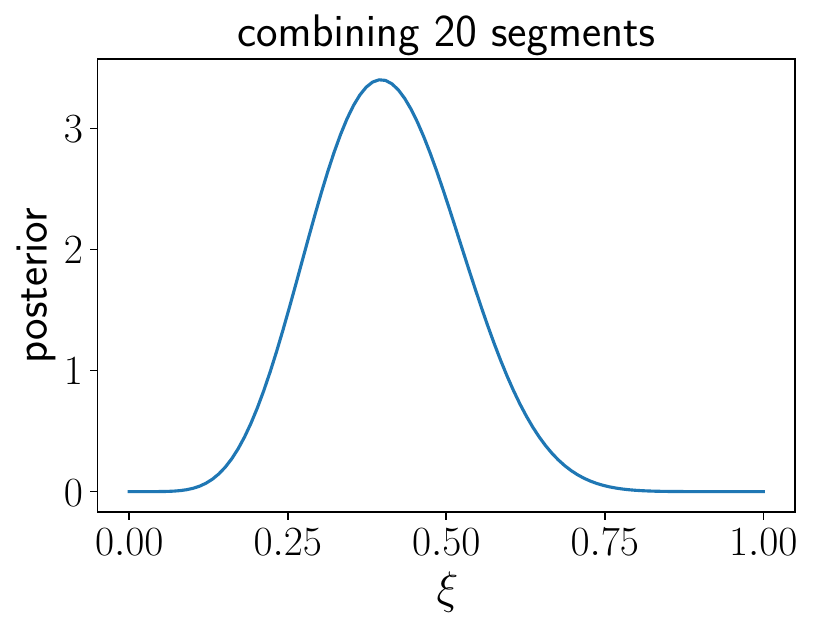}
\includegraphics[width=0.24\textwidth]{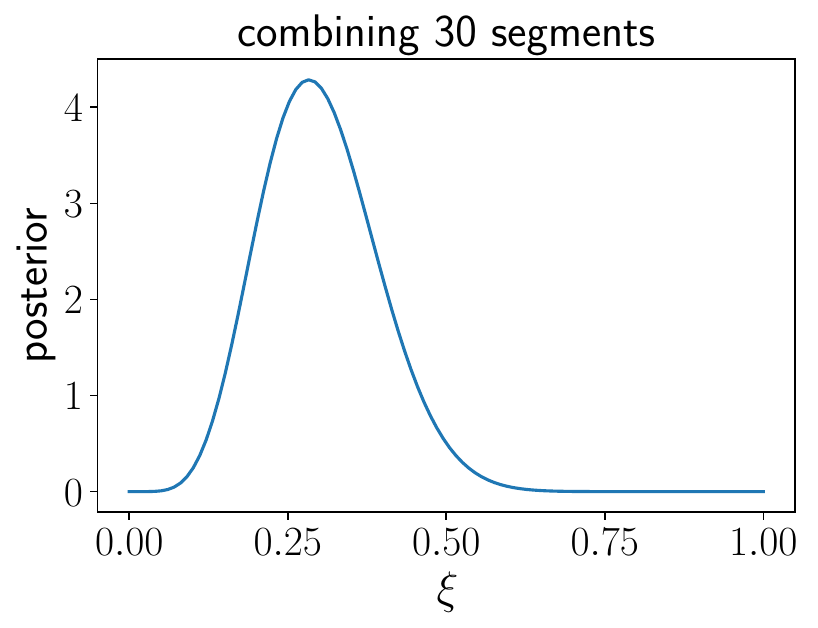}
\includegraphics[width=0.24\textwidth]{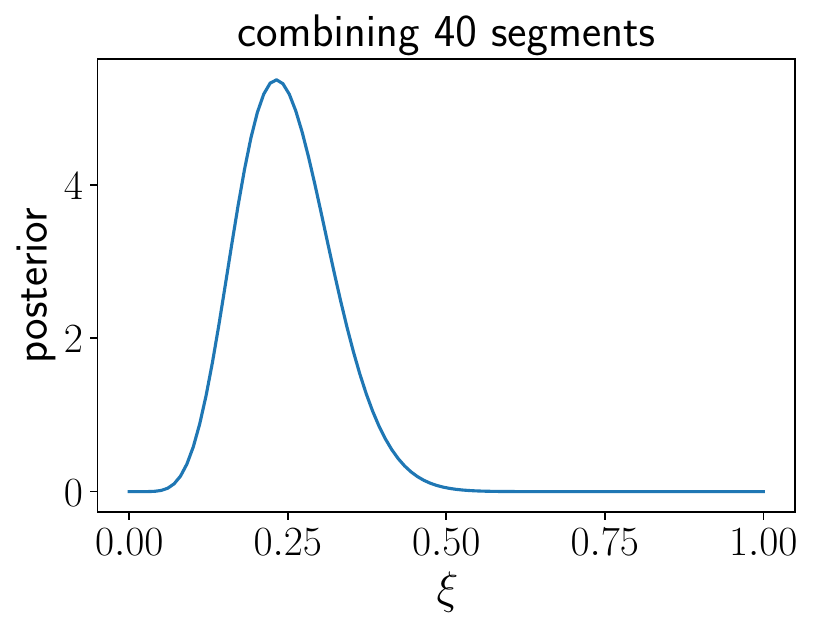}
\caption{Cumulative posterior distributions for $\xi$, obtained by 
combining the likelihood functions for the first $n$ segments of data.
The bottom-rightmost plot is also shown in Figure~\ref{f:posterior_xi}.}
\label{f:posteriors_xi_cum}
\end{center}
\end{figure}

\subsection{Comparison with the standard cross-correlation search}

Of course, one can also use the standard cross-correlation 
method to analyse this simulated data.
Although not optimal for a non-continuous background like 
this, the cross-correlation method performed rather well
for this simulation having 
${\rm SNR}_{\rm CC} = 8.9$.
This should be compared to
${\rm SNR}_{\rm bayes} = 15.3$
for the Bayesian search, where we converted the 
signal+noise to noise-only Bayes factor ${\cal B}_{10}(d)$ to a 
signal-to-noise ratio using \eqref{e:BF-SNR} and \eqref{e:Lambda}.
So the Bayesian search performed better than the 
standard cross-correlation search as expected, roughly
a factor of two in signal-to-noise ratio 
for this particular simulation.

More generally, the Bayesian method performs better 
than the standard cross-correlation method for the 
BBH background because it properly models the 
popcorn-like nature of the BBH merger signals.
It uses a mixture likelihood, allowing for different 
probability distributions for those 
segments that contain a signal and those that do not.
The standard cross-correlation method 
treats all segments on an equal footing, looking for 
excess correlated power which it can ascribe to the signal.
So if most of the segments contain only noise, as is the case
for BBH merger background, the standard cross-correlation
method is going to take longer to build up the 
signal-to-noise ratio needed to claim detection.
The standard cross-correlation method is basically 
measuring the product of the probability 
parameter $\xi$ and the cross-correlated power in an 
individual segment containing a signal.
The Bayesian method, on the other hand, is simply 
measuring $\xi$, and hence all segments---even those 
that contain just noise---are providing useful 
information.

In addition, the Bayesian method incorporates into its
signal model the fact that the background 
is produced by individual
BBH mergers, which are described by {\em deterministic} 
chirp waveforms (so tracks in time-frequency space).
This effectively reduces the time-frequency volume 
over which the Bayesian method has to search.
The standard cross-correlation method, on the other
hand, is very much a broadband search, defined by 
the shape of the power spectral density that one is 
searching for.
By searching over a larger time-frequency volume
than it has to, the standard cross-correlation search
has to contend with correspondingly more noise.

For these two reasons, one expects the proposed 
Bayesian search method to be more efficient than the standard 
cross-correlation method in detecting the popcorn
background produced by BBH merger signals.
Simulations~\cite{Smith-Thrane:2018} have shown a reduction in time to 
detection by roughly a factor of 1000.
This means that
when the advanced LIGO and Virgo detectors are operating 
at design sensitivity, 40 months of observation to detect
the BBH background at the 3-$\sigma$ level using the standard 
cross-correlation 
method (see Figure~\ref{f:BBH-BNS-SNR}) is reduced to 
$\sim\!1$~day using the Bayesian method described here.
So we might be detecting this background much sooner than we
originally thought.

\section{Final thoughts}

The purpose of these lecture notes was to introduce the reader 
to methods used to search for stochatic GW backgrounds.
By its very nature, an introduction is necessarily
incomplete; not all topics can be discussed.
As such, we did not discuss in any detail: (i) searches 
for stochastic backgrounds using the proposed space-based
interferometer LISA, (ii) search methods for anisotropic
or unpolarized backgrounds, and (iii) 
search methods for backgrounds predicted by 
alternative theories of gravity, etc.
The interested reader can find more detail about those topics
in~\cite{Romano-Cornish:2017} and references therein.

Although as of the time of writing these notes (summer 2019)
we have not yet detected a stochastic GW background, we know 
now that such a signal exists, and it's just a matter of time 
before we reach the sensitivity level needed to make a confident 
detection.
For the GWB produced by stellar-mass BBHs throughout 
the universe, we expect to reach this level
by the time the advanced LIGO and Virgo detectors are operating
at design sensitivity (in a couple of years time).
But recall that this time-to-detection estimate is 
conservative, since it assumes that we are using the standard 
cross-correlation method, which is not optimal for 
non-continuous backgrounds.
As mentioned in Section~\ref{s:nonstationary}, there is a good 
chance that we will detect this background earlier using an 
optimal Bayesian method~\cite{Smith-Thrane:2018}
that properly models the non-continuous, 
popcorn-like nature of the BBH mergers.
But then again, searches using pulsar timing arrays might make 
the first detection 
of a stochastic background~\cite{Siemens-et-al:2013,
Rosado:2015epa, Taylor-et-al:2016a}, 
although for a different class of source---inspiraling SMBHs 
in the centers of merging galaxies. 
Either way, it will be an exciting time.

Unlike the detection of the individually resolvable
BBH and BNS mergers that advanced LIGO and Virgo 
has detected, we won't be able to say on 
one particular day that we've 
definitely detected a stochastic GW background.
Rather we will first see evidence for a background at the 
3-$\sigma$ level; and then a year or two later, we will have
evidence at the 4 or 5-$\sigma$ level.
One of the nice things about stochatic backgrounds is that
they are persistent signals (even if popcorn-like), 
so the longer we observe them, the greater our confidence in
detecting them.
And once we've confidently detected a stochastic background, 
the fun part of characterizing what we have seen begins.
As mentioned in Section~\ref{s:different_types}, different 
GW sources will produce different types of backgrounds,
so we will need to tease apart their different contributions.
And as the sensitivity of GW detectors improve, we will be 
able to observe additional structure in a background, e.g., 
anisotropies~\cite{Cusin-et-al:2017, Jenkins-Sakellariadou:2018} 
that were not resolvable before.
Needless to say, there is plenty of work and interesting
science ahead of us.

\section*{Acknowledgements}
\label{s:acknowledgements}

I acknowledge support from National Science Foundation (NSF)
award PHY-1505861; a subward from the University of Wisconsin-Milwaukee
for the NSF NANOGrav Physics Frontier Center (NSF PFC-1430284); 
a subward from the University of Minnesota for NASA grant 80NSSC19K0318; 
and start-up funds from Texas Tech University.
I am also grateful to the organizers, fellow lecturers, and of course
the students at the 2018 Les Houches Summer School on Gravitational 
Waves for providing an extremely stimulating environment in which to 
give these lectures; and I thank the Observatoire Cote D'Azur in Nice, 
France for its hospitality, where most of these lecture notes were written.
Finally, special thanks goes out to  Alex Jenkins for carefully 
reading through several early drafts of these lecture notes.
He provided very helpful feedback and comments that have improved 
the presentation of the material.
(This document has been assigned LIGO Document Control Center number
LIGO-P1900196.) 

\appendix
\section{Exercises}
\label{s:exercises}

\subsection{Rate estimate of stellar-mass binary black hole mergers}
\label{exer:1}

Estimate the total rate (number of events per time) of 
stellar-mass binary black hole mergers throughout the universe 
by multiplying LIGO's O1 local rate 
estimate $R_0 \sim 10$~-~$200~{\rm Gpc}^{-3}\,{\rm yr}^{-1}$ by 
the comoving volume out to some large redshift, e.g., $z= 10$.
(For this calculation you can ignore any dependence of the 
rate density with redshift.)
You should find a merger rate of $\sim\!1$~per minute to a few 
per hour.
{\em Hint}: You will need to do numerically evaluate the
following integral for proper distance today as a function 
of source redshift:
\be
d_0(z) = \frac{c}{H_0}\int_0^z\frac{\D z'}{E(z')}\,,
\qquad
E(z)\equiv \sqrt{\Omega_{\rm m}(1+z)^3 + \Omega_\Lambda}\,,
\ee
with 
\be
\Omega_{\rm m}=0.31\,,
\qquad
\Omega_\Lambda=0.69\,,
\qquad
H_0 = 68~{\rm km}\,{\rm s}^{-1}\, {\rm Mpc}^{-1}\,.
\ee
Doing that integral, you should find what's shown in
Figure~\ref{f:d0vsz}, which you can then evaluate at
$z=10$ to convert $R_0$ (number of events per 
comoving volume per time) to total rate (number of 
events per time) for sources out to redshift $z=10$.
\begin{figure}[htbp!]
\begin{center}
\includegraphics[width=0.5\textwidth]{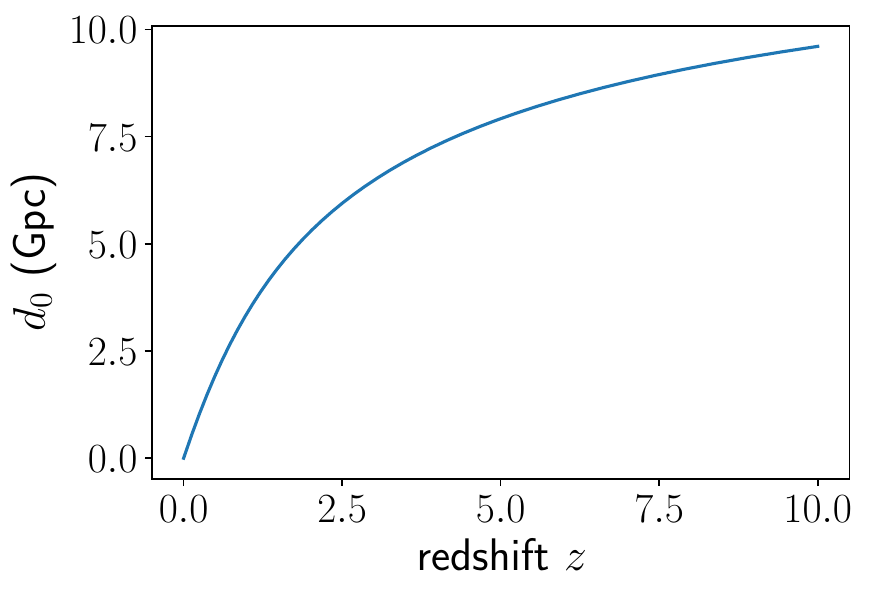}
\caption{}
\label{f:d0vsz}
\end{center}
\end{figure}

\subsection{Relating $S_h(f)$ and $\Omega_{\rm gw}(f)$}
\label{exer:2}

Derive the relationship 
\be
S_h(f) = \frac{3 H_0^2}{2\pi^2}\frac{\Omega_{\rm gw}(f)}{f^3}
\ee
between the strain power spectral density $S_h(f)$ and the 
dimensionless fractional energy density spectrum $\Omega_{\rm gw}(f)$.
({\em Hint}: You will need to use the various definitions of these
quantities and also 
\be
\rho_{\rm gw} =\frac{c^2}{32\pi G}\langle \dot h_{ab}(t,\vec x)\dot h^{ab}(t,\vec x)\rangle\,,
\ee
which expresses the energy-density in gravitational-waves to 
the metric perturbations $h_{ab}(t,\vec x)$.)

\subsection{Cosmology and the ``Phinney formula" for astrophysical backgrounds}
\label{exer:3}

(a) Using the Friedmann equation
\be
\left(\frac{\dot a}{a}\right)^2
=H_0^2\left(\frac{\Omega_{\rm m}}{a^{3}} + \Omega_\Lambda\right)
\ee
for a spatially-flat FRW spacetime with matter and 
cosmological constant, and the relationship 
\be
1+z = \frac{1}{a(t)}\,,
\qquad a(t_0)\equiv 1\quad(t_0\equiv {\rm today})\,,
\ee
between redshift $z$ and scale factor $a(t)$,
derive 
\be
\frac{\D t}{\D z} =-\frac{1}{(1+z)H_0 E(z)}\,,
\qquad
E(z) = \sqrt{\Omega_{\rm m}(1+z)^3 + \Omega_\Lambda}\,.
\ee
(b) Using this result for $\D t/\D z$, show that 
\be
\Omega_{\rm gw}(f)= \frac{f}{\rho_{\rm c}H_0}
\int_0^\infty \D z\>R(z)\,\frac{1}{(1+z)E(z)}
\left(\frac{\D E_{\rm gw}}{\D f_{\rm s}}\right)\bigg|_{f_{\rm s}=f(1+z)}
\ee
in terms of the rate density $R(z)$ as measured in 
the source frame 
(number of events per comoving volume per time interval
in the source frame).
({\em Hint}: The expression for $\D t/\D z$ from part
(a) will allow 
you to go from the ``Phinney formula" for
$\Omega_{\rm gw}(f)$ written in terms of the number 
density $n(z)$,
\be
\Omega_{\rm gw}(f)= \frac{1}{\rho_c}\int_0^\infty \D z\>
n(z)\,\frac{1}{1+z}\left(f_{\rm s}\,
\frac{\D E_{\rm gw}}{\D f_{\rm s}}\right)\bigg|_{f_{\rm s}=f(1+z)}\,,
\ee
to one in terms of the rate density 
$R(z)$, where $n(z)\,\D z=R(z)\,|\D t|_{t=t(z)}$.
Note: Both of the above expressions for $\Omega_{\rm gw}(f)$
assume that there is only one type of source, described by 
some set of average source parameters.  
If there is more than one type of source, one must sum
the contributions of each source to $\Omega_{\rm gw}(f)$.)

\subsection{Optimal filtering for the cross-correlation statistic}
\label{exer:4}

Verify the form 
\be
\tilde Q(f)\propto \frac{\Gamma_{12}(f)H(f)}
{P_1(f)P_2(f)}\,,
\ee
of the optimal filter function in the weak-signal limit,
where $H(f)$ is the assumed spectral shape of the 
gravitational-wave background,
$\Gamma_{12}(f)$ is the overlap function, and $P_1(f)$, $P_2(f)$ 
are the power spectral densities of the outputs of the 
two detectors (which are approximately equal to 
$P_{n_1}(f)$, $P_{n_2}(f)$, respectively).
Recall that the optimal filter $\tilde Q(f)$ maximizes
the signal-to-noise ratio of the cross-correlation 
statistic.
({\em Hint}: Introduce an inner product on the space of
functions of frequency $A(f)$, $B(f)$:
\be
(A,B)\equiv\int df A(f) B^*(f) P_1(f) P_2(f)\,.
\ee
This inner product
has all of the properties of the familiar dot product
of vectors in 3-dimensional space.
The signal-to-noise ratio of the cross-correlation
statistic can be written in terms of this inner product.)

\subsection{Maximum-likelihood estimators for single and multiple
parameters}
\label{exer:5}

(a) Show that the maximum-likelihood estimator $\hat a$ of 
the single parameter $a$ in the likelihood function
\be
p(d|a, \sigma) \propto
\exp\left[-\frac{1}{2}\sum_{i=1}^N \frac{(d_i-a)^2}{\sigma_i^2}\right]
\ee
is given by the noise-weighted average
\be
\hat a={\sum_i \frac{d_i}{\sigma_i^2}}\bigg/{\sum_j \frac{1}{\sigma_j^2}}\,.
\ee
(b) Extend the previous calculation to the likelihood
\be
p(d|A, C) \propto
\exp\left[-\frac{1}{2}(d-MA)^\dagger C^{-1} (d-MA)\right]\,,
\ee
where $A\equiv A_\alpha$ is a vector of parameters,
$C\equiv C_{ij}$ is the noise covariance matrix, and 
$M\equiv M_{i\alpha}$ is the response matrix mapping 
$A_\alpha$ to data samples, $MA\equiv \sum_\alpha M_{i\alpha}A_\alpha$.
For this more general case you should find:
\be
\hat A = F^{-1} X\,,
\ee
where
\be
F \equiv M^\dagger C^{-1} M\,,\qquad
X \equiv M^\dagger C^{-1} d\,.
\ee
In general, the matrix $F$ (called the {\em Fisher} matrix)
is not invertible, so some sort of regularization is needed
to do the matrix inversion.

\subsection{Timing-residual response for a 1-arm, 1-way detector}
\label{exer:6}

Derive the timing residual reponse function
\be
R^A(f,\hat k) = 
\frac{1}{2}u^a u^b e^A_{ab}(\hat k)
\frac{1}{i2\pi f}
\frac{1}{1-\hat k\cdot \hat u}
\left[1-e^{-\frac{i2\pi fL}{c}(1-\hat k\cdot\hat u)}\right]
\ee
for a single-link (i.e., a one-arm, one-way detector like 
that for pulsar timing).
Here $\hat u$ is the direction of propagation of the
electromagnetic pulse, and $\hat k$ is the direction of
propagation of the GW (the direction to the GW source is
$\hat n\equiv -\hat k$, and the direction to the pulsar is
$\hat p\equiv -\hat u$).
The origin of coordinates is taken to be at the position 
of the detector.

\subsection{Overlap function for colocated electric dipole antennae}
\label{exer:7}

Show that the overlap function for a pair of (short)
colocated electric dipole antennae pointing in directions 
$\hat u_1$ and $\hat u_2$ (Figure~\ref{f:dipole-orf})
is given by 
\be
\Gamma_{12} 
\propto
\hat u_1\cdot\hat u_2 
\equiv\cos\zeta
\ee
for the case of an unpolarized, isotropic electromagnetic field.
({\em Hint}: ``short" means that the phase of the electric 
field can be taken to be constant over of the lengths of 
the dipole antennae, 
so that the reponse of antenna $I=1,2$ to the field is
given by $r_I(t)=\hat u_I\cdot\vec E(t, \vec x_0)$, where
$\vec x_0$ is the common location of the two antenna.)
 
\subsection{Maximum-likelihood estimators for the standard 
cross-correlation statistic}
\label{exer:8}

Verify that 
\be
\hat C_{11}\equiv \frac{1}{N}\sum_{i=1}^N d_{1i}^2\,,
\qquad
\hat C_{22}\equiv \frac{1}{N}\sum_{i=1}^N d_{2i}^2\,,
\qquad
\hat C_{12}\equiv \frac{1}{N}\sum_{i=1}^N d_{1i} d_{2i}
\ee
are maximum-likelihood estimators of 
\be
S_1\equiv S_{n_1}+S_h\,,
\quad
S_2\equiv S_{n_2}+S_h\,,
\quad
S_h\,,
\ee
for the case of $N$ samples of a white GWB in uncorrelated
white detector noise, for a pair of colocated and coaligned 
detectors.
Recall that the likelihood function is
\be
p(d|S_{n_1}, S_{n_2}, S_h) =\frac{1}{\sqrt{{\rm det}(2\pi C)}}
\exp\left[-\frac{1}{2}d^T C^{-1} d\right]\,,
\ee
where
\be
C
= \left[
\begin{array}{cc}
(S_{n_1} +S_h)\,\unit_{N\times N} & S_h\,\unit_{N\times N}
\\
S_h\,\unit_{N\times N} & (S_{n_2} +S_h)\,\unit_{N\times N}
\\
\end{array}
\right]
\label{e:C_marginalized}
\ee
and 
\be
d^T C^{-1} d
\equiv \sum_{I,J=1}^2\sum_{i,j=1}^N
d_{Ii} \left(C^{-1}\right)_{Ii,Jj} d_{Jj}\,.
\label{e:argexp}
\ee

\subsection{Derivation of the maximum-likelihood ratio detection statistic}
\label{exer:9}

Verify that twice the log of the maximum-likelihood
ratio for the standard stochastic likelihood function
goes like the square of the (power) signal-to-noise ratio,
\be
2\ln \Lambda_{\rm ML}(d) \simeq
\frac{\hat C_{12}^2}{\hat C_{11}\hat C_{22}/N}\,,
\ee
in the weak-signal approximation.
({\em Hint:} For simplicity, do the calculation in the context 
of $N$ samples of a white GWB in uncorrelated 
white detector noise, for a pair of colocated and coaligned
detectors, using the results of Exercise~\ref{exer:8}.)

\subsection{Standard likelihood marginalizing over 
stochastic signal prior}
\label{exer:10}

Derive the standard form of the likelihood function
for stochastic background searches 
\be
p(d|S_{n_1}, S_{n_2}, S_h)
=\frac{1}{\sqrt{{\rm det}(2\pi C)}}
\exp\left[-\frac{1}{2} \sum_{I,J=1}^2 d_I \left(C^{-1}\right)_{IJ} d_J\right]\,,
\ee
where
\be
C\equiv \left[
\begin{array}{cc}
S_{n_1}+S_h & S_h\\
S_h & S_{n_2} + S_h\\
\end{array}
\right]\,,
\ee
by marginalizing 
\be
p_n(d- h|S_{n_1},S_{n_2}) =
\frac{1}{2\pi\sqrt{S_{n_1}S_{n_2}}}
\exp\left[-\frac{1}{2}\left\{
\frac{(d_1- h)^2}{S_{n_1}} + \frac{(d_2- h)^2}{S_{n_2}}
\right\}\right]
\ee
over the signal samples $h$ for the {\em stochastic} signal prior 
\be
p(h|S_h) = \frac{1}{\sqrt{2\pi S_h}}\exp\left[
-\frac{1}{2}\frac{h^2}{S_h}\right]\,.
\ee
In other words, show that
\be
p(d|S_{n_1}, S_{n_2}, S_h) 
=\int_{-\infty}^\infty \D h\>
p_n(d-h|S_{n_1}, S_{n_2}) p(h|S_h)\,.
\ee
({\em Hint}: You'll have to complete the square in the argument
of the exponential in the marginalization integral.)

\newpage
\bibliography{refs}

\end{document}